\documentclass[twoside,twocolumn,9pt]{article}
\usepackage{extsizes}
\usepackage[super,sort&compress,comma]{natbib} 
\usepackage[version=4]{mhchem}
\usepackage[left=1.5cm, right=1.5cm, top=1.785cm, bottom=2.0cm]{geometry}
\usepackage{balance}
\usepackage{mathptmx}
\usepackage{tablefootnote}
\usepackage{sectsty}
\usepackage[flushleft]{threeparttable}
\usepackage{stfloats}
\usepackage{graphicx} 
\usepackage{booktabs}
\usepackage{subcaption}
\usepackage{caption}
\usepackage{longtable}
\usepackage{lastpage}
\usepackage{soul}
\usepackage{siunitx}
\DeclareSIUnit{\molecule}{molecule}
\DeclareSIUnit{\cal}{cal}
\usepackage[format=plain,justification=justified,singlelinecheck=false,font={stretch=1.125,small,sf},labelfont=bf,labelsep=space]{caption}
\usepackage{float}
\usepackage{fancyhdr}
\usepackage{fnpos}
\usepackage[english]{babel}
\addto{\captionsenglish}{%
  
}
\usepackage{array}
\usepackage{droidsans}
\usepackage{charter}
\usepackage[T1]{fontenc}
\usepackage[usenames,dvipsnames]{xcolor}
\usepackage{setspace}
\usepackage[compact]{titlesec}
\usepackage{xr-hyper}
\usepackage{hyperref}

\usepackage{epstopdf}%This line makes .eps figures into .pdf - please comment out if not required.

\definecolor{cream}{RGB}{222,217,201}

\begin{document}

\pagestyle{fancy}
\thispagestyle{plain}
\fancypagestyle{plain}{
%%%HEADER%%%
\renewcommand{\headrulewidth}{0pt}
}
%%%END OF HEADER%%%

%%%PAGE SETUP - Please do not change any commands within this section%%%
\makeFNbottom
\makeatletter
\renewcommand\LARGE{\@setfontsize\LARGE{15pt}{17}}
\renewcommand\Large{\@setfontsize\Large{12pt}{14}}
\renewcommand\large{\@setfontsize\large{10pt}{12}}
\renewcommand\footnotesize{\@setfontsize\footnotesize{7pt}{10}}
\makeatother

\renewcommand{\thefootnote}{\fnsymbol{footnote}}
\renewcommand\footnoterule{\vspace*{1pt}% 
\color{cream}\hrule width 3.5in height 0.4pt \color{black}\vspace*{5pt}} 
\setcounter{secnumdepth}{5}

\makeatletter 
\renewcommand\@biblabel[1]{#1}            
\renewcommand\@makefntext[1]% 
{\noindent\makebox[0pt][r]{\@thefnmark\,}#1}
\makeatother 
\renewcommand{\figurename}{\small{Fig.}~}
\sectionfont{\sffamily\Large}
\subsectionfont{\normalsize}
\subsubsectionfont{\bf}
\setstretch{1.125} %In particular, please do not alter this line.
\setlength{\skip\footins}{0.8cm}
\setlength{\footnotesep}{0.25cm}
\setlength{\jot}{10pt}
\titlespacing*{\section}{0pt}{4pt}{4pt}
\titlespacing*{\subsection}{0pt}{15pt}{1pt}
%%%END OF PAGE SETUP%%%

%%%FOOTER%%%
\fancyfoot{}
\fancyfoot[LO,RE]{\vspace{-7.1pt}\includegraphics[height=9pt]{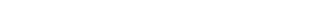}}
\fancyfoot[CO]{\vspace{-7.1pt}\hspace{11.9cm}\includegraphics{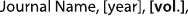}}
\fancyfoot[CE]{\vspace{-7.2pt}\hspace{-13.2cm}\includegraphics{head_foot/RF}}
\fancyfoot[RO]{\footnotesize{\sffamily{1--\pageref{LastPage} ~\textbar  \hspace{2pt}\thepage}}}
\fancyfoot[LE]{\footnotesize{\sffamily{\thepage~\textbar\hspace{4.65cm} 1--\pageref{LastPage}}}}
\fancyhead{}
\renewcommand{\headrulewidth}{0pt} 
\renewcommand{\footrulewidth}{0pt}
\setlength{\arrayrulewidth}{1pt}
\setlength{\columnsep}{6.5mm}
\setlength\bibsep{1pt}
%%%END OF FOOTER%%%

%%%FIGURE SETUP - please do not change any commands within this section%%%
\makeatletter 
\newlength{\figrulesep} 
\setlength{\figrulesep}{0.5\textfloatsep} 

\newcommand{\topfigrule}{\vspace*{-1pt}% 
\noindent{\color{cream}\rule[-\figrulesep]{\columnwidth}{1.5pt}} }

\newcommand{\botfigrule}{\vspace*{-2pt}% 
\noindent{\color{cream}\rule[\figrulesep]{\columnwidth}{1.5pt}} }

\newcommand{\dblfigrule}{\vspace*{-1pt}% 
\noindent{\color{cream}\rule[-\figrulesep]{\textwidth}{1.5pt}} }

\makeatother
%%%END OF FIGURE SETUP%%%

%%%TITLE, AUTHORS AND ABSTRACT%%%
\twocolumn[
  \begin{@twocolumnfalse}
{\includegraphics[height=30pt]{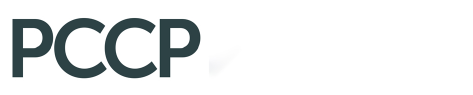}\hfill\raisebox{0pt}[0pt][0pt]{\includegraphics[height=55pt]{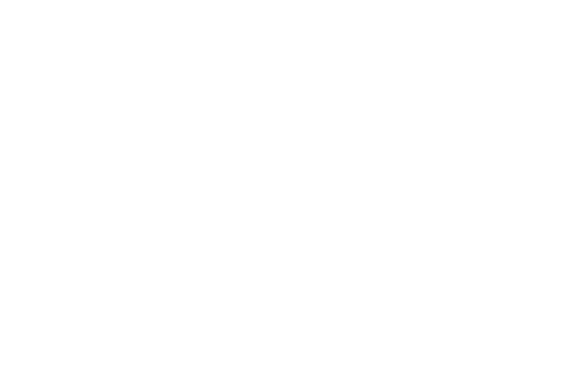}}\\[1ex]
\includegraphics[width=18.5cm]{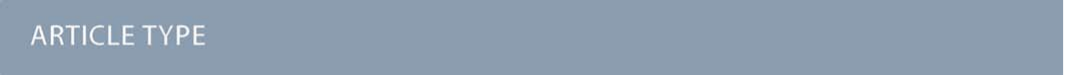}}\par
\vspace{1em}
\sffamily
\begin{tabular}{m{4.5cm} p{13.5cm} }

\includegraphics{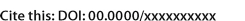} & \noindent\LARGE{\textbf{Ab Initio Conformational Analysis of $\alpha$/$\beta$-D-Xylopyranose at Pyrolysis Conditions$^\dag$}} \\%Article title goes here instead of the text "This is the title"
\vspace{0.3cm} & \vspace{0.3cm} \\

 & \noindent\large{Bernardo Ballotta,$^{\ast}$\textit{$^{ab}$} Jacopo Lupi,\textit{$^{ab}$} Leandro Ayarde-Henríquez,\textit{$^{ab}$} and Stephen Dooley \textit{$^{ab}$}} \\
 %Author names go here instead of "Full name", etc.

\includegraphics{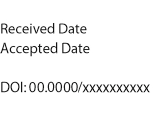} & \noindent\normalsize{Xylopyranose is the principal monosaccharide unit of hemicellulose, one of the three major biopolymers of lignocellulosic biomass. Understanding its decomposition mechanism is increasingly relevant for thermochemical biorefinery research such as pyrolysis. Significant efforts have been made to study its chemical and structural properties using both computational and experimental methods. However, due to its high structural flexibility and numerous hydroxyl groups, various metastable conformers arise.
In this work, we performed a computational exploration of the conformational space of both anomeric forms, $\alpha$ and $\beta$, of D-xylopyranose using the semi-empirical GFN2-xTB method in conjunction with metadynamics and density functional theory simulations for structural optimization and vibrational analysis. Xylopyranose conformers free energy and enthalpy variations are analyzed across temperatures typical of fast biomass pyrolysis (298-1068 K), with the Boltzmann population distribution of the most populated conformers determined. This study provides a detailed computational analysis of the conformational space and thermochemistry of xylopyranose. Additionally, 44 and 59 conformers of the $\alpha$ and $\beta$ anomers were found, for both of which a selection of 10 conformers based on Boltzmann population distribution analysis is performed to reduce the conformational space for \textit{ab initio} studies of the pyrolysis reaction kinetics.}

\end{tabular}

 \end{@twocolumnfalse} \vspace{0.6cm}

  ]
%%%END OF TITLE, AUTHORS AND ABSTRACT%%%

%%%FONT SETUP - please do not change any commands within this section
\renewcommand*\rmdefault{bch}\normalfont\upshape
\rmfamily
\section*{}
\vspace{-1cm}

%%%FOOTNOTES%%%

\footnotetext{\textit{$^{a}$~School of Physics, Trinity College Dublin, Dublin 2, Ireland; E-mail: bernardo.ballotta@tcd.ie}}
\footnotetext{\textit{$^{b}$~AMBER, Advance Materials and BioEngineering Research Centre, Dublin 2, Ireland. }}

%Please use \dag to cite the ESI in the main text of the article.
%If you article does not have ESI please remove the the \dag symbol from the title and the footnotetext below.
\footnotetext{\dag~Electronic supplementary information (ESI) available. See DOI: 10.1039/cXCP00000x/}
%additional addresses can be cited as above using the lower-case letters, c, d, e... If all authors are from the same address, no letter is required

%%%END OF FOOTNOTES%%%

%%%MAIN TEXT%%%%

\section{Introduction}

\begin{figure}[ht]
\centering
  \includegraphics[height=6cm]{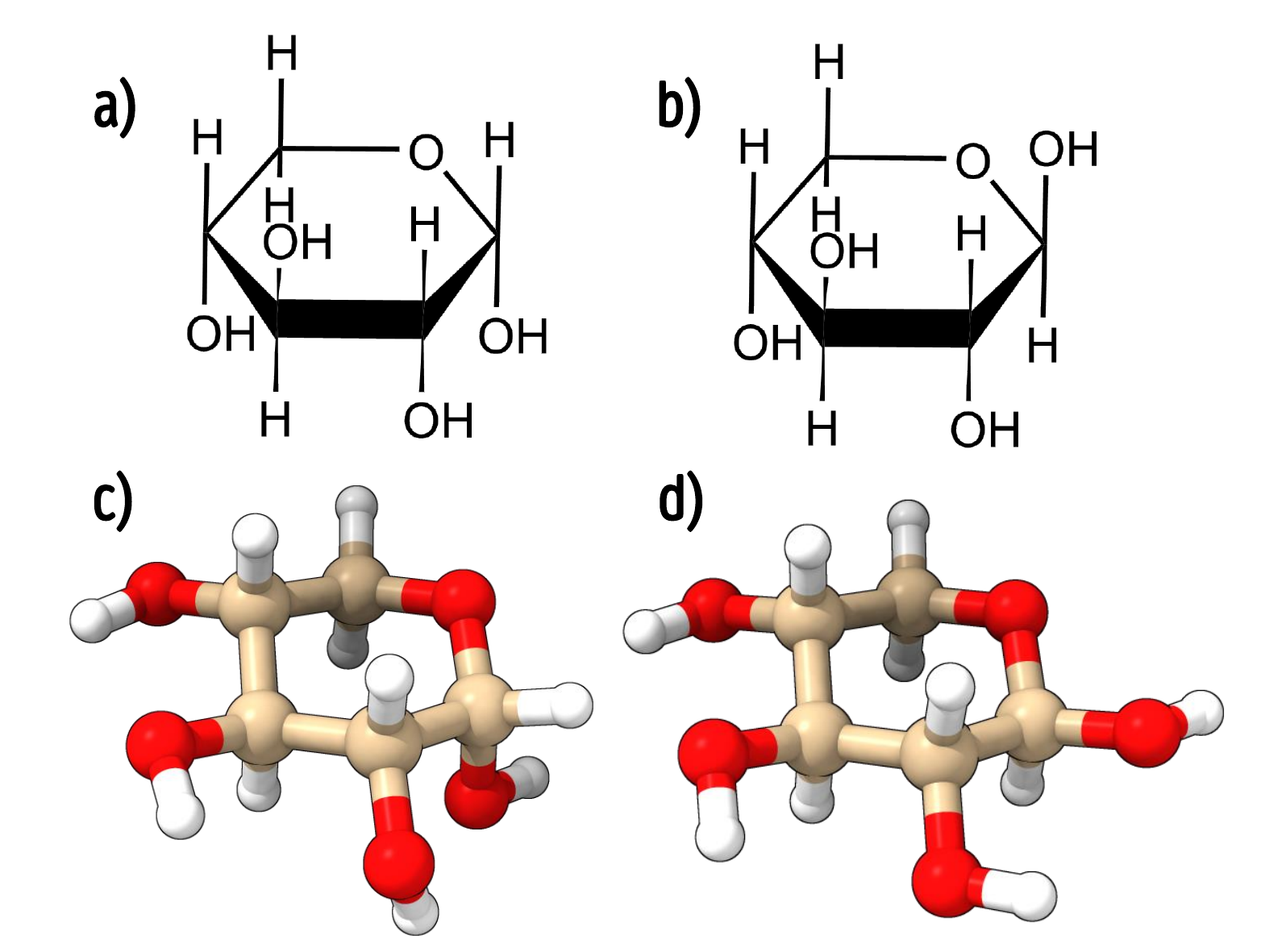}
  \caption{Haworth projections (a, b) and 3D chair configurations (c, d) of $\alpha$ (a, c) and $\beta$ (b, d) anomers of xylopyranose.}
  \label{fig:alphabeta}
\end{figure}

Xylopyranose is an aldopentose monosaccharide with five carbon atoms and an aldehyde functional group. 
Xylopyranose is the aldopentose analogue of glucose, differing by the absence of a hydroxymethyl group.
In solution, aldopentoses like xylose cyclize to form five-membered (furanose) and six-membered (pyranose) rings\cite{sinnot2007}. 
Xylose predominantly exists in the pyranose form in both the condensed and gas phases\cite{angyal1969,hordvik1971}. 
The pyranose structures have two anomers, $\alpha$ and $\beta$, distinguished by the position of the hydroxyl group at the chiral carbon. 
In the planar Haworth projection, the $\alpha$ anomer has the hydroxyl group pointing downwards, in an axial position, while in the $\beta$ anomer, it points upwards, in an equatorial position (see Figure \ref{fig:alphabeta}). 
Given its relevance in monosaccharide chemistry and its high torsional flexibility, xylopyranose has been the subject of numerous experimental and computational studies to determine its equilibrium structure, anomeric ratio, and the effect of solvents on its conformation. 

Schmidt et al.\cite{schmidt1996} employed molecular dynamics (MD) and free energy perturbation (FEP) simulations to analyze the anomeric equilibrium of D-xylopyranose in gas-phase and aqueous solution. 
This molecule served as a simple model for studying the anomeric effect in sugars. 
Free energy calculations indicated a small difference favoring the $\alpha$ form (\SI{0.15}{\kilo\cal\per\mol}), while experimental values slightly favored the $\beta$ form (in \SI{-0.38}{\kilo\cal\per\mol}). 
Thermodynamic integration (TI) showed that the free energy difference arises from a balance between an internal term favoring the $\alpha$ anomer and a solvation term favoring the $\beta$ anomer. Hydrogen bonding analysis from MD simulations explained the solvation preference for the $\beta$ anomer, attributed to improved hydrogen bonding of the anomeric hydroxyl group and increased accessible surface area. 
These findings emphasized the importance of both intramolecular and intermolecular interactions in determining the conformational preferences and anomeric equilibria of D-xylopyranose in gas-phase and solution phase\cite{schmidt1996}.

\begin{figure}[h]
    \centering
    \includegraphics[width=0.90\linewidth]{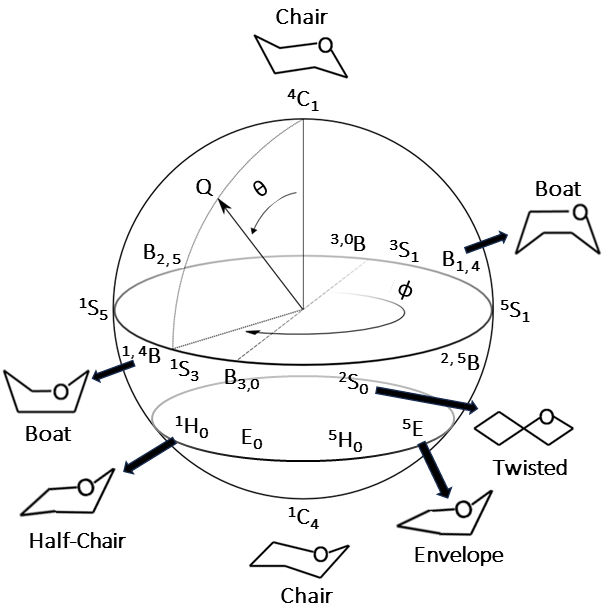}
    \caption{Schematic view of the major conformations of a six-membered ring using Cremer–Pople coordinates.}
    \label{fig:cremerpoplesphere}
\end{figure}

Subsequently, H\"{o}\"{o}g et al.\cite{hoog2001} conducted further FEP simulations to investigate the anomeric equilibrium of D-xylopyranose in gas-phase. 
The MD simulations employed the canonical ensemble to calculate the Helmholtz free energy change ($\Delta$A) using FEP and TI methods. H\"{o}\"{o}g et al.\cite{hoog2001}, the relative free energy difference between the $\beta$- and $\alpha$-anomers of D-xylopyranose in gas-phase was found to be very small, indicating minimal intrinsic energy preference between the anomeric forms. This suggests that the anomeric configuration of free D-xylopyranose is not significantly influenced by intrinsic molecular energetics but is rather affected by external factors such as solvent interactions.
Mayes et al.\cite{mayes2014} provided a comprehensive library of low-energy local minima and puckering interconversion transition states for five biologically relevant pyranose sugars, including $\beta$-D-xylopyranose, based on a thorough theoretical investigation of the 38 IUPAC puckering geometries.
Iglesias-Fernández et al.\cite{iglesias2015} computed the conformational free energy landscape of $\beta$-D-xylopyranose using \textit{ab initio} metadynamics with Cremer–Pople puckering coordinates as collective variables. Figure \ref{fig:cremerpoplesphere} shows a graphical representation of the Cremer-Pople sphere\cite{cremer1975} and the main conformations of a six-membered ring by means of the Boeyens classification\cite{boeyens1978}.

Pe\~{n}a et al.\cite{pena2013} investigated the conformational behavior of $\alpha$/$\beta$-D-xylopyranose by vaporizing crystalline samples using laser ablation and analyzing them with Fourier transform microwave spectroscopy. 
This study revealed two $\alpha$-D-xylopyranose conformers, stabilized by the anomeric effect and hydrogen bond networks. 
The spectroscopic analysis tracked fine structural changes due to the arrangements of OH groups in these conformers.

Moreover, in the last decades, xylopyranose has been used as a model to investigate hemicellulose's reactivity during pyrolysis, a promising technology for decarbonizing fuel and commercializing sustainable bioproducts\cite{zhou2017,zhou2018, goussougli2021}.
Thanks to the advancement of theoretical methods and computer hardware, quantum chemistry simulations are playing increasingly important roles in the understanding of lignocellulosic biomass pyrolysis as well as the experimental limitations\cite{ayarde2024}. 
Xylopyranose is considered the most appropriate structural motif in xylan, a key component of hemicellulose, one of the most abundant components of lignocellulose \cite{mettler2012, pinheiro2019,KAN2016}.  
Recently, Lupi et al.\cite{lupi2024} employed advanced quantum chemistry simulations to compute the potential energy surfaces (PESs), electronic energies and thermal rate constants derived from transition state theory of $\beta$-D-xylopyranose initial thermal decomposition steps. A kinetic model was proposed to elucidate the initial stages of the $\beta$-D-xylopyranose pyrolysis, revealing the role of ring opening to acyclic D-xylose in forming important pyrolytic compounds such as furfural and glycolaldehyde\cite{lupi2024}.
In the literature, many reviews detailing the experimental specificities of lignocellulosic biomass pyrolysis have been reported \cite{burnham2015}. As for the pyrolysis mechanism of biomass, most of the reviews focus on the pyrolysis kinetics, especially on global kinetic models \cite{HAMEED2019}.
Despite extensive research, many studies on pyrolysis employ gross approximations, especially regarding the kinetic modelling, making results of questionable physicality and often chemically unreal\cite{wang2015, HUANG2016, HU2019}.
Among those, one major oversight is the choice of the conformational isomer or set of rotamers to employ in a computational study of the chemical reactivity using multiconformational approaches to compute single-step reactions of the $\beta$-D-xylopyranose decomposition rate constants.
Recent literature highlights the impact of conformational space on reaction kinetics in both gaseous and condensed phases\cite{bao2016,bao2017}. 
Comprehensive analysis of the entire conformational space is computationally prohibitive, necessitating approximations like selecting a cut-off value of the Boltzmann population distribution.

In this work, we present the conformational analysis of $\alpha$- and $\beta$-D-xylopyranose using metadynamics-based computational methods and density functional theory (DFT) simulations. 
We provide enthalpic and free energy contributions and Boltzmann population distributions, over the temperature range relevant to fast pyrolysis.
Furthermore, from the analysis of the variation of the population distribution as a function of temperature, we propose two subsets of conformers for the $\alpha$ and $\beta$ anomers of D-xylopyranose that from the results obtained are the most abundant and therefore kinetically relevant in the analyzed temperature range. These subsets of conformers should be the starting point for future \textit{ab initio} kinetic studies based on multipath or multiconformational approaches that could lead to more accurate kinetic models\cite{viegas2018,viegas2023,yang2019,moller2016}.
The article is organized as follows: In Sec. \ref{sec:comp}, the computational methods are described. In Sec. \ref{sec:thermo}, we discuss the results of molecular thermodynamics, thermochemistry and  Boltzmann population analysis. Sec. \ref{sec:concl} provides conclusions and perspectives.

\section{Computational Details}\label{sec:comp}

\begin{figure*}[!b]
    \centering
    \includegraphics[scale=0.35]{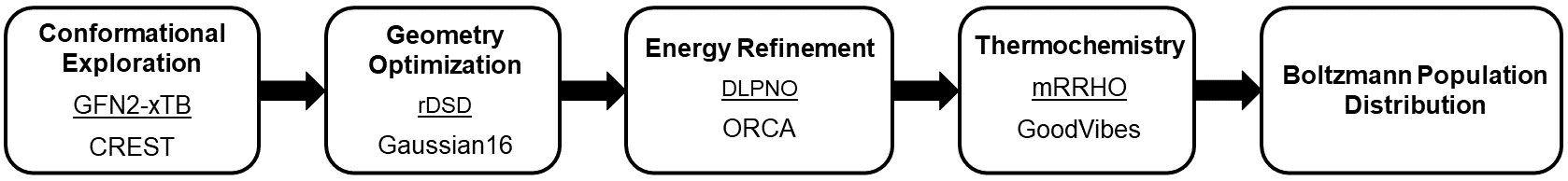}
    \caption{Step-by-step schematic representation of the computational methodology, functionals, models and codes used.}
    \label{fig:methods}
\end{figure*}

In this section, a detailed description of the computational procedure is provided.
Figure \ref{fig:methods} shows the overall computational methodology explained in each subsection.

\subsection{Conformational Sampling}

The conformational space of D-xylopyranose was explored using the methodology implemented in the CREST program\cite{crest} which combines the GFN2-xTB\cite{bannwarth2019} Density-Functional-Based Tight Binding (DFTB) method with Root-Mean-Square-Deviation (RMSD)-based metadynamics \cite{grimme2019} to efficiently sample the PES\cite{barducci2011,bussi2020,pietrucci2017}.
The initial geometry of D-xylopyranose was optimized at the M06-2X/6-311++G(d,p) level of theory \cite{zhao2008,clark1983} as it has been widely used in previous works as a standard for xylopyranose like structures \cite{hu2017}. 
The iMTD-GC algorithm implemented in CREST\cite{crest} enhances conformer identification by integrating GFN2-xTB calculations with RMSD-based metadynamics. 
A history-dependent biasing potential is applied using previous minima on the PES as collective variables:

\begin{equation}
    V_{\text{bias}} = \sum_{i=1}^{n} k_i \exp\left(-\alpha \Delta_i^2\right)
\end{equation}

where \( n \) is the number of reference structures, \( k_i \) are the pushing strengths, and \( \alpha \) determines the potential's shape. Additional forces derived from this potential act as guiding forces in the metadynamics simulations, penalizing molecular dynamics snapshots that are too similar to reference structures. 
This facilitates extensive PES exploration and high-barrier crossings.

Genetic Z-matrix crossing (GC) complements this workflow by generating new structures through the projection of structural elements from existing conformers onto a reference. 
Using internal (Z-matrix, \( R \)) coordinates, new structures are created by:

\begin{equation}
    R_{\text{new}} = R_{\text{ref}} + R_i - R_j
\end{equation}

where \( R_i \) and \( R_j \) label coordinate pairs, and \( R_{\text{new}} \) is the newly generated structure. 
This structure then undergoes full geometry optimization.

\subsection{Electronic Structure and Thermochemistry}

All conformer structures generated by CREST are subsequently optimized and the harmonic frequencies were calculated at the revDSD-PBEP86 \cite{santra2019minimally}-D3(BJ) \cite{grimme2011} double-hybrid functional theoretical level in combination with the jun-cc-pVTZ \cite{papajak2011} basis set (hereafter rDSD).
Several studies have shown that this combination of functional and basis set provides accurate geometrical structures \cite{ceselin2021} and vibrational frequencies \cite{barone2020}.
Single-point energy calculations were conducted on top of DFT geometries to further refine electronic energies at the DLPNO-CCSD(T) \cite{riplinger2013} (with F12 explicit correlation correction) \cite{pavosevic2017} level of theory in conjunction with the cc-pVTZ-F12 basis set \cite{peterson2008} (hereafter DLPNO). 
DLPNO calculations were performed employing the ORCA code \cite{neese2022} and using the tightPNO cut-off.
The Root Mean Square Deviation (RMSD) program was used to check for the presence of possible repeating structures following DFT reoptimization of the geometries \cite{kabsch1976,WALKER1991,RMSD}.
The neglect of anharmonicity tends to overestimate the vibrational zero-point energy (ZPE).
Vibrational scaling factors are applied to correct systematic overestimations of vibrational frequencies due to the harmonic oscillator approximation, basis set limitations, and incomplete electron correlation. These factors, derived empirically by comparing calculated and experimental frequencies, adjust the computed vibrational modes, improving the accuracy of ZPEs, thermal corrections, and derived thermodynamic properties. Without these corrections, the calculated thermochemical data would be less reliable, leading to errors in predictions of enthalpy, entropy, and free energy.
Therefore, in order to obtain accurate ZPEs, avoiding the calculation of perturbative anharmonic corrections, ZPE and frequencies are scaled by 0.982.
The scale factor for the vibrational frequencies has been computed using the FREQ program \cite{alecu2010,yu2017}.
All DFT calculations were carried out using the Gaussian16 code \cite{g16}.
Quasi-harmonic entropies were calculated using Grimme’s approximation implemented in GoodVibes code \cite{luchini2022}. 
Indeed, the rigid-rotor harmonic oscillator (RRHO) model is inadequate for significant anharmonic motions, often due to internal torsional hindrance. 
Pitzer and Gwinn have extensively studied the effect of these motions on molecular entropy, demonstrating that the RRHO model underestimates the contributions of hindered internal rotations\cite{pitzer1942}.
Although the one-dimensional hindered rotor model provides some corrections, its applicability is limited. To address this limitation, Grimme’s modified RRHO (mRRHO) approach \cite{grimme2012} is preferred, as it incorporates anharmonic corrections, especially for low-frequency vibrational modes, offering broader applicability for complex systems\cite{east1997}.
The mRRHO approximation accounts for anharmonic effects in low-frequency modes, especially those corresponding to hindered internal rotations. The vibrational contributions to Gibbs free energy and enthalpy in the mRRHO approximation are expressed as:
\[
\Delta G^{\text{mRRHO}} = \Delta E + \Delta ZPE + \sum_i \left[ E_i - T S_i^{\text{mRRHO}} \right],
\]
\[
\Delta H^{\text{mRRHO}} = \Delta E + \Delta ZPE + \sum_i H_i^{\text{mRRHO}},
\]
where \(\Delta E\) is the electronic energy difference between conformers, \(\Delta ZPE\) refers to the difference in ZPE between conformers, \(E_i\) represents the vibrational energy of mode \(i\), \( S_i^{\text{mRRHO}} \) and \( H_i^{\text{mRRHO}} \) incorporate anharmonic corrections for hindered internal rotations, and \(T\) is the temperature. These corrections are expressed through \( S_{\text{rot}}^{\text{corr}} \) and \( H_{\text{rot}}^{\text{corr}} \), which account for the deviations from harmonic behavior. The corrected entropy \( S_{\text{mRRHO}} \) is given by:
\[
S_{\text{mRRHO}} = S_{\text{RRHO}} + S_{\text{rot}}^{\text{corr}},
\]
where
\[
S_{\text{rot}}^{\text{corr}} = R \ln \left( \frac{\sigma}{q_{\text{rot}}^{\text{free}}} \right) + R \left[ \frac{\Theta_{\text{rot}}^{\text{hindered}}}{T} \right],
\]
with \(\sigma\) being the symmetry number, \(q_{\text{rot}}^{\text{free}}\) the partition function for a free rotor, \(\Theta_{\text{rot}}^{\text{hindered}}\) the rotational temperature for the hindered rotor, \(T\) is the temperature, and \(R\) is the gas constant.

Similarly, the enthalpy correction \( H_{\text{mRRHO}} \) is given by:
\[
H_{\text{mRRHO}} = H_{\text{RRHO}} + H_{\text{rot}}^{\text{corr}},
\]
where
\[
H_{\text{rot}}^{\text{corr}} = R T \left[ \frac{\Theta_{\text{rot}}^{\text{hindered}}}{\sinh \left( \frac{\Theta_{\text{rot}}^{\text{hindered}}}{T} \right)} \right],
\]
correcting for the rotational contributions of the hindered internal rotations. These adjustments ensure that the mRRHO approximation provides a more accurate thermodynamic description, particularly for systems with flexible internal torsions \cite{grimme2012}.
Thermochemical analysis was conducted using GoodVibes code\cite{luchini2022} to compute enthalpy, entropy, and free energy.

\subsection{Boltzmann Population Distribution Analysis}

Boltzmann population distribution analysis was used to determine the relative populations of conformational states for $\alpha$/$\beta$-D-xylopyranose, crucial for understanding which conformers are more populated in the range of temperatures accessible during fast pyrolysis \cite{klemm2005, karplus2002, leach2001}.

The relative population \(\frac{N_i}{N}\) of a conformational state \(i\) with energy \(E_i\) and degeneracy \(g_i\) is given by the following formula:

\[
\frac{N_i}{N} = \frac{g_i e^{-E_i / k_B T}}{\sum_i g_i e^{-E_i / k_B T}}
\]

where \(k_B\) is the Boltzmann constant, and \(T\) is the temperature. 
In this study, relative free energy changes with respect to the most stable conformer for both anomers are considered as the energy states of each conformer, in order to account for the enthalpic and entropic contributions that become relevant at high temperatures.
Detailed conformational analysis is crucial for identifying the lowest energy structure among various conformations in flexible molecular systems. 
Different conformations can significantly affect electronic properties, interaction sites and also chemical reactivity. 
The possible structure of $\alpha$/$\beta$-D-xylopyranose is complex due to the 4 different dihedral angles which are defined by the orientation of each of the  4 hydroxyl groups of the pyranose ring. Each of these has potential to be a significantly hindered rotation, and must be treated using redundant internal coordinates to identify the internal rotation (IR) modes.
\cite{pfaendtner20071,ayala1998identification,mcclurg1997hindered,mcclurg1999comment}. 
For this reason, its PES likely contains multiple local minima, complicating structural optimization. 
Different theoretical levels can yield distinct molecular structures. 
To avoid non-global minima, an accurate exploration of the PES is necessary to find the lowest energy structures. 
While \textit{ab initio} methods can accurately evaluate local minima, they are inefficient for sampling larger PES areas. 
Therefore, optimized strategies are required to explore extensive regions of the configuration space and carefully assess energy barriers between local minima.

\section{Results and Discussion}\label{sec:thermo}

This section is organized in subsections to show the results obtained in each step of the methodology explained previously in Section \ref{sec:comp}.
In Subsection \ref{sec:mol_term} the results obtained by sampling the conformational space of D-xylopyranose for both $\alpha$ and $\beta$ anomers is discussed and the energetics obtained for the DFT reoptimization and the DLPNO refinement of the conformational structures identified by CREST is shown.
Each identified conformer is classified using the Boeyens system based on the value of the Cremer-Pople puckering coordinates, Q, $\theta$ and $\phi$.
To identify the different conformations, five labels are used: chair, half-chair, boat, envelope and twisted-boat.
Subsequently, the results of the thermochemical analysis obtained by applying the mRRHO approximation implemented in GoodVibes is shown, to evaluate the variation of the conformer free energies at temperatures of relevance for pyrolysis. In Subsection \ref{sec:pop} the results of the Boltzmann population analysis for both anomers is discussed and shown to define a 90\% population cut-off value limiting the available conformational space.
In Subsection \ref{sec:rel_conf} the choice of two subgroups of conformers for both xylopyranoside anomers is discussed which, from the analysis conducted, appear to be the most populated and therefore most plausibly relevant in studies of multiconformational \textit{ab initio} reactivity.

\subsection{Molecular Thermodynamics}\label{sec:mol_term}

\subsubsection{$\alpha$ Anomer}
\begin{figure}[h!]
    \centering
    \includegraphics[scale=0.3]{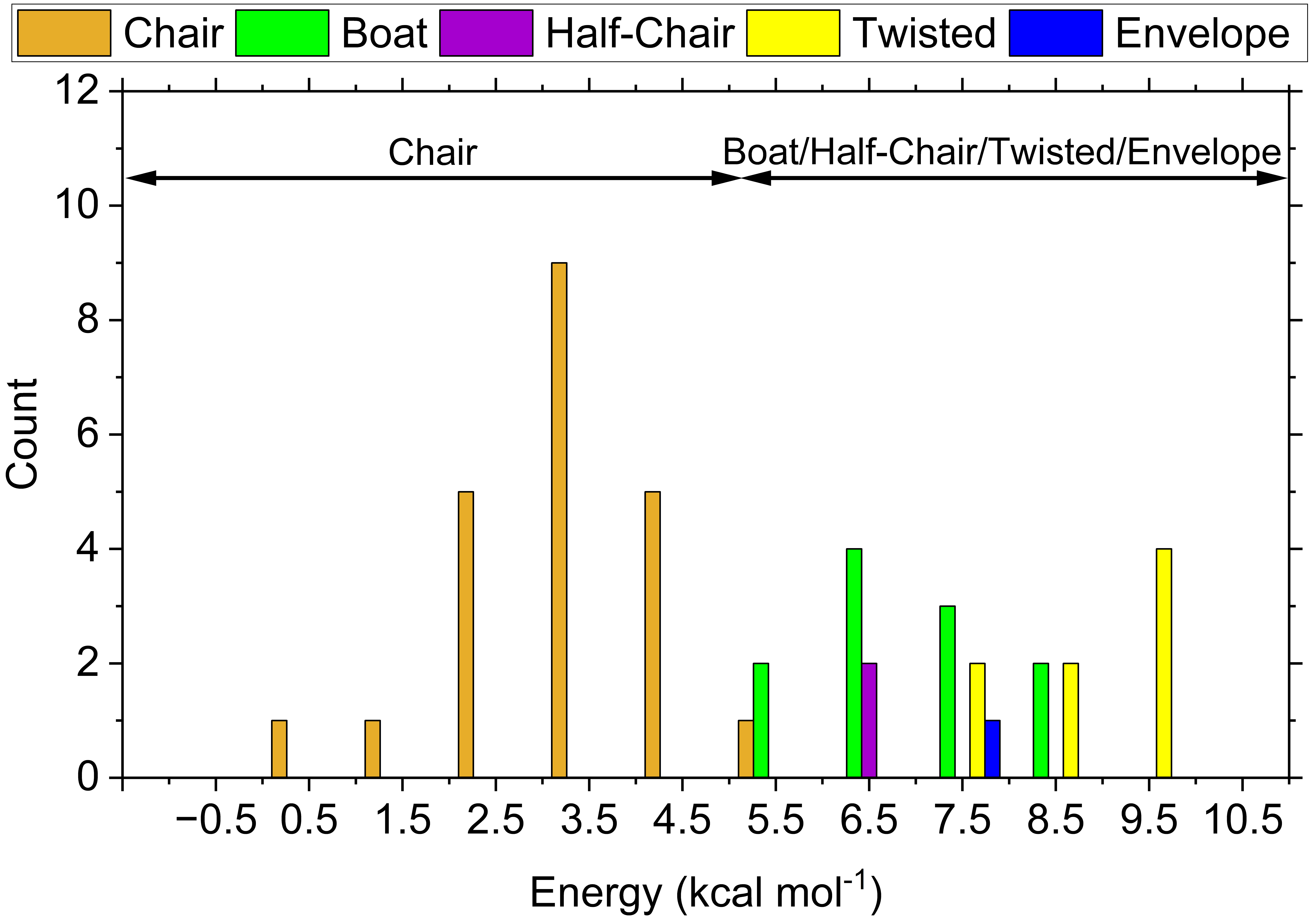}
    \caption{Frequency plot of the $\alpha$-D-xylopyranose ring conformations, showing the number of chair, boat, twisted, and half-chair conformations contained in each \SI{1}{\kilo\cal\per\mol} energy bin of the considered conformational space of approximately \SI{10}{\kilo\cal \per\mol}.}
    \label{fig:freq_alfa}
\end{figure}

The exploration through metadynamics identified 57 conformers for the $\alpha$ anomer consistent in total energy by approximately \SI{12}{\kilo\cal \per\mol}.
Following reoptimization and vibrational analysis of each CREST-generated structure, the number of conformers was reduced to 44.
The RMSD program was used to compare and identify repeating structures among the reoptimized conformers.
The DLPNO-ZPE corrected electronic energies for each conformer are reported in Table \ref{tab:alfa_energies}.
In SI Table \ref{tab:alfa_energies_SI} the comparison of the ZPE-corrected electronic energies calculated at the rDSD and DLPNO levels can be found.
The identified conformers span an energy range of approximately \SI{10}{\kilo\cal \per\mol}, indicating significant variation. Arbitrarily selecting any of these conformers could substantially underestimate the initial reaction barriers. Structurally, the conformers mainly differ in the orientation of the hydroxyl groups and their dihedral angles, as well as the conformation of the hexagonal ring, which can adopt boat, chair, half-chair, twisted-boat or envelope shapes. The stability of these conformers is strongly influenced by intramolecular interactions, including hydrogen bonds and steric effects.
As can be seen from Table \ref{tab:alfa_energies} and the frequency plot in Figure \ref{fig:freq_alfa}, the conformers at lower energies present a chair conformation while at higher energies the conformers predominantly take on boat or twisted boat conformations.
Specifically, 22 structures have the shape of a chair, 2 of a half-chair, 8 of a twisted boat, 11 of a boat and 1 of an envelope.

\begin{table*}[b]
\small
\centering
\caption{Zero-point corrected energies (in \SI{}{\kilo\cal\per\mol}) relative to the most thermodynamically stable $\alpha$ conformers computed at DLPNO level.}
\label{tab:alfa_energies}
\begin{tabular}{c c}
\begin{tabular}{@{}cccccc@{}}
\toprule
 \textbf{Conformer}  & \textbf{$\Delta$E$_{DLPNO}$} & \textbf{Conformation} & Q(\AA) & $\theta$($^{\circ}$) & $\phi$($^{\circ}$)\\ \midrule
    \textbf{$\alpha$-C1}    & 0.0 & Chair & 0.5464 & 175.10 & 75.95 \\
    \textbf{$\alpha$-C2}    & 1.67 & Chair & 0.5526 & 176.01 & 54.07 \\
    \textbf{$\alpha$-C3}    & 2.01 & Chair & 0.5079 & 9.00 & 108.27 \\
    \textbf{$\alpha$-C4}    & 2.17 & Chair & 0.5446 & 177.69 & 78.94 \\
    \textbf{$\alpha$-C5}    & 2.18 & Chair & 0.5158 & 8.92 & 110.75 \\
    \textbf{$\alpha$-C6}    & 2.72 & Chair & 0.5488 & 176.64 & 55.44 \\
    \textbf{$\alpha$-C7}    & 2.76 & Chair & 0.4975 & 10.35 & 109.46 \\
    \textbf{$\alpha$-C8}    & 3.05 & Chair & 0.5633 & 178.22 & 153.60 \\
    \textbf{$\alpha$-C9}    & 3.13 & Chair & 0.5488 & 177.12 & 53.16 \\
    \textbf{$\alpha$-C10}    & 3.17 & Chair & 0.4937 & 10.83 & 108.59 \\
    \textbf{$\alpha$-C11}    & 3.22 & Chair & 0.5418 & 175.12 & 93.78 \\
    \textbf{$\alpha$-C12}    & 3.27 & Chair & 0.5439 & 175.94 & 53.03 \\
    \textbf{$\alpha$-C13}    & 3.38 & Chair & 0.5561 & 176.64 & 184.78 \\
    \textbf{$\alpha$-C14}    & 3.38 & Chair & 0.4874 & 12.21 & 96.90 \\   
    \textbf{$\alpha$-C15}    & 3.76 & Chair & 0.5372 & 177.46 & 70.99 \\
    \textbf{$\alpha$-C16}    & 3.90 & Chair & 0.4799 & 12.90 & 95.28 \\
    \textbf{$\alpha$-C17}    & 4.06 & Chair & 0.5262 & 10.79 & 114.83 \\
    \textbf{$\alpha$-C18}    & 4.16 & Chair & 0.5116 & 8.07 & 110.12 \\
    \textbf{$\alpha$-C19}    & 4.37 & Chair & 0.5292 & 10.65 & 117.28 \\
    \textbf{$\alpha$-C20}    & 4.69 & Chair & 0.4901 & 9.93 & 107.95 \\
    \textbf{$\alpha$-C21}    & 4.97 & Chair & 0.4796 & 11.69 & 96.14 \\ 
    \textbf{$\alpha$-C22}    & 5.16 & Boat & 0.1086 & 112.48 & 184.44 \\
\end{tabular}

&

\begin{tabular}{@{}ccccccc@{}}
\toprule
\textbf{Conformer}  & \textbf{$\Delta$E$_{DLPNO}$} & \textbf{Conformation} & Q(\AA) & $\theta$($^{\circ}$) & $\phi$($^{\circ}$)\\ \midrule 
    \textbf{$\alpha$-C23}    & 5.64  & Chair & 0.5268 & 9.81 & 117.59 \\
    \textbf{$\alpha$-C24}    & 5.86  & Boat & 0.6430 & 94.88 & 200.30 \\
    \textbf{$\alpha$-C25}    & 6.29  & Boat & 0.3415 & 94.02 & 160.49 \\
    \textbf{$\alpha$-C26}    & 6.56  & Half-chair & 0.1127 & 123.93 & 270.28 \\
    \textbf{$\alpha$-C27}    & 6.77  & Boat & 0.6493 & 94.90 & 200.00 \\
    \textbf{$\alpha$-C28}    & 6.92  & Boat & 0.3827 & 93.98 & 159.28 \\
    \textbf{$\alpha$-C29}    & 6.96  & Boat & 0.6506 & 94.69 & 200.74 \\
    \textbf{$\alpha$-C30}    & 7.00  & Half-chair & 0.1076 & 128.36 & 265.20 \\
    \textbf{$\alpha$-C31}    & 7.02  & Envelope & 0.7698 & 87.02 & 324.84 \\
    \textbf{$\alpha$-C32}    & 7.06  & Boat & 0.6450 & 95.32 & 200.49 \\
    \textbf{$\alpha$-C33}    & 7.12  & Twisted & 0.4879 & 87.46 & 269.31 \\
    \textbf{$\alpha$-C34}    & 7.30  & Boat & 0.6138 & 93.30 & 204.69 \\
    \textbf{$\alpha$-C35}    & 7.56  & Boat & 0.6617 & 97.48 & 201.71 \\
    \textbf{$\alpha$-C36}    & 8.14  & Twisted & 0.0921 & 119.22 & 257.21 \\
    \textbf{$\alpha$-C37}    & 8.33  & Boat & 0.6574 & 97.46 & 202.13 \\
    \textbf{$\alpha$-C38}    & 8.34  & Twisted & 0.4872 & 86.50 & 274.11 \\
    \textbf{$\alpha$-C39}    & 8.43  & Twisted & 0.4796 & 88.59 & 260.19 \\
    \textbf{$\alpha$-C40}    & 9.01  & Boat & 0.6221 & 93.04 & 203.11 \\
    \textbf{$\alpha$-C41}    & 9.10  & Twisted & 0.4988 & 86.55 & 268.69 \\
    \textbf{$\alpha$-C42}    & 9.14  & Twisted & 0.4558 & 85.88 & 263.29 \\ 
    \textbf{$\alpha$-C43}    & 9.88  & Twisted & 0.4848 & 86.14 & 267.59 \\  
    \textbf{$\alpha$-C44}    & 9.97  & Twisted & 0.5817 & 89.50 & 282.64 \\ 
\end{tabular}
\end{tabular}
\end{table*}

\subsubsection{$\beta$ Anomer}
\begin{figure}[h!]
    \centering
    \includegraphics[scale=0.3]{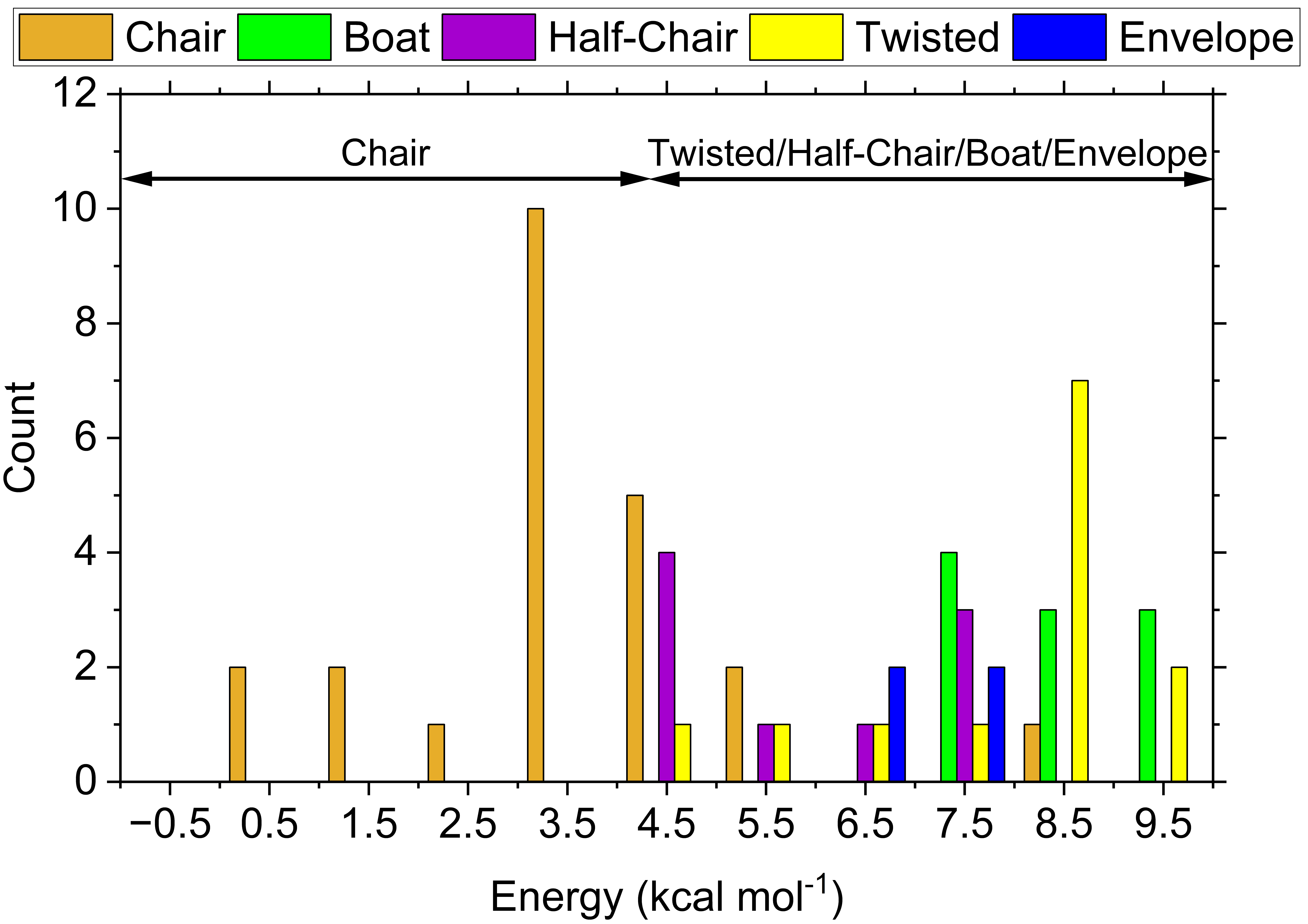}
    \caption{Frequency plot of the $\beta$-D-xylopyranose ring conformations, showing the number of chair, boat, twisted, and half-chair conformations contained in each \SI{1}{\kilo\cal \per\mol} energy bin of the considered conformational space of approximately \SI{10}{\kilo\cal\per\mol}.}
    \label{fig:freq_beta}
\end{figure}

CREST identified 73 conformers of D-xylopyranose within an energy range of approximately in \SI{12}{\kilo\cal\per\mol}. 
Metadynamics may overestimate the number of distinct conformers generating non-physical or unstable conformations.
Reoptimization using higher-level DFT methods, such as rDSD, provide a more accurate description of the PES, for which several initially distinct conformations may converge to the same local minimum resulting in a decreased number of unique conformers.
For this reason, subsequent reoptimization and vibrational analysis by means of rDSD reduced this number to 59 conformers. 
The DLPNO-ZPE corrected electronic energies for each conformer are listed in Table \ref{tab:beta_energies}.
Table \ref{tab:beta_energies_SI}, reported in SI, compares the ZPE-corrected electronic energies calculated at the rDSD and DLPNO levels for each conformer.
Similarly to the $\alpha$ anomer, the conformational diversity of $\beta$-D-xylopyranose primarily arises from variations of the ring conformation and the orientation of hydroxyl groups and their associated dihedral angles.
This diversity is further influenced by the flexibility of the hexagonal ring, which can adopt various conformations, most notably the boat and chair forms.
The chair conformation is generally more stable as it minimizes steric interactions and allows for favorable hydrogen bonding patterns.
In this conformation, the equatorial positioning of hydroxyl groups reduces steric hindrance, leading to more stable molecular structures.
In contrast, the boat conformation is typically less stable due to increased steric and torsional strains. 
However, specific intramolecular interactions, such as hydrogen bonding, can stabilize certain boat conformers by creating a network of interactions that counteract the destabilizing effects of these strains.
As also reported by Pe\~{n}a et al.\cite{pena2013}, intramolecular hydrogen bonding significantly influences the stability of D-xylopyranose conformers. 
These bonds, forming between hydroxyl groups in adjacent or nearby positions, result in a more compact and energetically favorable structure. 
The strength and number of hydrogen bonds vary with the orientation of the hydroxyl groups and the overall ring conformation.
Also for the $\beta$ anomer of D-xylopyranose it can be seen from Table \ref{tab:beta_energies} and Figure \ref{fig:freq_beta} that the conformers at lower energies present a chair conformation while at higher energies the conformers predominantly take on boat or twisted boat conformations.
More precisely, 23 structures have the shape of a chair, 9 of a half-chair, 13 of a twisted boat, 10 of a boat and 4 of an envelope.
In Table \ref{tab:alfa_beta_energies}, the relative energies in kcal mol$^{-1}$ of both the $\alpha$ and $\beta$ anomers of D-xylopyranose are presented. It can be observed that the thermodynamically stable conformers correspond to those reported in previous investigations by Schmidt et al.\cite{schmidt1996} and Peña et al.\cite{pena2013}.

\begin{table*}[b!]
\centering
\small
\caption{Zero-point corrected energies (in \SI{}{\kilo\cal\per\mol}) relative to the most thermodynamically stable $\beta$ conformers computed at DLPNO level.}
\label{tab:beta_energies}
\begin{tabular}{c c}
\begin{tabular}{@{}cccccc@{}}
\toprule
 \textbf{Conformer}  & \textbf{$\Delta$E$_{DLPNO}$} & \textbf{Conformation} & Q(\AA) & $\theta$($^{\circ}$) & $\phi$($^{\circ}$)\\ \midrule
      \textbf{$\beta$-C1}    & 0.00  & Chair & 0.5827 & 177.63 & 303.76 \\
      \textbf{$\beta$-C2}    & 0.67  & Chair & 0.5810 & 178.63 & 321.01 \\
      \textbf{$\beta$-C3}    & 1.83  & Chair & 0.0958 & 53.45 & 328.75 \\
      \textbf{$\beta$-C4}    & 1.89  & Chair & 0.2816 & 167.53 & 136.74 \\
      \textbf{$\beta$-C5}    & 2.42  & Chair & 0.2757 & 165.39 & 151.43 \\
      \textbf{$\beta$-C6}    & 3.02  & Chair & 0.5797 & 175.96 & 255.93 \\
      \textbf{$\beta$-C7}    & 3.24  & Chair & 0.5861 & 177.08 & 286.84 \\
      \textbf{$\beta$-C8}    & 3.35  & Chair & 0.1876 & 156.42 & 151.51 \\
      \textbf{$\beta$-C9}    & 3.47  & Chair & 0.0962 & 110.25 & 331.48 \\
      \textbf{$\beta$-C10}   & 3.52  & Chair & 0.1904 & 28.24 & 161.16 \\
      \textbf{$\beta$-C11}   & 3.62  & Chair & 0.5831 & 176.82 & 340.48 \\
      \textbf{$\beta$-C12}   & 3.92  & Chair & 0.1287 & 134.48 & 335.34 \\
      \textbf{$\beta$-C13}   & 3.93  & Chair & 0.5727 & 176.73 & 274.93 \\
      \textbf{$\beta$-C14}   & 3.95  & Chair & 0.5727 & 175.88 & 277.62 \\
      \textbf{$\beta$-C15}   & 3.97  & Chair & 0.5797 & 177.19 & 318.22 \\
      \textbf{$\beta$-C16}   & 3.99  & Chair & 0.5767 & 175.43 & 267.97 \\
      \textbf{$\beta$-C17}   & 4.01  & Chair & 0.7305 & 83.52 & 275.13 \\
      \textbf{$\beta$-C18}   & 4.06  & Chair & 0.5674 & 175.27 & 289.50 \\
      \textbf{$\beta$-C19}   & 4.08  & Half-chair & 0.2046 & 147.92 & 332.94 \\
      \textbf{$\beta$-C20}   & 4.12  & Half-chair & 0.1066 & 122.52 & 333.14 \\
      \textbf{$\beta$-C21}   & 4.14  & Half-chair & 0.1045 & 54.76 & 330.50 \\
      \textbf{$\beta$-C22}   & 4.17  & Half-chair & 0.1679 & 142.79 & 334.25 \\
      \textbf{$\beta$-C23}   & 4.23  & Chair & 0.2183 & 27.27 & 160.27 \\
      \textbf{$\beta$-C24}   & 4.52  & Chair & 0.5826 & 3.35 & 118.29 \\
      \textbf{$\beta$-C25}   & 4.58  & Chair & 0.5664 & 176.28 & 295.77 \\
      \textbf{$\beta$-C26}   & 4.70  & Half-chair & 0.1129 & 133.09 & 150.87 \\
      \textbf{$\beta$-C27}   & 5.32  & Twisted & 0.7263 & 86.08 & 276.10 \\
      \textbf{$\beta$-C28}   & 5.83  & Chair & 0.4342 & 7.99 & 39.16 \\
      \textbf{$\beta$-C29}   & 5.86  & Chair & 0.4437 & 172.50 & 216.81 \\
      \textbf{$\beta$-C30}   & 6.11  & Envelope & 0.7002 & 80.70 & 318.40 \\
\end{tabular}
&
\begin{tabular}{@{}cccccc@{}}
\toprule
 \textbf{Conformer}  & \textbf{$\Delta$E$_{DLPNO}$} & \textbf{Conformation} & Q(\AA) & $\theta$($^{\circ}$) & $\phi$($^{\circ}$)\\ \midrule    
      \textbf{$\beta$-C31}   & 6.31  & Half-chair & 0.1496 & 144.78 & 343.49 \\
      \textbf{$\beta$-C32}   & 6.47  & Envelope & 0.6918 & 82.16 & 321.40 \\ 
      \textbf{$\beta$-C33}   & 6.52  & Twisted & 0.7206 & 84.75 & 281.46 \\
      \textbf{$\beta$-C34}   & 7.11  & Half-chair & 0.7530 & 36.46 & 145.65 \\
      \textbf{$\beta$-C35}   & 7.21  & Half-chair & 0.7779 & 39.62 & 146.85 \\
      \textbf{$\beta$-C36}   & 7.31  & Twisted & 0.7318 & 86.32 & 274.87 \\
      \textbf{$\beta$-C37}   & 7.35  & Boat & 0.7038 & 92.74 & 212.26 \\
      \textbf{$\beta$-C38}   & 7.47  & Envelope & 0.6971 & 82.06 & 322.07 \\
      \textbf{$\beta$-C39}   & 7.64  & Boat& 0.7350 & 82.41 & 142.72 \\
      \textbf{$\beta$-C40}   & 7.76  & Boat & 0.7050 & 92.90 & 210.78 \\
      \textbf{$\beta$-C41}   & 7.92  & Boat & 0.6971 & 89.45 & 204.51 \\
      \textbf{$\beta$-C42}   & 7.92  & Envelope & 0.6978 & 91.10 & 47.39 \\
      \textbf{$\beta$-C43}   & 7.96  & Half-chair & 0.8449 & 139.25 & 324.27 \\
      \textbf{$\beta$-C44}   & 8.02  & Twisted & 0.7145 & 91.01 & 88.12 \\
      \textbf{$\beta$-C45}   & 8.06  & Twisted & 0.5789 & 77.34 & 109.63 \\
      \textbf{$\beta$-C46}   & 8.12  & Twisted & 0.5647 & 78.10 & 107.81 \\
      \textbf{$\beta$-C47}   & 8.37  & Twisted & 0.1481 & 94.60 & 63.05 \\
      \textbf{$\beta$-C48}   & 8.45  & Boat & 0.6850 & 92.23 & 214.70 \\
      \textbf{$\beta$-C49}   & 8.57  & Twisted & 0.3771 & 75.21 & 105.25 \\
      \textbf{$\beta$-C50}   & 8.59  & Boat & 0.7435 & 82.25 & 144.48 \\
      \textbf{$\beta$-C51}   & 8.77  & C & 0.9749 & 152.14 & 324.56 \\
      \textbf{$\beta$-C52}   & 8.86  & Twisted & 0.5286 & 78.43 & 104.50 \\
      \textbf{$\beta$-C53}   & 8.92  & Boat & 0.6726 & 91.39 & 212.22 \\
      \textbf{$\beta$-C54}   & 8.95  & Twisted & 0.4383 & 76.30 &104.95 \\
      \textbf{$\beta$-C55}   & 8.97  & Boat & 0.6779 & 91.36 & 213.38 \\
      \textbf{$\beta$-C56}   & 9.08  & Boat & 0.7272 & 82.01 & 142.01 \\
      \textbf{$\beta$-C57}   & 9.18  & Boat & 0.6731 & 88.15 & 205.72 \\
      \textbf{$\beta$-C58}   & 9.25  & Twisted & 0.2448 & 91.80 & 74.25 \\
      \textbf{$\beta$-C59}   & 9.26  & Twisted & 0.4676 & 76.62 & 105.42 \\ 
         &   &   &   &  &  \\       
\end{tabular}
\end{tabular}
\end{table*}

\begin{table*}[b!]
\centering
\small
\caption{Zero-point corrected energies (in \SI{}{\kilo\cal\per\mol}) and absolute energies (in \SI{}{\hartree}) relative to the most thermodynamically stable rotamer among the $\alpha$ and $\beta$ anomers computed at DLPNO level.}
\label{tab:alfa_beta_energies}
\begin{tabular}{c c c}
\begin{tabular}{@{}ccc@{}}
\toprule
 \textbf{Conformer}  & \textbf{$\Delta$E$_{DLPNO}$} & \textbf{E$_{DLPNO}$} \\ \midrule
      \textbf{$\alpha$-C1}    & 0.00  &  -571.8930608 \\
      \textbf{$\beta$-C1}    & 0.64  &  -571.8920477 \\
      \textbf{$\beta$-C2}    &  1.31 &  -571.8909757 \\
      \textbf{$\alpha$-C2}    & 1.67  &  -571.8904041 \\
      \textbf{$\alpha$-C3}    & 2.02  &  -571.8898553 \\
      \textbf{$\alpha$-C4}    & 2.18  &  -571.8895993 \\
      \textbf{$\alpha$-C5}    & 2.19  & -571.8895889  \\
      \textbf{$\beta$-C3}    & 2.49  &  -571.8891035 \\
      \textbf{$\beta$-C4}    & 2.55  & -571.8890106  \\
      \textbf{$\alpha$-C6}   & 2.73  &  -571.8887336 \\
      \textbf{$\alpha$-C7}   & 2.77  &  -571.8886702 \\
      \textbf{$\alpha$-C8}   & 3.06  & -571.8881994  \\
      \textbf{$\beta$-C5}   & 3.08  &  -571.8881792 \\
      \textbf{$\alpha$-C9}   & 3.14  &  -571.8880805 \\
      \textbf{$\alpha$-C10}   & 3.18  & -571.888015  \\
      \textbf{$\alpha$-C11}   & 3.23  &  -571.8879271 \\
      \textbf{$\alpha$-C12}   & 3.28  & -571.8878493  \\
      \textbf{$\alpha$-C13}   & 3.39  &  -571.8876807 \\
      \textbf{$\alpha$-C14}   & 3.40  &  -571.8876692 \\
      \textbf{$\beta$-C6}   & 3.68  & -571.8872195  \\
      \textbf{$\alpha$-C15}   & 3.77  &  -571.8870745 \\
      \textbf{$\beta$-C7}   & 3.78  &  -571.8870627 \\
      \textbf{$\beta$-C8}   & 3.89  & -571.8868927  \\
      \textbf{$\alpha$-C16}   & 3.91  &  -571.8868469 \\
      \textbf{$\beta$-C9}   & 4.00  &  -571.8867123 \\
      \textbf{$\alpha$-C17}   & 4.08  & -571.8865829  \\
      \textbf{$\beta$-C10}   & 4.13  & -571.8865084  \\
      \textbf{$\alpha$-C18}   & 4.18  & -571.8864299  \\
      \textbf{$\beta$-C11}   &  4.18 &  -571.8864193 \\
      \textbf{$\beta$-C12}   &  4.29 & -571.8862563  \\
      \textbf{$\alpha$-C19}   & 4.39  &  -571.8860945 \\
      \textbf{$\beta$-C13}   &  4.58 &  -571.8857905 \\
       \textbf{$\beta$-C14}   & 4.59  & -571.8857689  \\
      \textbf{$\beta$-C15}   &  4.64 &  -571.8856979 \\
             \textbf{$\beta$-C16}   &  4.64 & -571.8856952  \\

\end{tabular}
&
\begin{tabular}{@{}ccc@{}}
\toprule
 \textbf{Conformer}  & \textbf{$\Delta$E$_{DLPNO}$} & \textbf{E$_{DLPNO}$} \\ \midrule
      \textbf{$\beta$-C17}   &  4.67 & -571.8856482  \\
       \textbf{$\beta$-C18}   & 4.71  &  -571.8855826 \\
      \textbf{$\alpha$-C20}   & 4.71  & -571.8855825  \\
      \textbf{$\beta$-C19}   & 4.73  &  -571.8855489 \\
      \textbf{$\beta$-C20}   & 4.77  &  -571.8854874 \\
      \textbf{$\beta$-C21}   & 4.82  &  -571.8854129 \\
      \textbf{$\beta$-C22}   & 4.88  & -571.8853091  \\
      \textbf{$\alpha$-C21}   & 4.99  & -571.8851443  \\
      \textbf{$\alpha$-C22}   & 5.18  & -571.88484  \\
      \textbf{$\beta$-C23}   & 5.18  &  -571.8848354 \\
      \textbf{$\beta$-C24}   & 5.24  & -571.8847511  \\
      \textbf{$\beta$-C25}   & 5.35  &  -571.8845617 \\
      \textbf{$\alpha$-C23}   & 5.66  & -571.8840726  \\
      \textbf{$\beta$-C26}   &  5.69 &  -571.8840272 \\
      \textbf{$\alpha$-C24}   & 5.88  &  -571.8837234 \\
      \textbf{$\beta$-C27}   & 5.98  & -571.8835707  \\
      \textbf{$\alpha$-C25}   & 6.12  &  -571.8833525 \\
      \textbf{$\alpha$-C26}   & 6.31  & -571.8830405  \\
      \textbf{$\beta$-C28}   & 6.49  & -571.8827556  \\
      \textbf{$\beta$-C29}   & 6.52  & -571.8827087  \\
      \textbf{$\alpha$-C27}   &  6.58 &  -571.882613 \\
      \textbf{$\beta$-C30}   & 6.77  &  -571.8823101 \\
      \textbf{$\alpha$-C28}   & 6.80  & -571.8822677  \\
      \textbf{$\alpha$-C29}   & 6.95  & -571.8820344  \\
      \textbf{$\beta$-C31}   & 6.97  & -571.8819999  \\
      \textbf{$\alpha$-C30}   & 6.99  &  -571.8819733 \\
      \textbf{$\alpha$-C31}   & 7.03  & -571.8818961  \\
      \textbf{$\alpha$-C32}   &  7.05 &  -571.8818782 \\
      \textbf{$\alpha$-C33}   & 7.09  & -571.8818029  \\
      \textbf{$\beta$-C32}   & 7.14  &  -571.8817274 \\
      \textbf{$\alpha$-C34}   & 7.15  &  -571.8817178 \\
      \textbf{$\beta$-C33}   & 7.19  &  -571.8816542 \\
      \textbf{$\alpha$-C35}   & 7.32  &  -571.8814346 \\
      \textbf{$\alpha$-C36}   & 7.59  &  -571.8810075 \\
       \textbf{$\beta$-C34}   & 7.77  &  -571.8807225 \\
\end{tabular}
&
\begin{tabular}{@{}ccc@{}}
\toprule
 \textbf{Conformer}  & \textbf{$\Delta$E$_{DLPNO}$} & \textbf{E$_{DLPNO}$} \\ \midrule
      \textbf{$\beta$-C35}   & 7.87  &  -571.8805714 \\
      \textbf{$\beta$-C36}   & 7.98  &  -571.8803934 \\
      \textbf{$\beta$-C37}   & 8.02  &  -571.8803347 \\
      \textbf{$\beta$-C38}   & 8.14  &  -571.8801441 \\
      \textbf{$\alpha$-C37}   & 8.18  &  -571.8800821 \\
      \textbf{$\beta$-C39}   &  8.31 &  -571.8798714 \\
      \textbf{$\alpha$-C38}   &  8.36 &  -571.8797938 \\
      \textbf{$\alpha$-C39}   &  8.37 &  -571.8797775 \\
      \textbf{$\beta$-C40}   & 8.43  &  -571.8796808 \\
      \textbf{$\alpha$-C40}   & 8.47  &  -571.879624 \\
      \textbf{$\beta$-C41}   & 8.59  &  -571.8794293 \\
      \textbf{$\beta$-C42}   & 8.60  &  -571.8794147 \\
      \textbf{$\beta$-C43}   &  8.64 &  -571.8793525 \\
      \textbf{$\beta$-C44}   & 8.69  &  -571.8792746 \\
      \textbf{$\beta$-C45}   & 8.73  &  -571.8792063 \\
      \textbf{$\beta$-C46}   & 8.79  &  -571.8791106 \\
      \textbf{$\beta$-C47}   & 9.05  &  -571.8787014 \\
      \textbf{$\alpha$-C41}   & 9.05  &  -571.8786992 \\
      \textbf{$\beta$-C48}   & 9.13  &  -571.8785707 \\
      \textbf{$\alpha$-C42}   &  9.14 &  -571.8785538 \\
      \textbf{$\alpha$-C43}   &  9.15 &  -571.8785335 \\
      \textbf{$\beta$-C49}   & 9.24  & -571.8783958 \\
      \textbf{$\beta$-C50}   &  9.27 &  -571.8783476 \\
      \textbf{$\beta$-C51}   & 9.44  & -571.8780735  \\
      \textbf{$\beta$-C52}   & 9.55  & -571.8779086  \\
      \textbf{$\beta$-C53}   & 9.59  & -571.8778352  \\
      \textbf{$\beta$-C54}   & 9.63  & -571.8777808  \\
      \textbf{$\beta$-C55}   & 9.65  &  -571.87774 \\
      \textbf{$\beta$-C56}   &  9.76 &  -571.8775754 \\
      \textbf{$\beta$-C57}   &  9.87 & -571.8773926  \\
      \textbf{$\alpha$-C44}   & 9.92  &  -571.8773178 \\
      \textbf{$\beta$-C58}   & 9.93  &  -571.8773005 \\
      \textbf{$\beta$-C59}   & 9.94  &  -571.877288 \\
      & & \\
            & & \\
\end{tabular}
\end{tabular}
\end{table*}

\subsubsection{$\alpha$ and $\beta$ Thermochemical Analysis}

\begin{figure*}[ht!]
 \centering
 \includegraphics[scale=0.6]{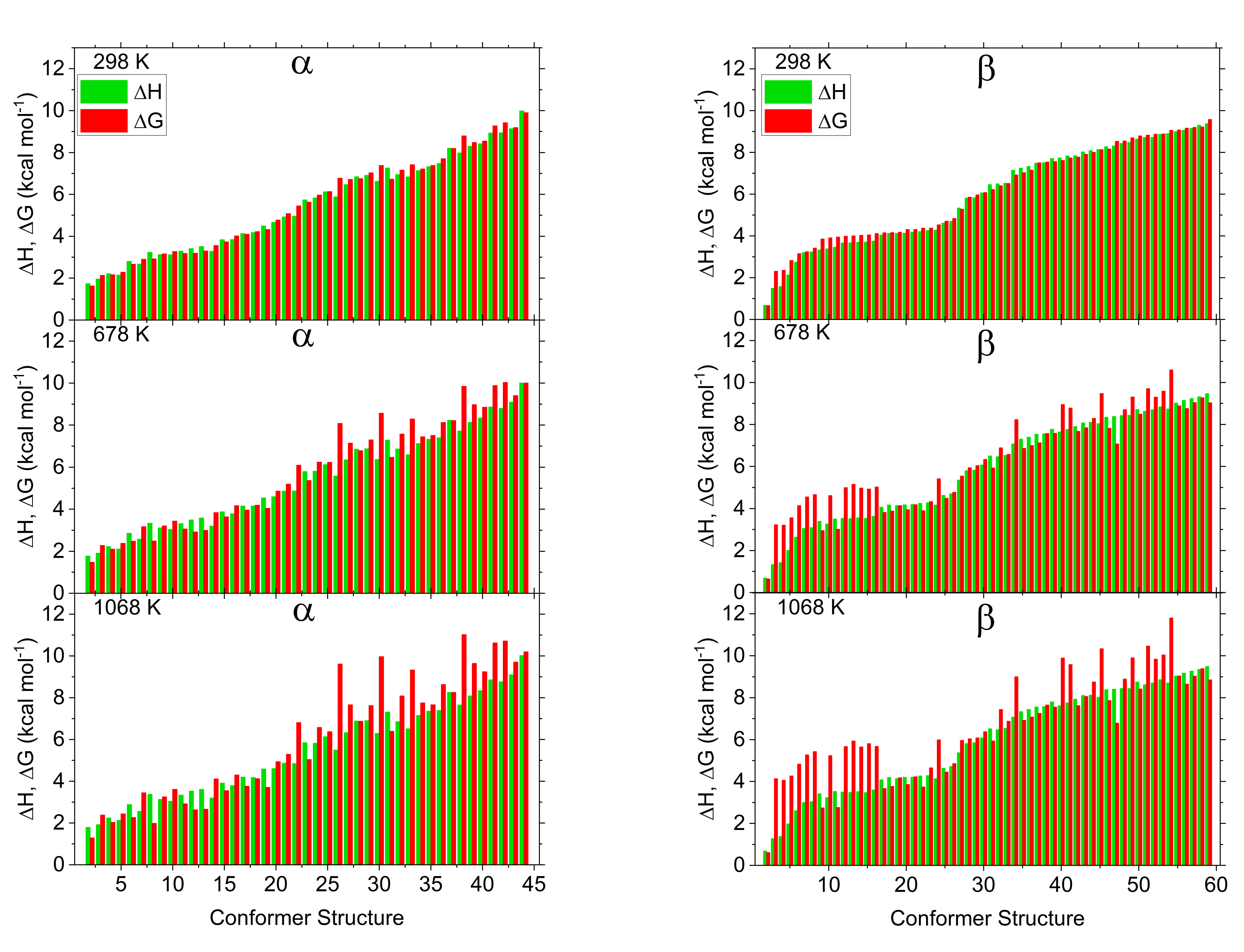}
 \caption{Relative enthalpies and free energies in \SI{}{\kilo\cal\per\mol}, calculated relative to the respective lowest energy conformer, for all $\alpha$ (left panels) and $\beta$ (right panels) anomers of D-xylopyranose at three different temperatures: 298, 678, and 1068 K.}
 \label{fig:betaHG}
\end{figure*}

A thermochemical analysis is conducted using the methodology detailed in Section \ref{sec:comp} to better understand the enthalpic and entropic contributions within a temperature range relevant to fast pyrolysis (298 to 1068 K). 
The results, depicted in Figure \ref{fig:betaHG}, display the relative variations of enthalpic and entropic contributions for all conformers of the $\alpha$ and $\beta$ anomers of D-xylopyranose at 298, 678, and 1068 K. 
At 298 K, the relative stability among the conformers aligns with the order reported for the electronic energies reported in the Tables \ref{tab:alfa_energies}, \ref{tab:beta_energies}. 
At 678 K, minor variations in enthalpic factors are observed across all conformers for both anomers. 
However, larger variations in total relative free energy, primarily due to entropic contributions, are noted for specific subsets of both anomers.
This observation emphasizes the role of free energy, in assessing the relative stability of D-xylopyranose conformers. 
Free energy is crucial as it incorporates both enthalpic and entropic corrections to the electronic and zero-point energy components. 
At lower temperatures, some conformers may appear more stable due to their enthalpy contributions. 
However, as temperature increases, entropic factors become more significant, altering the relative stability of these conformers. 
Table \ref{tab:conf_struc} shows the conformer structures most affected by entropic effects at temperatures of 1068 K, with free energy changes greater than 1 kcal mol$^{-1}$, classified by conformations.
It can be observed that for $\alpha$-D-xylopyranose there are nine conformers with variations greater than \SI{1}{\kilo\cal\per\mol} with two as half-chair, two as boat and five as twisted conformations with relative energies greater than \SI{6.5}{\kilo\cal\per\mol} as reported in Table \ref{tab:alfa_energies}.
As for $\beta$-D-xylopyranose, 23 conformers show high free energy variations throughout the relative energy range up to about \SI{9}{\kilo\cal\per\mol}.
It can be observed that for $\beta$-D-xylopyranose most of these structures are 14 chair in the lower energy range up to about \SI{4}{\kilo\cal\per\mol}, while for higher energies the conformations with higher energy changes are three boat and six twisted.
Therefore, determining and analyzing relative free energy is essential for understanding conformational stability, especially when aiming to gain insights into the behavior of D-xylopyranose at a wide range of temperatures.
The SI tables \ref{tab:alpha_HG_SI}, \ref{tab:beta_HG_SI} report the numerical values of the relative changes in enthalpy and free energy for both anomers.

\begin{table}[h!]
\centering
\footnotesize
\caption{Comparison of energy differences and anomeric ratios between the $\alpha$- and $\beta$-anomers of D-xylopyranose in gas-phase and solution at room temperature from Peña et al. (2013), Schmidt et al. (1996), Höög et al. (2001) and Mayes et al.(2014). The methodology used in each work is also reported.}
\begin{tabular}{c c c c c}
\hline
\textbf{Condition} & \textbf{Study} & \textbf{Method} & \textbf{$\Delta G_{\alpha-\beta}$} & \textbf{$\alpha$/$\beta$}\\
& & &\textbf{(kcal mol$^{-1}$)} & \textbf{(\%)} \\ 
\hline
Gas-phase & Peña et al.\cite{pena2013} & DFT & 0.54 (favoring $\alpha$) & 56/44 \\
%\cline{2-4}
& Schmidt et al.\cite{schmidt1996} & MD-FEP & 0.05 (favoring $\alpha$) & 52/48 \\
%\cline{2-4}
& Höög \& Widmalm\cite{hoog2001} & MD-FEP-TI & -0.80 (favoring $\beta$) & 33/67 \\
%\cline{2-4}
& This work & MTD-DFT & 0.50 (favoring $\alpha$) & 61/39 \\
\hline
Solution & Schmidt et al.\cite{schmidt1996} & MD-FEP &  -0.30 (favoring $\beta$) & 35/65 \\
%\cline{2-4}
& Höög \& Widmalm\cite{hoog2001} & MD-FEP-TI &  -0.40 (favoring $\beta$) & 33/67 \\
%\cline{2-4}
& Mayes et al.\cite{mayes2014} & MD-DFT &  -0.30 (favoring $\beta$) & 40/60 \\
\hline
\end{tabular}
\label{tab:anomer-comparison}
\end{table}

Table \ref{tab:anomer-comparison} reports the values of the free energy difference between alpha and beta anomers ($\Delta G_{\alpha-\beta}$), the anomeric ratio ($\alpha$/$\beta$), calculated in this work and a collection of data obtained in works reported in the literature both in gas-phase and in aqueous solution.
The studies show that in aqueous solution, the $\beta$-anomer is more stable than the $\alpha$-anomer, with a free energy difference of \SI{0.38}{\kilo\cal \per\mol}, consistent with an anomeric ratio of 65\% for $\beta$ and 35\% for $\alpha$ \cite{schmidt1996}. In gas-phase, the situation is reversed, with a slight preference for the $\alpha$-anomer due to the anomeric effect, with a free energy difference of \SI{0.15}{\kilo\cal \per\mol} \cite{pena2013}. 
This solvent-driven shift towards the $\beta$-anomer in solution is attributed to enhanced hydrogen bonding at the anomeric hydroxyl group, as corroborated by molecular dynamics simulations.

\begin{table*}[ht!]
\begin{threeparttable}
\footnotesize
 
    \centering
    \begin{tabular} {c c c c c} 
    \hline
         & \textbf{Chair}  & \textbf{Half-Chair} & \textbf{Boat} & \textbf{Twisted} \\
    \hline
      \textbf{$\alpha$-D-xylopyranose}   &   & \textbf{$\alpha$-C26}, \textbf{$\alpha$-C30} & \textbf{$\alpha$-C27}, \textbf{$\alpha$-C32}  & \textbf{$\alpha$-C33}, \textbf{$\alpha$-C38}, \textbf{$\alpha$-C39}, \textbf{$\alpha$-C41}, \textbf{$\alpha$-C42} \\
      \textbf{$\beta$-D-xylopyranose}   &  \textbf{$\beta$-C2} to \textbf{$\beta$-C7}, \textbf{$\beta$-C9}, \textbf{$\beta$-C11} to \textbf{$\beta$-C15}, \textbf{$\beta$-C23}, \textbf{$\beta$-C51} &  & \textbf{$\beta$-C40}, \textbf{$\beta$-C41}, \textbf{$\beta$-C53} & \textbf{$\beta$-C33}, \textbf{$\beta$-C45}, \textbf{$\beta$-C47}, \textbf{$\beta$-C49}, \textbf{$\beta$-C52}, \textbf{$\beta$-C54} 
    \end{tabular}
        \caption{List of $\alpha$,$\beta$-D-xylopyranose conformers with free energy  at 1068K relative to free energy at 298K that is greater than \SI{1}{\kilo\cal\per\mol}.}
    \begin{tablenotes}
      \footnotesize
      \item Note that for $\alpha$ anomers, the nine conformers showing large free energy changes are all non-chair. Whereas for the $\beta$ anomers the conformers that adjust in relative free energy with temperature are also chair conformers.
    \end{tablenotes}    
    \label{tab:conf_struc}
  \end{threeparttable}
\end{table*}

These simulations highlight the balance between internal molecular forces favoring the $\alpha$-anomer in isolation and solvation forces favoring the $\beta$-anomer in solution \cite{schmidt1996}. Across both gas-phase and solution, the $\beta$-pyranose form is shown to be more thermodynamically stable in solution, aligning with findings from previous computational and experimental studies. These investigations underline the crucial role of solvent in influencing the conformational preferences of D-xylopyranose and suggest that the stability of the $\beta$-anomer is a characteristic feature of aldopentoses, driven by cooperative hydrogen bonding networks \cite{pena2013}.
Höög \& Widmalm\cite{hoog2001} performed free energy simulations of D-xylopyranose and methyl D-xylopyranoside in gas-phase and aqueous solution. Their gas-phase simulations favored the $\beta$-anomer, with a substantial energy difference of \SI{-0.8}{\kilo\cal\per\mol} and a ratio of 33:67. In solution, the $\beta$-anomer was also preferred, but with a smaller free energy difference of \SI{-0.4}{\kilo\cal\per\mol} and a similar anomeric ratio.
Mayes et al.\cite{mayes2014} applied quantum mechanical methods to examine the energy landscape of several sugars, including D-xylopyranose. 
In solution, their calculations showed a preference for the $\beta$-anomer, with a free energy difference of \SI{-0.3}{\kilo\cal\per\mol} and an anomeric ratio of 40:60. Their work highlights the impact of solvation on stabilizing the $\beta$-anomer in solution.

\begin{figure*}[ht!]
    \centering
    \includegraphics[scale=0.68]{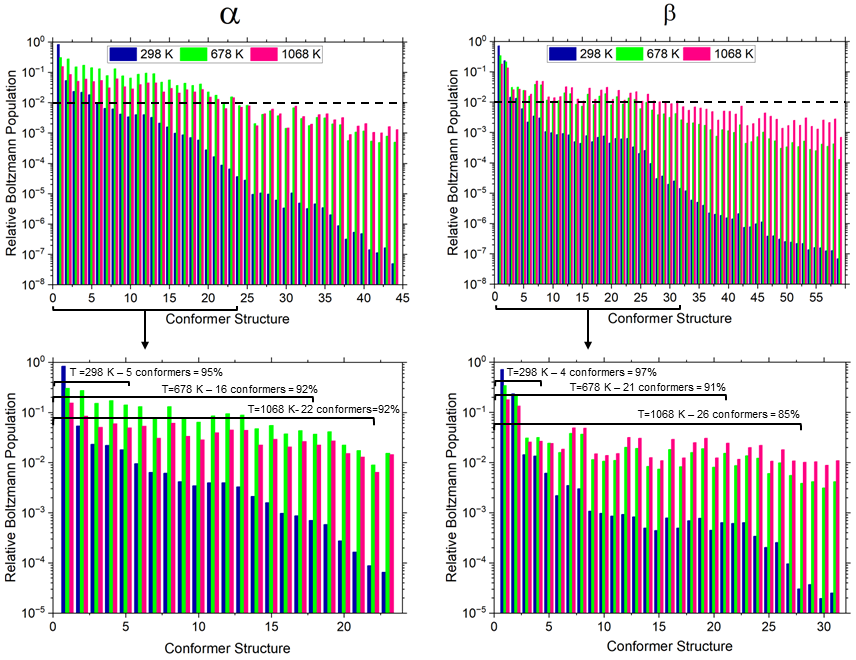}
    \caption{Relative Boltzmann population distributions of both $\alpha$ and $\beta$ anomers conformational spaces at 298, 678 and 1068 K. In the lower panels the population distributions are represented for the conformational space reduced following the application of the cut-off value. The horizontal lines represent the conformational subsets at the three temperatures listed in which the defined percentages of the population are contained.}
    \label{fig:relab}    
\end{figure*}

\subsection{Population Distribution}\label{sec:pop}
After determining the Gibbs free energy changes at 298 K, 678 K, and 1068 K, the population distribution was calculated relative to the lowest energy conformers identified, labeled $\alpha$-C1 and $\beta$-C1 for the $\alpha$ and $\beta$ anomers of D-xylopyranose, respectively.
The Figure \ref{fig:relab} illustrates the relative Boltzmann population trends for all conformers of the $\alpha$ and $\beta$ anomers. At 298 K, 95\% of the population is concentrated in the five lowest-energy conformers for $\alpha$, while the 97\% among the four lowest-energy conformers for $\beta$. 
At higher temperatures, 678 K and 1068 K, approximately 20\% and 35\% of the population, respectively, redistributes to higher-energy conformers.
By applying population cut-off (discarding conformers with a relative population percentage lower than 1\%), the number of conformers is reduced to 23 for $\alpha$  and 31 for $\beta$, as depicted in the lower part of Figure \ref{fig:relab}. 
This cut-off results in a neglectible population deviation of about 0.02\% at 298 K, 1.63\% at 678 K, and 5.5\% at 1068 K for both anomers.
Figure \ref{fig:relab} also highlights the ranges of structures where approximately 90\% of the population is renormalized with respect to the reduced conformational space at the three reported temperatures.
In Table \ref{tab:popboltz} are reported the relative population percentages at the three indicated temperatures with the more populated subset of conformers.

\begin{table}[h!]
%\scriptsize
    \centering
    \footnotesize
    \begin{tabular}{c c}
    {\Large\textbf{$\alpha$}} &    {\Large\textbf{$\beta$}} \\
\begin{tabular}{c c c}
  \hline
  \textbf{T (K)} & \textbf{\%} & \textbf{Conformers}\\
  \hline
   298   &  95   & \textbf{$\alpha$-C1} to \textbf{$\alpha$-C5} \\
   678   &  92   & \textbf{$\alpha$-C1} to \textbf{$\alpha$-C16} \\
   1068  &  92   & \textbf{$\alpha$-C1} to \textbf{$\alpha$-C22} \\  
\end{tabular}

         & 

\begin{tabular}{c c c}
  \hline
  \textbf{T (K)} & \textbf{\%} & \textbf{Conformers}\\
  \hline
   298   &  97   & \textbf{$\alpha$-C1} to \textbf{$\alpha$-C4} \\
   678   &  91   & \textbf{$\alpha$-C1} to \textbf{$\alpha$-C21} \\
   1068  &  85   & \textbf{$\alpha$-C1} to \textbf{$\alpha$-C26} \\  

\end{tabular}

\end{tabular}
\caption{Percentage of the population and conformers of the list in which it is concentrated at the three temperatures 298, 678 and 1068 K, as reported in Figure \ref{fig:relab}.}\label{tab:popboltz}
\end{table}

\subsection{Relevant Conformers}\label{sec:rel_conf}

\begin{figure*}[!ht]
    \centering
    \includegraphics[scale=0.59]{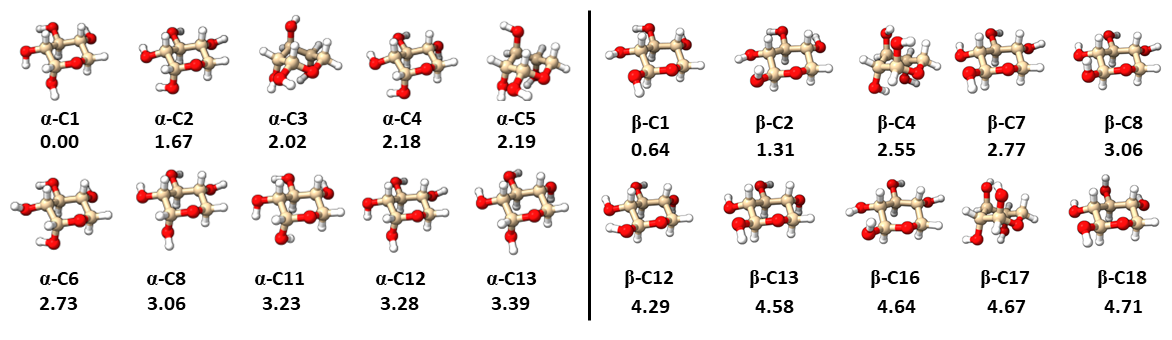}
    \caption{3D molecular structures of the most populated conformers for $\alpha$ and $\beta$ anomers of D-xylopyranose and Zero-point corrected DLPNO energies in kcal mol$^{-1}$ relative to the most stable conformer \textbf{$\alpha$-C1}.}\label{fig:3D}
\end{figure*}

\begin{figure*}[ht!]
    \centering
    % First row: Two figures side by side
    \begin{subfigure}[b]{0.38\textwidth}
        \centering
        \includegraphics[width=\textwidth]{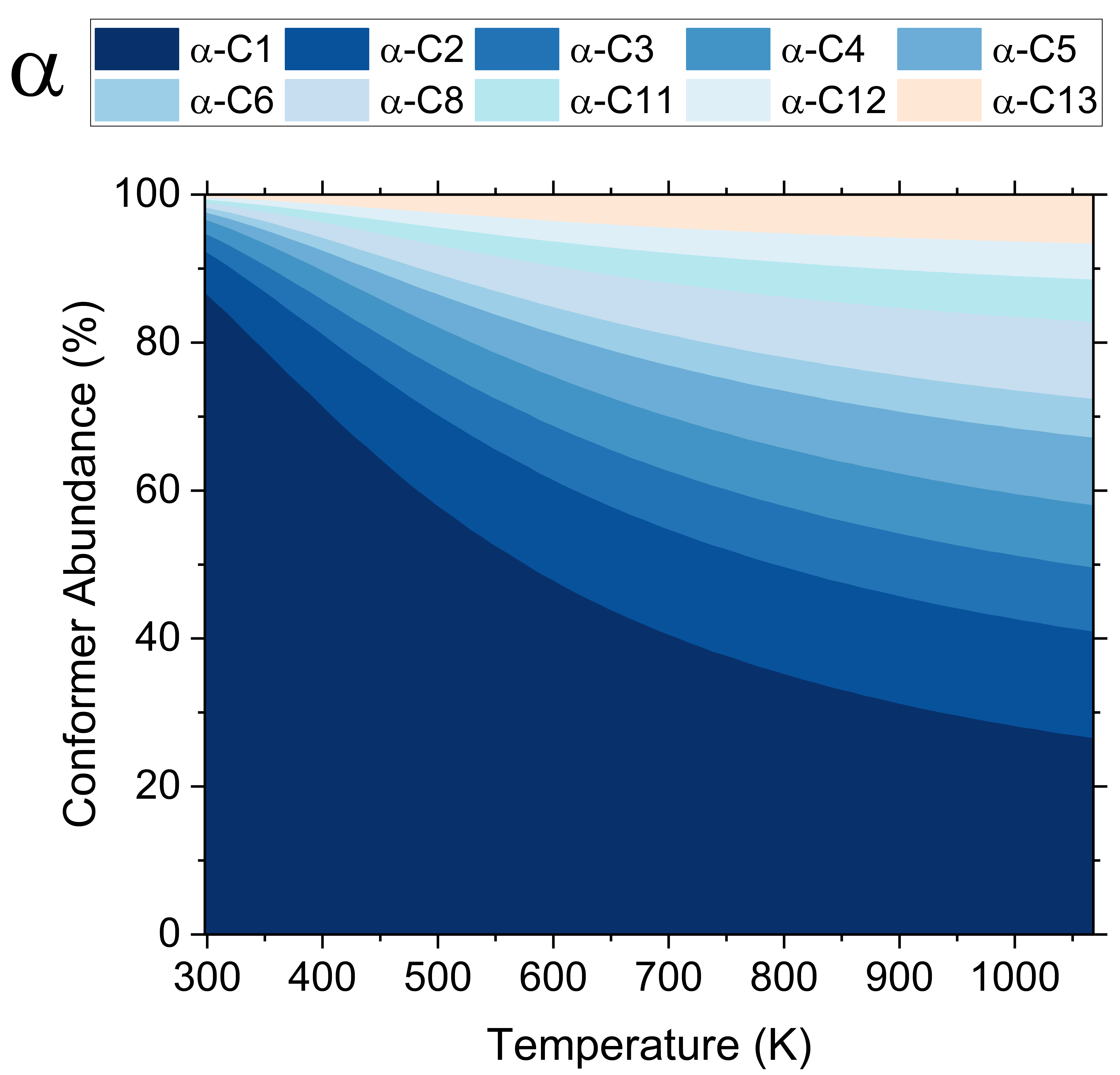}
        \label{fig:comp_plot_alfa}
    \end{subfigure}
    \hfill
    \begin{subfigure}[b]{0.40\textwidth}      
    \centering
       \includegraphics[width=\textwidth]{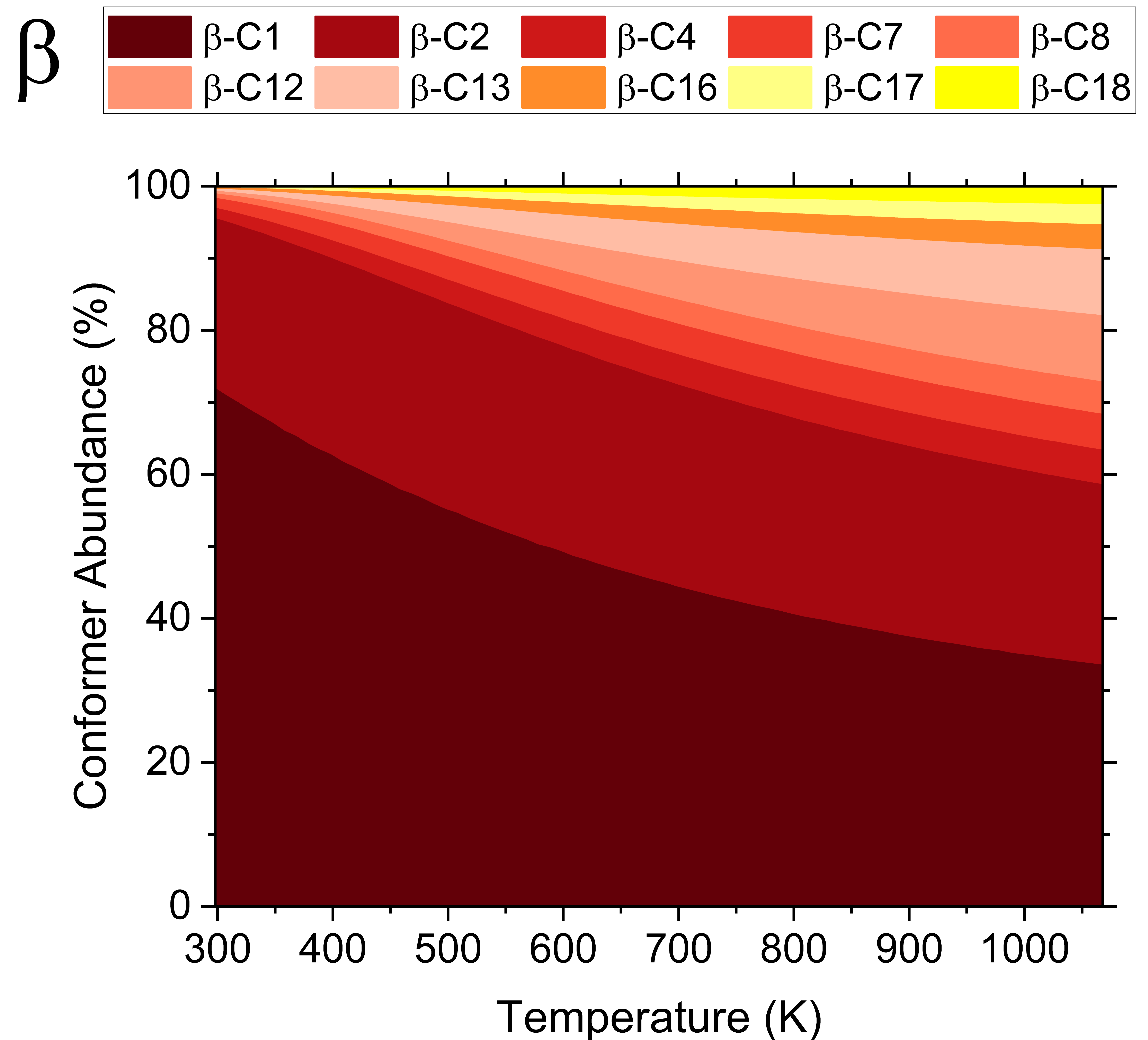}
        \label{fig:comp_plot_beta}
    \end{subfigure}
    
    % Second row: One figure below
%    \vspace{1cm} % Adjusts the vertical spacing
    \begin{subfigure}[b]{0.46\textwidth}
%        \centering
        \includegraphics[width=\textwidth]{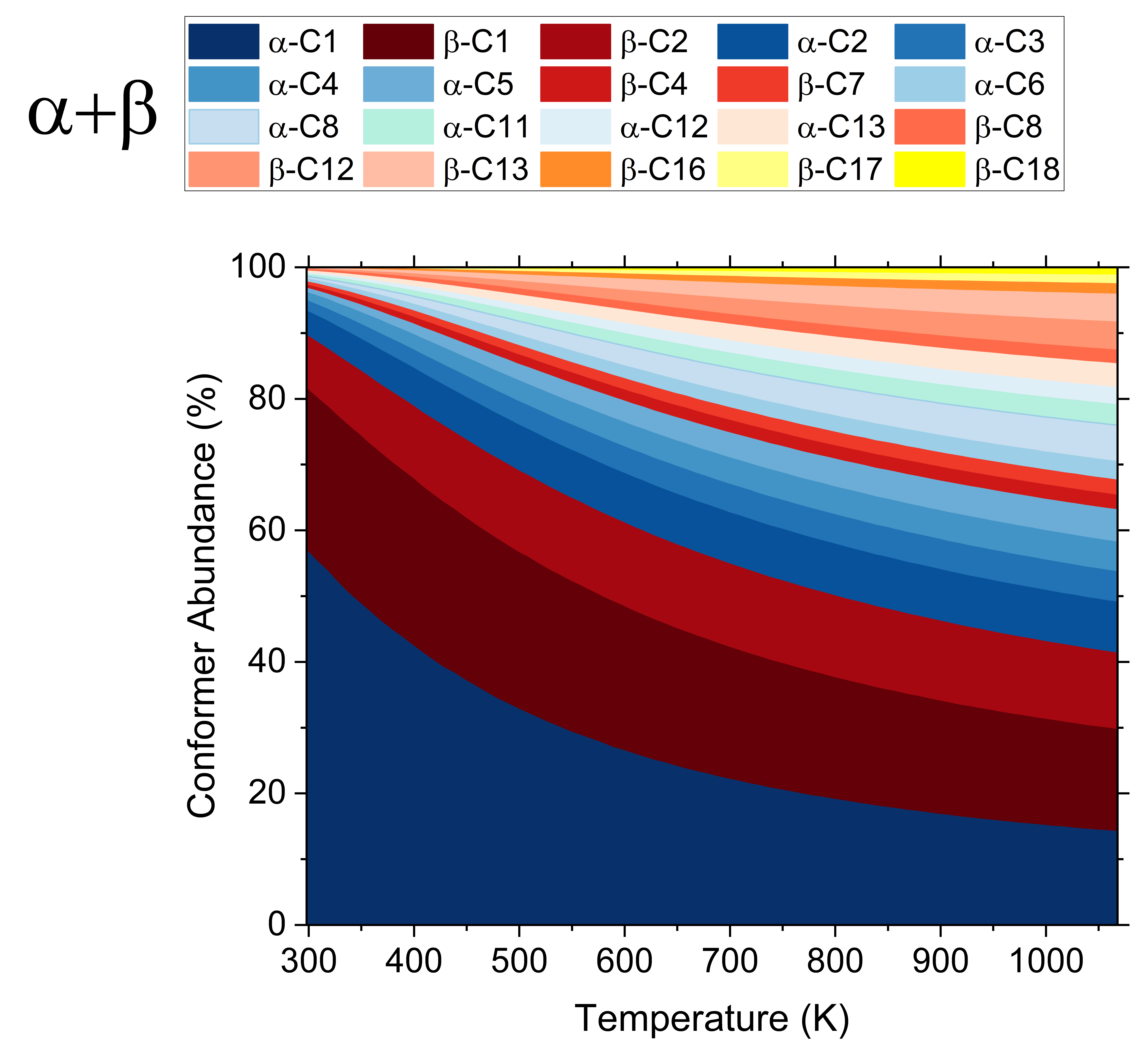}
        \label{fig:3}
    \end{subfigure}
    
    % Single caption for all figures
    \caption{Relative abundances as a function of temperature of the subset of 10 most abundant conformers in the reduced conformational space for $\alpha$, $\beta$ and the combination of both ($\alpha$+$\beta$).}
    \label{fig:comp_plot_alfa+beta}
\end{figure*}

Figure \ref{fig:3D} reports the molecular structures of the ten conformers highlighted in the composition plots for both D-xylopyranose anomers (the Cartesian coordinates of the optimized geometries are reported in the SI).
From the reduced set of conformers represented in Figure \ref{fig:relab}, the trends of the relative abundances of the 10 most populated conformers for both anomers ($\alpha$ and $\beta$) and the combination of them ($\alpha$+$\beta$) in the range of temperatures from 298 to 1068 K have been graphically depicted in Figure \ref{fig:comp_plot_alfa+beta}. The total relative population remains above 80\% when considering these subsets of 10 conformers.
For the $\alpha$ anomer, the population is uniformly redistributed among higher-energy conformers at elevated temperatures, while for the $\beta$ anomer, the two lowest energy conformers are more populated than the other conformers.
This difference is primarily due to the relative stability of the two lowest energy conformers in the $\beta$ anomer. 
Beyond $\beta$-C4, large variations in free energy (approximately \SI{4}{\kilo\cal\per\mol}) result in lower occupancies at higher temperatures. 
In contrast, for the $\alpha$ anomer, the lowest energy conformers exhibit minimal variations in free energy, maintaining a constant relative population percentage across the investigated temperature range. 
Significant variations in free energy for the $\alpha$ anomer are observed only in the highest energy conformers, which are already sparsely populated.

\section{Conclusions}\label{sec:concl}

This study presents a comprehensive conformational analysis of both anomeric forms of D-xylopyranose, the basic structural unit of hemicellulose, in gas-phase using semi-empirical GFN2-xTB tight binding calculations combined with metadynamic simulations to explore the conformational space across 298-1068 K, temperatures relevant to pyrolysis. The methodology employed performs subsequent refinement of geometries and frequencies at the rDSD level of theory and refinement of electronic energies with  DLPNO.

To our knowledge, such in-depth computational conformational analysis of the $\alpha$ and $\beta$ anomers of D-xylopyranose has not previously been performed.
The analysis identifies 57 conformers for the $\alpha$ anomer and 73 conformers for the $\beta$ anomer each representing PES minima structures, spanning a relative energy range of approximately \SI{10}{\kilo\cal\per\mol}. The relative molecular enthalpies and free energies of all conformers by means of the  modified rigid rotor harmonic oscillator approximation (mRRHO) which accounts for the anharmonic effects due to low-frequency modes is evaluated at 298 K, 673K and 1068 K. At all conditions the lowest enthalpy and entropy conformers are chair conformations, with competition between chair, boat and half-chair conformations for the next rank of most stable conformers.

At 0 K, of all conformers, the lowest in energy is an $\alpha$ anomer in the chair configuration which is \SI{0.64}{\kilo\cal\per\mol} lower in energy than the next lowest conformer, which is a $\beta$ anomer also in the chair configuration. The twenty one lowest energy $\alpha$ anomer conformers are all chair conformations, with the lowest energy non-chair conformation, a boat, being \SI{5.16}{\kilo\cal\per\mol} higher in energy that the lowest energy conformer.

The $\beta$ anomer data sets is differentiated from the $\alpha$ anomer data set in that it shows more competition between conformer variety in the low energy manifold. The sixteen lowest energy $\beta$ anomers are all chairs, with five half-chairs, one twisted, and one envelope, but no boat conformation making up the thirty lowest energy conformers. It is notable that lowest energy non-chair conformer, $\beta$-C17, is a twisted conformation which despite its highly strained nature, is just \SI{4.01}{\kilo\cal\per\mol} higher in energy than the overall lowest energy $\beta$ conformer.

A thermochemical analysis of the molecular enthalpy and free energy is conducted at pyrolysis reference conditions of 298 K, 678 K and 1068 K showing that free energy variations due to anharmonicity effects caused by low-frequency modes significantly influence the relative stability of the conformers for both $\alpha$ and $\beta$ anomers.

Of the $\alpha$ anomers, it is shown that the less stable conformers, specifically those that are higher in free energy by  6 kcal mol$^{-1}$ than the lowest free energy conformer, show large variation in free energy with respect to temperature. Whereas those conformers at lowest energy, do not show significant temperature dependent free energy change behaviour.

This is in notable contrast to the situation exhibited by the $\beta$ anomer, where the significant temperature dependent free energy change behaviour is observed for the most stable conformers.

The variation in relative stability was further elucidated through a Boltzmann population analysis with respect to temperature.

For both anomers, population cut-off value were imposed in order to reduce the conformational space and identify the most populated conformers at the three temperatures of interest for pyrolysis, i.e. 298, 673 and 1068 K.

For the $\alpha$ conformers it is shown that at 1068 K relative to 298 K, the population is redistributed to higher energy conformers, where 22 conformers are required to describe >85\% of the population, whereas, 26 conformers are required to describe >85\% of the $\beta$ anomer population.

As the conformation space is complex, an analysis is performed identify 20 conformers across both anomers which always describe 90\% of the population across 298-1068 K. Future studies should select from this reduced population for detailed study pertaining to thermochemical processes such as pyrolysis.

Our data show that the anomeric ratio derived from evaluation of the Boltzmann population at 298 K is 61:39 for $\alpha/\beta$. This is consistent with prior literature in that most suggestion also propose the $\alpha$ anomers to be favoured. However these studies report closer to a 50:50 for $\alpha/\beta$ split.

Furthermore, future additional investigations into the effect of solvent on the conformational space of \(\alpha\)- and \(\beta\)-D-xylopyranose could yield more relevant information to better define their conformational space under thermodynamically relevant conditions for pyrolysis. We recommend using these conformers for multiconformational \textit{ab initio} kinetic studies of characteristic pyrolysis decomposition pathways.

The data thus infer, that representative conformer for detailed study should be carefully chosen such that reliable results fairly representative of actual events would be revealed by detailed studies such as potential energy surface investigations or kinetic studies.

%uncchoosing a highly unstable and poorly populated conformational isomer as the initial reagent could also lead to an incorrect calculation of the formation rates of the final products, overestimating or underestimating their relative abundances making the kinetic model unreliable and unusable.
%The application of condensed reactivity indicators, such as Fukui functions, will also provide valuable insights for predicting the reactivity of these systems and how solvents may influence their reactivity\cite{ayarde2024}.
%We propose these conformers as starting points for \textit{ab initio} kinetic studies, as they are within a limited energy range that minimizes errors in initial activation barriers. 
%We also want to emphasize the importance and necessity of justifying the choice of the conformational isomer in studies concerning reactivity. 

\section*{Author contributions}
B. Ballotta - conceptualization, data curation, methodology, formal analysis, investigation, visualization, writing - original draft, writing review \& editing, resources. 
J. Lupi - conceptualization, data curation, methodology, investigation, writing - original draft, supervision, resources. 
L. Ayarde-Henríquez - conceptualization, supervision, validation, resources. 
S. Dooley - conceptualization, data curation, visualization, writing - original draft, resources, supervision, funding acquisition.

\section*{Conflicts of interest}
There are no conflicts of interest to declare.

\section*{Acknowledgements}
This publication has emanated from research conducted with the financial support of the European Union through the European Research Council, Mod-L-T, action number 101002649, and Science Foundation Ireland (SFI) under grant number 12/RC/2278\_2. This publication is also cofunded under the European Regional Development Fund under the AMBER award. TThe Irish Centre for High-End Computing (ICHEC) and Luxembourg national supercomputer MeluXina have provided computational facilities. B.B. and S.D. thank Dr. Manik Ghosh for engaging in invaluable discussions that have significantly improved the quality of the draft version of the paper.\\
%%%END OF MAIN TEXT%%%


%\balance


%%%REFERENCES%%%
\bibliography{xylo} %You need to replace "rsc" on this line with the name of your .bib file

\providecommand*{\mcitethebibliography}{\thebibliography}
\csname @ifundefined\endcsname{endmcitethebibliography}
{\let\endmcitethebibliography\endthebibliography}{}
\begin{mcitethebibliography}{64}
\providecommand*{\natexlab}[1]{#1}
\providecommand*{\mciteSetBstSublistMode}[1]{}
\providecommand*{\mciteSetBstMaxWidthForm}[2]{}
\providecommand*{\mciteBstWouldAddEndPuncttrue}
  {\def\EndOfBibitem{\unskip.}}
\providecommand*{\mciteBstWouldAddEndPunctfalse}
  {\let\EndOfBibitem\relax}
\providecommand*{\mciteSetBstMidEndSepPunct}[3]{}
\providecommand*{\mciteSetBstSublistLabelBeginEnd}[3]{}
\providecommand*{\EndOfBibitem}{}
\mciteSetBstSublistMode{f}
\mciteSetBstMaxWidthForm{subitem}
{(\emph{\alph{mcitesubitemcount}})}
\mciteSetBstSublistLabelBeginEnd{\mcitemaxwidthsubitemform\space}
{\relax}{\relax}

\bibitem[Sinnot(2007)]{sinnot2007}
M.~Sinnot, \emph{Carbohydrate Chemistry and Biochemistry: Structureand
  Metabolism}, RSC Publishing. UK, 2007\relax
\mciteBstWouldAddEndPuncttrue
\mciteSetBstMidEndSepPunct{\mcitedefaultmidpunct}
{\mcitedefaultendpunct}{\mcitedefaultseppunct}\relax
\EndOfBibitem
\bibitem[Angyal(1969)]{angyal1969}
S.~J. Angyal, \emph{Angew. Chem.}, 1969, \textbf{8}, 157--166\relax
\mciteBstWouldAddEndPuncttrue
\mciteSetBstMidEndSepPunct{\mcitedefaultmidpunct}
{\mcitedefaultendpunct}{\mcitedefaultseppunct}\relax
\EndOfBibitem
\bibitem[Hordvik(1971)]{hordvik1971}
A.~Hordvik, \emph{Acta Chem. Scand}, 1971, \textbf{25}, 2175--2182\relax
\mciteBstWouldAddEndPuncttrue
\mciteSetBstMidEndSepPunct{\mcitedefaultmidpunct}
{\mcitedefaultendpunct}{\mcitedefaultseppunct}\relax
\EndOfBibitem
\bibitem[Schmidt \emph{et~al.}(1996)Schmidt, Karplus, and Brady]{schmidt1996}
R.~K. Schmidt, M.~Karplus and J.~W. Brady, \emph{J. Am. Chem. Soc.}, 1996,
  \textbf{118}, 541--546\relax
\mciteBstWouldAddEndPuncttrue
\mciteSetBstMidEndSepPunct{\mcitedefaultmidpunct}
{\mcitedefaultendpunct}{\mcitedefaultseppunct}\relax
\EndOfBibitem
\bibitem[Höög and Widmalm(2001)]{hoog2001}
C.~Höög and G.~Widmalm, \emph{J. Phys. Chem. B}, 2001, \textbf{105},
  6375--6379\relax
\mciteBstWouldAddEndPuncttrue
\mciteSetBstMidEndSepPunct{\mcitedefaultmidpunct}
{\mcitedefaultendpunct}{\mcitedefaultseppunct}\relax
\EndOfBibitem
\bibitem[Mayes \emph{et~al.}(2014)Mayes, Broadbelt, and Beckham]{mayes2014}
H.~B. Mayes, L.~J. Broadbelt and G.~T. Beckham, \emph{J. Am. Chem. Soc.}, 2014,
  \textbf{136}, 1008--1022\relax
\mciteBstWouldAddEndPuncttrue
\mciteSetBstMidEndSepPunct{\mcitedefaultmidpunct}
{\mcitedefaultendpunct}{\mcitedefaultseppunct}\relax
\EndOfBibitem
\bibitem[Iglesias-Fernández \emph{et~al.}(2015)Iglesias-Fernández, Raich,
  Ardèvol, and Rovira]{iglesias2015}
J.~Iglesias-Fernández, L.~Raich, A.~Ardèvol and C.~Rovira, \emph{Chem. Sci.},
  2015, \textbf{6}, 1167--1177\relax
\mciteBstWouldAddEndPuncttrue
\mciteSetBstMidEndSepPunct{\mcitedefaultmidpunct}
{\mcitedefaultendpunct}{\mcitedefaultseppunct}\relax
\EndOfBibitem
\bibitem[Cremer and Pople(1975)]{cremer1975}
D.~Cremer and J.~Pople, \emph{J. Am. Chem. Soc.}, 1975, \textbf{97},
  1354--1358\relax
\mciteBstWouldAddEndPuncttrue
\mciteSetBstMidEndSepPunct{\mcitedefaultmidpunct}
{\mcitedefaultendpunct}{\mcitedefaultseppunct}\relax
\EndOfBibitem
\bibitem[Boeyens(1978)]{boeyens1978}
J.~C. Boeyens, \emph{J. Chem. Crystallogr.}, 1978, \textbf{8}, 317--320\relax
\mciteBstWouldAddEndPuncttrue
\mciteSetBstMidEndSepPunct{\mcitedefaultmidpunct}
{\mcitedefaultendpunct}{\mcitedefaultseppunct}\relax
\EndOfBibitem
\bibitem[Peña \emph{et~al.}(2013)Peña, Mata, Martín, Cabezas, Daly, and
  Alonso]{pena2013}
I.~Peña, S.~Mata, A.~Martín, C.~Cabezas, A.~M. Daly and J.~L. Alonso,
  \emph{Phys. Chem. Chem. Phys.}, 2013, \textbf{15}, 18243--18248\relax
\mciteBstWouldAddEndPuncttrue
\mciteSetBstMidEndSepPunct{\mcitedefaultmidpunct}
{\mcitedefaultendpunct}{\mcitedefaultseppunct}\relax
\EndOfBibitem
\bibitem[Zhou \emph{et~al.}(2017)Zhou, Li, Mabon, and Broadbelt]{zhou2017}
X.~Zhou, W.~Li, R.~Mabon and L.~J. Broadbelt, \emph{Energy Technol.}, 2017,
  \textbf{5}, 52--79\relax
\mciteBstWouldAddEndPuncttrue
\mciteSetBstMidEndSepPunct{\mcitedefaultmidpunct}
{\mcitedefaultendpunct}{\mcitedefaultseppunct}\relax
\EndOfBibitem
\bibitem[Zhou \emph{et~al.}(2018)Zhou, Li, Mabon, and Broadbelt]{zhou2018}
X.~Zhou, W.~Li, R.~Mabon and L.~J. Broadbelt, \emph{Energy Environ. Sci.},
  2018, \textbf{11}, 1240--1260\relax
\mciteBstWouldAddEndPuncttrue
\mciteSetBstMidEndSepPunct{\mcitedefaultmidpunct}
{\mcitedefaultendpunct}{\mcitedefaultseppunct}\relax
\EndOfBibitem
\bibitem[Goussougli \emph{et~al.}(2021)Goussougli, Sirjean, Glaude, and
  Fournet]{goussougli2021}
M.~Goussougli, B.~Sirjean, P.-A. Glaude and R.~Fournet, \emph{Phys. Chem. Chem.
  Phys.}, 2021, \textbf{23}, 2605--2621\relax
\mciteBstWouldAddEndPuncttrue
\mciteSetBstMidEndSepPunct{\mcitedefaultmidpunct}
{\mcitedefaultendpunct}{\mcitedefaultseppunct}\relax
\EndOfBibitem
\bibitem[Ayarde-Henríquez \emph{et~al.}(2024)Ayarde-Henríquez, Lupi, and
  Dooley]{ayarde2024}
L.~Ayarde-Henríquez, J.~Lupi and S.~Dooley, \emph{Phys. Chem. Chem. Phys.},
  2024, \textbf{26}, 12820--12837\relax
\mciteBstWouldAddEndPuncttrue
\mciteSetBstMidEndSepPunct{\mcitedefaultmidpunct}
{\mcitedefaultendpunct}{\mcitedefaultseppunct}\relax
\EndOfBibitem
\bibitem[Mettler \emph{et~al.}(2012)Mettler, Vlachos, and
  Dauenhauer]{mettler2012}
M.~S. Mettler, D.~G. Vlachos and P.~J. Dauenhauer, \emph{Energy Environ. Sci.},
  2012, \textbf{5}, 7797--7809\relax
\mciteBstWouldAddEndPuncttrue
\mciteSetBstMidEndSepPunct{\mcitedefaultmidpunct}
{\mcitedefaultendpunct}{\mcitedefaultseppunct}\relax
\EndOfBibitem
\bibitem[Pinheiro~Pires \emph{et~al.}(2019)Pinheiro~Pires, Arauzo, Fonts,
  Domine, Fernández~Arroyo, Garcia-Perez, Montoya, Chejne, Pfromm, and
  Garcia-Perez]{pinheiro2019}
A.~P. Pinheiro~Pires, J.~Arauzo, I.~Fonts, M.~E. Domine, A.~Fernández~Arroyo,
  M.~E. Garcia-Perez, J.~Montoya, F.~Chejne, P.~Pfromm and M.~Garcia-Perez,
  \emph{Energy Fuels}, 2019, \textbf{33}, 4683--4720\relax
\mciteBstWouldAddEndPuncttrue
\mciteSetBstMidEndSepPunct{\mcitedefaultmidpunct}
{\mcitedefaultendpunct}{\mcitedefaultseppunct}\relax
\EndOfBibitem
\bibitem[Kan \emph{et~al.}(2016)Kan, Strezov, and Evans]{KAN2016}
T.~Kan, V.~Strezov and T.~J. Evans, \emph{Renew. Sustain. Energy Rev.}, 2016,
  \textbf{57}, 1126--1140\relax
\mciteBstWouldAddEndPuncttrue
\mciteSetBstMidEndSepPunct{\mcitedefaultmidpunct}
{\mcitedefaultendpunct}{\mcitedefaultseppunct}\relax
\EndOfBibitem
\bibitem[Lupi \emph{et~al.}(2024)Lupi, Ayarde-Henríquez, Kelly, and
  Dooley]{lupi2024}
J.~Lupi, L.~Ayarde-Henríquez, M.~Kelly and S.~Dooley, \emph{J. Phys. Chem. A},
  2024, \textbf{128}, 1009--1024\relax
\mciteBstWouldAddEndPuncttrue
\mciteSetBstMidEndSepPunct{\mcitedefaultmidpunct}
{\mcitedefaultendpunct}{\mcitedefaultseppunct}\relax
\EndOfBibitem
\bibitem[Burnham \emph{et~al.}(2015)Burnham, Zhou, and Broadbelt]{burnham2015}
A.~K. Burnham, X.~Zhou and L.~J. Broadbelt, \emph{Energy Fuels}, 2015,
  \textbf{29}, 2906--2918\relax
\mciteBstWouldAddEndPuncttrue
\mciteSetBstMidEndSepPunct{\mcitedefaultmidpunct}
{\mcitedefaultendpunct}{\mcitedefaultseppunct}\relax
\EndOfBibitem
\bibitem[Hameed \emph{et~al.}(2019)Hameed, Sharma, Pareek, Wu, and
  Yu]{HAMEED2019}
S.~Hameed, A.~Sharma, V.~Pareek, H.~Wu and Y.~Yu, \emph{Biomass Bioenergy},
  2019, \textbf{123}, 104--122\relax
\mciteBstWouldAddEndPuncttrue
\mciteSetBstMidEndSepPunct{\mcitedefaultmidpunct}
{\mcitedefaultendpunct}{\mcitedefaultseppunct}\relax
\EndOfBibitem
\bibitem[Wang \emph{et~al.}(2015)Wang, Liu, Li, and Xu]{wang2015}
M.~Wang, C.~Liu, Q.~Li and X.~Xu, \emph{J. Mol. Model.}, 2015, \textbf{21},
  1--10\relax
\mciteBstWouldAddEndPuncttrue
\mciteSetBstMidEndSepPunct{\mcitedefaultmidpunct}
{\mcitedefaultendpunct}{\mcitedefaultseppunct}\relax
\EndOfBibitem
\bibitem[Huang \emph{et~al.}(2016)Huang, He, Wu, and Tong]{HUANG2016}
J.~Huang, C.~He, L.~Wu and H.~Tong, \emph{Chem. Phys. Lett.}, 2016,
  \textbf{658}, 114--124\relax
\mciteBstWouldAddEndPuncttrue
\mciteSetBstMidEndSepPunct{\mcitedefaultmidpunct}
{\mcitedefaultendpunct}{\mcitedefaultseppunct}\relax
\EndOfBibitem
\bibitem[Hu \emph{et~al.}(2019)Hu, Lu, Zhang, Wu, Li, Dong, and Yang]{HU2019}
B.~Hu, Q.~Lu, Z.-X. Zhang, Y.-T. Wu, K.~Li, C.-Q. Dong and Y.-P. Yang,
  \emph{Combust. Flame}, 2019, \textbf{206}, 177--188\relax
\mciteBstWouldAddEndPuncttrue
\mciteSetBstMidEndSepPunct{\mcitedefaultmidpunct}
{\mcitedefaultendpunct}{\mcitedefaultseppunct}\relax
\EndOfBibitem
\bibitem[Bao \emph{et~al.}(2016)Bao, Sripa, and Truhlar]{bao2016}
J.~L. Bao, P.~Sripa and D.~G. Truhlar, \emph{Phys. Chem. Chem. Phys.}, 2016,
  \textbf{18}, 1032--1041\relax
\mciteBstWouldAddEndPuncttrue
\mciteSetBstMidEndSepPunct{\mcitedefaultmidpunct}
{\mcitedefaultendpunct}{\mcitedefaultseppunct}\relax
\EndOfBibitem
\bibitem[Bao and Truhlar(2017)]{bao2017}
J.~L. Bao and D.~G. Truhlar, \emph{Chem. Soc. Rev.}, 2017, \textbf{46},
  7548--7596\relax
\mciteBstWouldAddEndPuncttrue
\mciteSetBstMidEndSepPunct{\mcitedefaultmidpunct}
{\mcitedefaultendpunct}{\mcitedefaultseppunct}\relax
\EndOfBibitem
\bibitem[Viegas(2018)]{viegas2018}
L.~P. Viegas, \emph{J. Phys. Chem. A}, 2018, \textbf{122}, 9721--9732\relax
\mciteBstWouldAddEndPuncttrue
\mciteSetBstMidEndSepPunct{\mcitedefaultmidpunct}
{\mcitedefaultendpunct}{\mcitedefaultseppunct}\relax
\EndOfBibitem
\bibitem[Viegas(2023)]{viegas2023}
L.~P. Viegas, \emph{J. Phys. Org. Chem.}, 2023, \textbf{36}, e4470\relax
\mciteBstWouldAddEndPuncttrue
\mciteSetBstMidEndSepPunct{\mcitedefaultmidpunct}
{\mcitedefaultendpunct}{\mcitedefaultseppunct}\relax
\EndOfBibitem
\bibitem[Yang \emph{et~al.}(2019)Yang, Lin, Zhou, Hu, Gai, Zhao, long, and
  Zhang]{yang2019}
Z.~Yang, X.~Lin, J.~Zhou, M.~Hu, Y.~Gai, W.~Zhao, B.~long and W.~Zhang,
  \emph{RSC Adv.}, 2019, \textbf{9}, 40437--40444\relax
\mciteBstWouldAddEndPuncttrue
\mciteSetBstMidEndSepPunct{\mcitedefaultmidpunct}
{\mcitedefaultendpunct}{\mcitedefaultseppunct}\relax
\EndOfBibitem
\bibitem[Møller \emph{et~al.}(2016)Møller, Otkjær, Hyttinen, Kurtén, and
  Kjaergaard]{moller2016}
K.~H. Møller, R.~V. Otkjær, N.~Hyttinen, T.~Kurtén and H.~G. Kjaergaard,
  \emph{J. Phys. Chem. A}, 2016, \textbf{120}, 10072--10087\relax
\mciteBstWouldAddEndPuncttrue
\mciteSetBstMidEndSepPunct{\mcitedefaultmidpunct}
{\mcitedefaultendpunct}{\mcitedefaultseppunct}\relax
\EndOfBibitem
\bibitem[Pracht \emph{et~al.}(2020)Pracht, Bohle, and Grimme]{crest}
P.~Pracht, F.~Bohle and S.~Grimme, \emph{Phys. Chem. Chem. Phys.}, 2020,
  \textbf{22}, 7169--7192\relax
\mciteBstWouldAddEndPuncttrue
\mciteSetBstMidEndSepPunct{\mcitedefaultmidpunct}
{\mcitedefaultendpunct}{\mcitedefaultseppunct}\relax
\EndOfBibitem
\bibitem[Bannwarth \emph{et~al.}(2019)Bannwarth, Ehlert, and
  Grimme]{bannwarth2019}
C.~Bannwarth, S.~Ehlert and S.~Grimme, \emph{J. Chem. Theory Comput.}, 2019,
  \textbf{15}, 1652--1671\relax
\mciteBstWouldAddEndPuncttrue
\mciteSetBstMidEndSepPunct{\mcitedefaultmidpunct}
{\mcitedefaultendpunct}{\mcitedefaultseppunct}\relax
\EndOfBibitem
\bibitem[Grimme(2019)]{grimme2019}
S.~Grimme, \emph{J. Chem. Theory Comput.}, 2019, \textbf{15}, 2847--2862\relax
\mciteBstWouldAddEndPuncttrue
\mciteSetBstMidEndSepPunct{\mcitedefaultmidpunct}
{\mcitedefaultendpunct}{\mcitedefaultseppunct}\relax
\EndOfBibitem
\bibitem[Barducci \emph{et~al.}(2011)Barducci, Bonomi, and
  Parrinello]{barducci2011}
A.~Barducci, M.~Bonomi and M.~Parrinello, \emph{Wiley Interdiscip. Rev. Comput.
  Mol. Sci.}, 2011, \textbf{1}, 826--843\relax
\mciteBstWouldAddEndPuncttrue
\mciteSetBstMidEndSepPunct{\mcitedefaultmidpunct}
{\mcitedefaultendpunct}{\mcitedefaultseppunct}\relax
\EndOfBibitem
\bibitem[Bussi and Laio(2020)]{bussi2020}
G.~Bussi and A.~Laio, \emph{Nat. Rev. Phys.}, 2020, \textbf{2}, 200--212\relax
\mciteBstWouldAddEndPuncttrue
\mciteSetBstMidEndSepPunct{\mcitedefaultmidpunct}
{\mcitedefaultendpunct}{\mcitedefaultseppunct}\relax
\EndOfBibitem
\bibitem[Pietrucci(2017)]{pietrucci2017}
F.~Pietrucci, \emph{Rev. Phys.}, 2017, \textbf{2}, 32--45\relax
\mciteBstWouldAddEndPuncttrue
\mciteSetBstMidEndSepPunct{\mcitedefaultmidpunct}
{\mcitedefaultendpunct}{\mcitedefaultseppunct}\relax
\EndOfBibitem
\bibitem[Zhao and Truhlar(2008)]{zhao2008}
Y.~Zhao and D.~G. Truhlar, \emph{Theor. Chem. Acc.}, 2008, \textbf{120},
  215--241\relax
\mciteBstWouldAddEndPuncttrue
\mciteSetBstMidEndSepPunct{\mcitedefaultmidpunct}
{\mcitedefaultendpunct}{\mcitedefaultseppunct}\relax
\EndOfBibitem
\bibitem[Clark \emph{et~al.}(1983)Clark, Chandrasekhar, Spitznagel, and
  Schleyer]{clark1983}
T.~Clark, J.~Chandrasekhar, G.~W. Spitznagel and P.~V.~R. Schleyer, \emph{J.
  Comput. Chem.}, 1983, \textbf{4}, 294--301\relax
\mciteBstWouldAddEndPuncttrue
\mciteSetBstMidEndSepPunct{\mcitedefaultmidpunct}
{\mcitedefaultendpunct}{\mcitedefaultseppunct}\relax
\EndOfBibitem
\bibitem[Hu \emph{et~al.}(2017)Hu, Lu, Jiang, Dong, Cui, Dong, and
  Yang]{hu2017}
B.~Hu, Q.~Lu, X.-y. Jiang, X.-c. Dong, M.-s. Cui, C.-q. Dong and Y.-p. Yang,
  \emph{Energy \& Fuels}, 2017, \textbf{31}, 8291--8299\relax
\mciteBstWouldAddEndPuncttrue
\mciteSetBstMidEndSepPunct{\mcitedefaultmidpunct}
{\mcitedefaultendpunct}{\mcitedefaultseppunct}\relax
\EndOfBibitem
\bibitem[Santra \emph{et~al.}(2019)Santra, Sylvetsky, and
  Martin]{santra2019minimally}
G.~Santra, N.~Sylvetsky and J.~M. Martin, \emph{J. Phys. Chem. A}, 2019,
  \textbf{123}, 5129--5143\relax
\mciteBstWouldAddEndPuncttrue
\mciteSetBstMidEndSepPunct{\mcitedefaultmidpunct}
{\mcitedefaultendpunct}{\mcitedefaultseppunct}\relax
\EndOfBibitem
\bibitem[Grimme \emph{et~al.}(2011)Grimme, Ehrlich, and Goerigk]{grimme2011}
S.~Grimme, S.~Ehrlich and L.~Goerigk, \emph{J. Comput. Chem.}, 2011,
  \textbf{32}, 1456--1465\relax
\mciteBstWouldAddEndPuncttrue
\mciteSetBstMidEndSepPunct{\mcitedefaultmidpunct}
{\mcitedefaultendpunct}{\mcitedefaultseppunct}\relax
\EndOfBibitem
\bibitem[Papajak \emph{et~al.}(2011)Papajak, Zheng, Xu, Leverentz, and
  Truhlar]{papajak2011}
E.~Papajak, J.~Zheng, X.~Xu, H.~R. Leverentz and D.~G. Truhlar, \emph{J. Chem.
  Theory Comput.}, 2011, \textbf{7}, 3027--3034\relax
\mciteBstWouldAddEndPuncttrue
\mciteSetBstMidEndSepPunct{\mcitedefaultmidpunct}
{\mcitedefaultendpunct}{\mcitedefaultseppunct}\relax
\EndOfBibitem
\bibitem[Ceselin \emph{et~al.}(2021)Ceselin, Barone, and Tasinato]{ceselin2021}
G.~Ceselin, V.~Barone and N.~Tasinato, \emph{J. Chem. Theory Comput.}, 2021,
  \textbf{17}, 7290--7311\relax
\mciteBstWouldAddEndPuncttrue
\mciteSetBstMidEndSepPunct{\mcitedefaultmidpunct}
{\mcitedefaultendpunct}{\mcitedefaultseppunct}\relax
\EndOfBibitem
\bibitem[Barone \emph{et~al.}(2020)Barone, Ceselin, Fusè, and
  Tasinato]{barone2020}
V.~Barone, G.~Ceselin, M.~Fusè and N.~Tasinato, \emph{Front. Chem.}, 2020,
  \textbf{8}, \relax
\mciteBstWouldAddEndPuncttrue
\mciteSetBstMidEndSepPunct{\mcitedefaultmidpunct}
{\mcitedefaultendpunct}{\mcitedefaultseppunct}\relax
\EndOfBibitem
\bibitem[Riplinger \emph{et~al.}(2013)Riplinger, Sandhoefer, Hansen, and
  Neese]{riplinger2013}
C.~Riplinger, B.~Sandhoefer, A.~Hansen and F.~Neese, \emph{J. Chem. Phys.},
  2013, \textbf{139}, 134101\relax
\mciteBstWouldAddEndPuncttrue
\mciteSetBstMidEndSepPunct{\mcitedefaultmidpunct}
{\mcitedefaultendpunct}{\mcitedefaultseppunct}\relax
\EndOfBibitem
\bibitem[Pavošević \emph{et~al.}(2017)Pavošević, Peng, Pinski, Riplinger,
  Neese, and Valeev]{pavosevic2017}
F.~Pavošević, C.~Peng, P.~Pinski, C.~Riplinger, F.~Neese and E.~F. Valeev,
  \emph{J. Chem. Phys.}, 2017, \textbf{146}, 174108\relax
\mciteBstWouldAddEndPuncttrue
\mciteSetBstMidEndSepPunct{\mcitedefaultmidpunct}
{\mcitedefaultendpunct}{\mcitedefaultseppunct}\relax
\EndOfBibitem
\bibitem[Peterson \emph{et~al.}(2008)Peterson, Adler, and Werner]{peterson2008}
K.~A. Peterson, T.~B. Adler and H.-J. Werner, \emph{J. Chem. Phys.}, 2008,
  \textbf{128}, 084102\relax
\mciteBstWouldAddEndPuncttrue
\mciteSetBstMidEndSepPunct{\mcitedefaultmidpunct}
{\mcitedefaultendpunct}{\mcitedefaultseppunct}\relax
\EndOfBibitem
\bibitem[Neese(2022)]{neese2022}
F.~Neese, \emph{Wiley Interdiscip. Rev. Comput. Mol. Sci.}, 2022, \textbf{12},
  e1606\relax
\mciteBstWouldAddEndPuncttrue
\mciteSetBstMidEndSepPunct{\mcitedefaultmidpunct}
{\mcitedefaultendpunct}{\mcitedefaultseppunct}\relax
\EndOfBibitem
\bibitem[Kabsch(1976)]{kabsch1976}
W.~Kabsch, \emph{Acta Crystallogr. A}, 1976, \textbf{32}, 922--923\relax
\mciteBstWouldAddEndPuncttrue
\mciteSetBstMidEndSepPunct{\mcitedefaultmidpunct}
{\mcitedefaultendpunct}{\mcitedefaultseppunct}\relax
\EndOfBibitem
\bibitem[Walker \emph{et~al.}(1991)Walker, Shao, and Volz]{WALKER1991}
M.~W. Walker, L.~Shao and R.~A. Volz, \emph{Comput. Vis. Image Underst.}, 1991,
  \textbf{54}, 358--367\relax
\mciteBstWouldAddEndPuncttrue
\mciteSetBstMidEndSepPunct{\mcitedefaultmidpunct}
{\mcitedefaultendpunct}{\mcitedefaultseppunct}\relax
\EndOfBibitem
\bibitem[Kromann(2019)]{RMSD}
J.~C. Kromann, \emph{{Calculate Root-mean-square deviation (RMSD) of Two
  Molecules Using Rotation}}, 2019,
  \url{https://github.com/charnley/rmsd.git}\relax
\mciteBstWouldAddEndPuncttrue
\mciteSetBstMidEndSepPunct{\mcitedefaultmidpunct}
{\mcitedefaultendpunct}{\mcitedefaultseppunct}\relax
\EndOfBibitem
\bibitem[Alecu \emph{et~al.}(2010)Alecu, Zheng, Zhao, and Truhlar]{alecu2010}
I.~M. Alecu, J.~Zheng, Y.~Zhao and D.~G. Truhlar, \emph{J. Chem. Theory
  Comput.}, 2010, \textbf{6}, 2872--2887\relax
\mciteBstWouldAddEndPuncttrue
\mciteSetBstMidEndSepPunct{\mcitedefaultmidpunct}
{\mcitedefaultendpunct}{\mcitedefaultseppunct}\relax
\EndOfBibitem
\bibitem[Yu \emph{et~al.}(2017)Yu, Fiedler, Alecu, and Truhlar]{yu2017}
H.~S. Yu, L.~J. Fiedler, I.~Alecu and D.~G. Truhlar, \emph{Comput. Phys.
  Commun.}, 2017, \textbf{210}, 132--138\relax
\mciteBstWouldAddEndPuncttrue
\mciteSetBstMidEndSepPunct{\mcitedefaultmidpunct}
{\mcitedefaultendpunct}{\mcitedefaultseppunct}\relax
\EndOfBibitem
\bibitem[Frisch \emph{et~al.}(2016)Frisch, Trucks, Schlegel, Scuseria, Robb,
  Cheeseman, Scalmani, Barone, Petersson, Nakatsuji, Li, Caricato, Marenich,
  Bloino, Janesko, Gomperts, Mennucci, Hratchian, Ortiz, Izmaylov, Sonnenberg,
  Williams-Young, Ding, Lipparini, Egidi, Goings, Peng, Petrone, Henderson,
  Ranasinghe, Zakrzewski, Gao, Rega, Zheng, Liang, Hada, Ehara, Toyota, Fukuda,
  Hasegawa, Ishida, Nakajima, Honda, Kitao, Nakai, Vreven, Throssell,
  Montgomery, Peralta, Ogliaro, Bearpark, Heyd, Brothers, Kudin, Staroverov,
  Keith, Kobayashi, Normand, Raghavachari, Rendell, Burant, Iyengar, Tomasi,
  Cossi, Millam, Klene, Adamo, Cammi, Ochterski, Martin, Morokuma, Farkas,
  Foresman, and Fox]{g16}
M.~J. Frisch, G.~W. Trucks, H.~B. Schlegel, G.~E. Scuseria, M.~A. Robb, J.~R.
  Cheeseman, G.~Scalmani, V.~Barone, G.~A. Petersson, H.~Nakatsuji, X.~Li,
  M.~Caricato, A.~V. Marenich, J.~Bloino, B.~G. Janesko, R.~Gomperts,
  B.~Mennucci, H.~P. Hratchian, J.~V. Ortiz, A.~F. Izmaylov, J.~L. Sonnenberg,
  D.~Williams-Young, F.~Ding, F.~Lipparini, F.~Egidi, J.~Goings, B.~Peng,
  A.~Petrone, T.~Henderson, D.~Ranasinghe, V.~G. Zakrzewski, J.~Gao, N.~Rega,
  G.~Zheng, W.~Liang, M.~Hada, M.~Ehara, K.~Toyota, R.~Fukuda, J.~Hasegawa,
  M.~Ishida, T.~Nakajima, Y.~Honda, O.~Kitao, H.~Nakai, T.~Vreven,
  K.~Throssell, J.~A. Montgomery, {Jr.}, J.~E. Peralta, F.~Ogliaro, M.~J.
  Bearpark, J.~J. Heyd, E.~N. Brothers, K.~N. Kudin, V.~N. Staroverov, T.~A.
  Keith, R.~Kobayashi, J.~Normand, K.~Raghavachari, A.~P. Rendell, J.~C.
  Burant, S.~S. Iyengar, J.~Tomasi, M.~Cossi, J.~M. Millam, M.~Klene, C.~Adamo,
  R.~Cammi, J.~W. Ochterski, R.~L. Martin, K.~Morokuma, O.~Farkas, J.~B.
  Foresman and D.~J. Fox, \emph{Gaussian 16 {R}evision {C}.01}, 2016, Gaussian
  Inc. Wallingford CT\relax
\mciteBstWouldAddEndPuncttrue
\mciteSetBstMidEndSepPunct{\mcitedefaultmidpunct}
{\mcitedefaultendpunct}{\mcitedefaultseppunct}\relax
\EndOfBibitem
\bibitem[Luchini \emph{et~al.}(2022)Luchini, Paton, Alegre-Requena,
  Rodríguez-Guerra, Berquist, Chen, IFunes, Velmiskina, froessler, Mayes,
  Vejaykummar, and sibo]{luchini2022}
G.~Luchini, R.~Paton, J.~Alegre-Requena, J.~Rodríguez-Guerra, E.~Berquist,
  J.~Chen, IFunes, J.~Velmiskina, froessler, H.~Mayes, S.~S.~S. Vejaykummar and
  sibo, \emph{patonlab/GoodVibes: Bug Fixes \& Updated References}, 2022,
  \url{https://doi.org/10.5281/zenodo.6977304}\relax
\mciteBstWouldAddEndPuncttrue
\mciteSetBstMidEndSepPunct{\mcitedefaultmidpunct}
{\mcitedefaultendpunct}{\mcitedefaultseppunct}\relax
\EndOfBibitem
\bibitem[Pitzer and Gwinn(1942)]{pitzer1942}
K.~S. Pitzer and W.~D. Gwinn, \emph{J. Chem. Phys.}, 1942, \textbf{10},
  428--440\relax
\mciteBstWouldAddEndPuncttrue
\mciteSetBstMidEndSepPunct{\mcitedefaultmidpunct}
{\mcitedefaultendpunct}{\mcitedefaultseppunct}\relax
\EndOfBibitem
\bibitem[Grimme(2012)]{grimme2012}
S.~Grimme, \emph{Chemistry – A European Journal}, 2012, \textbf{18},
  9955--9964\relax
\mciteBstWouldAddEndPuncttrue
\mciteSetBstMidEndSepPunct{\mcitedefaultmidpunct}
{\mcitedefaultendpunct}{\mcitedefaultseppunct}\relax
\EndOfBibitem
\bibitem[East and Radom(1997)]{east1997}
A.~L.~L. East and L.~Radom, \emph{J. Chem. Phys.}, 1997, \textbf{106},
  6655--6674\relax
\mciteBstWouldAddEndPuncttrue
\mciteSetBstMidEndSepPunct{\mcitedefaultmidpunct}
{\mcitedefaultendpunct}{\mcitedefaultseppunct}\relax
\EndOfBibitem
\bibitem[Klemm \emph{et~al.}(2005)Klemm, Heublein, Fink, and Bohn]{klemm2005}
D.~Klemm, B.~Heublein, H.-P. Fink and A.~Bohn, \emph{Angew. Chem.}, 2005,
  \textbf{44}, 3358--3393\relax
\mciteBstWouldAddEndPuncttrue
\mciteSetBstMidEndSepPunct{\mcitedefaultmidpunct}
{\mcitedefaultendpunct}{\mcitedefaultseppunct}\relax
\EndOfBibitem
\bibitem[Karplus and McCammon(2002)]{karplus2002}
M.~Karplus and J.~A. McCammon, \emph{Nat. Struct. Biol.}, 2002, \textbf{9},
  646--652\relax
\mciteBstWouldAddEndPuncttrue
\mciteSetBstMidEndSepPunct{\mcitedefaultmidpunct}
{\mcitedefaultendpunct}{\mcitedefaultseppunct}\relax
\EndOfBibitem
\bibitem[Leach(2001)]{leach2001}
A.~R. Leach, \emph{Molecular Modelling: Principles and Application}, Pearson
  education, 2001\relax
\mciteBstWouldAddEndPuncttrue
\mciteSetBstMidEndSepPunct{\mcitedefaultmidpunct}
{\mcitedefaultendpunct}{\mcitedefaultseppunct}\relax
\EndOfBibitem
\bibitem[Pfaendtner \emph{et~al.}(2007)Pfaendtner, Yu, and
  Broadbelt]{pfaendtner20071}
J.~Pfaendtner, X.~Yu and L.~J. Broadbelt, \emph{Theor. Chem. Acc.}, 2007,
  \textbf{118}, 881--898\relax
\mciteBstWouldAddEndPuncttrue
\mciteSetBstMidEndSepPunct{\mcitedefaultmidpunct}
{\mcitedefaultendpunct}{\mcitedefaultseppunct}\relax
\EndOfBibitem
\bibitem[Ayala and Schlegel(1998)]{ayala1998identification}
P.~Y. Ayala and H.~B. Schlegel, \emph{J. Chem. Phys.}, 1998, \textbf{108},
  2314--2325\relax
\mciteBstWouldAddEndPuncttrue
\mciteSetBstMidEndSepPunct{\mcitedefaultmidpunct}
{\mcitedefaultendpunct}{\mcitedefaultseppunct}\relax
\EndOfBibitem
\bibitem[McClurg \emph{et~al.}(1997)McClurg, Flagan, and
  Goddard~III]{mcclurg1997hindered}
R.~B. McClurg, R.~C. Flagan and W.~A. Goddard~III, \emph{J. Chem. Phys.}, 1997,
  \textbf{106}, 6675--6680\relax
\mciteBstWouldAddEndPuncttrue
\mciteSetBstMidEndSepPunct{\mcitedefaultmidpunct}
{\mcitedefaultendpunct}{\mcitedefaultseppunct}\relax
\EndOfBibitem
\bibitem[McClurg(1999)]{mcclurg1999comment}
R.~B. McClurg, \emph{J. Chem. Phys.}, 1999, \textbf{111}, 7163--7164\relax
\mciteBstWouldAddEndPuncttrue
\mciteSetBstMidEndSepPunct{\mcitedefaultmidpunct}
{\mcitedefaultendpunct}{\mcitedefaultseppunct}\relax
\EndOfBibitem
\end{mcitethebibliography}
\bibliographystyle{rsc} %the RSC's .bst file

\end{document}

\maketitle

\newpage

\tableofcontents

\newpage
\renewcommand{\thepage}{S\arabic{page}} 
\renewcommand{\thesection}{S\arabic{section}}  
\renewcommand{\thetable}{S\arabic{table}}  
\renewcommand{\thefigure}{S\arabic{figure}}	

\section{Cartesian coordinates}
Here are reported the cartesian coordinates of each optimized structures of the alpha and beta anomers of the xylopyranose.\\
\subsection{$\alpha$ Anomers}

%\centering

\begin{center}
   \textbf{$\alpha$-C1}
\end{center}

\begin{table}[H]
\caption{Cartesian coordinates in {\AA} of the $\alpha$-C1 structure optimized at revDSD-PBEP86-D3(BJ)/jun-cc-pVTZ level of theory. }
\begin{center}
\begin{tabular}{cD{.}{.}{8} D{.}{.}{8} D{.}{.}{8}}
    \toprule
    \multicolumn{1}{c}{Atom} & \multicolumn{1}{c}{X} & \multicolumn{1}{c}{Y} & \multicolumn{1}{c}{Z} \\
    \midrule
    O & 0.619769 & -1.668939 & -0.704117 \\
    O & -1.257073 & 1.972283 & -0.085253 \\
    O & 1.633132 & 1.804865 & -0.066799 \\
    O & -2.620295 & -0.472617 & 0.449909 \\
    O & 1.856040 & -0.686109 & 1.032640 \\
    C & -0.548687 & 0.774476 & 0.203058 \\
    C & 0.797548 & 0.745078 & -0.508512 \\
    C & -1.398890 & -0.390570 & -0.263845 \\
    C & 1.467785 & -0.616433 & -0.327839 \\
    C & -0.636617 & -1.682790 & -0.018219 \\
    H & -0.390590 & 0.678370 & 1.285386 \\
    H & 0.632697 & 0.907171 & -1.575837 \\
    H & -1.586696 & -0.276528 & -1.339755 \\
    H & 2.340077 & -0.704598 & -0.979366 \\
    H & -0.489922 & -1.822926 & 1.056620 \\
    H & -1.185746 & -2.534616 & -0.414740 \\
    H & -0.662364 & 2.707691 & 0.102411 \\
    H & 2.004808 & 1.525281 & 0.779511 \\
    H & -3.022398 & 0.403020 & 0.420222 \\
    H & 2.420722 & -1.457297 & 1.146652 \\
    \bottomrule
    \end{tabular}
\end{center}
\end{table}

\newpage

\begin{center}
   \textbf{$\alpha$-C2}
\end{center}

\begin{table}[H]
\caption{Cartesian coordinates in {\AA} of the $\alpha$-C2 structure optimized at revDSD-PBEP86-D3(BJ)/jun-cc-pVTZ level of theory. }
\begin{center}
    \begin{tabular}{cD{.}{.}{8} D{.}{.}{8} D{.}{.}{8}}
    \toprule
    \multicolumn{1}{c}{Atom} & \multicolumn{1}{c}{X} & \multicolumn{1}{c}{Y} & \multicolumn{1}{c}{Z} \\
    \midrule
    O & 0.582031 & -1.690071 & -0.699067 \\
    O & -1.061260 & 2.060917 & -0.137429 \\
    O & 1.744965 & 1.660186 & 0.016085 \\
    O & -2.602025 & -0.259481 & 0.486303 \\
    O & 1.836459 & -0.862371 & 1.035001 \\
    C & -0.512831 & 0.787893 & 0.182141 \\
    C & 0.834067 & 0.704796 & -0.512033 \\
    C & -1.394703 & -0.359461 & -0.266641 \\
    C & 1.465921 & -0.675501 & -0.312486 \\
    C & -0.658945 & -1.668511 & 0.003039 \\
    H & -0.366493 & 0.700869 & 1.265944 \\
    H & 0.688031 & 0.866305 & -1.586447 \\
    H & -1.598008 & -0.257058 & -1.339077 \\
    H & 2.334262 & -0.787747 & -0.967495 \\
    H & -0.501953 & -1.788934 & 1.078377 \\
    H & -1.225964 & -2.521933 & -0.370541 \\
    H & -1.929347 & 2.108593 & 0.278380 \\
    H & 1.282314 & 2.505725 & 0.046948 \\
    H & -3.269962 & -0.819581 & 0.080177 \\
    H & 2.184711 & -0.014971 & 1.342482 \\
    \bottomrule
    \end{tabular}
\end{center}
\end{table}

\begin{center}
   \textbf{$\alpha$-C3}
\end{center}

\begin{table}[H]
\caption{Cartesian coordinates in {\AA} of the $\alpha$-C3 structure optimized at revDSD-PBEP86-D3(BJ)/jun-cc-pVTZ level of theory. }
\begin{center}
    \begin{tabular}{cD{.}{.}{8} D{.}{.}{8} D{.}{.}{8}}
    \toprule
    \multicolumn{1}{c}{Atom} & \multicolumn{1}{c}{X} & \multicolumn{1}{c}{Y} & \multicolumn{1}{c}{Z} \\
    \midrule
    O & 0.725189 & -1.209358 & -0.674231 \\
    O & -1.133064 & 1.758031 & -0.984268 \\
    O & 0.695045 & 0.338685 & 1.755033 \\
    O & -1.610059 & -1.195239 & 1.151932 \\
    O & 2.579501 & -0.009241 & -0.163213 \\
    C & -0.979619 & 0.945790 & 0.177325 \\
    C & 0.518798 & 0.922251 & 0.470463 \\
    C & -1.511841 & -0.476098 & -0.061853 \\
    C & 1.248715 & 0.097768 & -0.585886 \\
    C & -0.641438 & -1.195683 & -1.085902 \\
    H & -1.495780 & 1.377933 & 1.040466 \\
    H & 0.906026 & 1.946438 & 0.448353 \\
    H & -2.528545 & -0.406026 & -0.464182 \\
    H & 1.172725 & 0.591337 & -1.564875 \\
    H & -0.945433 & -2.238170 & -1.163849 \\
    H & -0.734562 & -0.710876 & -2.066106 \\
    H & -2.071569 & 1.919285 & -1.119125 \\
    H & 1.613672 & 0.041906 & 1.794898 \\
    H & -0.777205 & -1.057409 & 1.625534 \\
    H & 3.000102 & -0.691611 & -0.698030 \\
    \bottomrule
    \end{tabular}
\end{center}
\end{table}

\begin{center}
   \textbf{$\alpha$-C4}
\end{center}

\begin{table}[H]
\caption{Cartesian coordinates in {\AA} of the $\alpha$-C4 structure optimized at revDSD-PBEP86-D3(BJ)/jun-cc-pVTZ level of theory. }
\begin{center}
    \begin{tabular}{cD{.}{.}{8} D{.}{.}{8} D{.}{.}{8}}
    \toprule
    \multicolumn{1}{c}{Atom} & \multicolumn{1}{c}{X} & \multicolumn{1}{c}{Y} & \multicolumn{1}{c}{Z} \\
    \midrule
    O & 0.571918 & -1.685950 & -0.716288 \\
    O & -1.039754 & 2.076082 & -0.135216 \\
    O & 1.769179 & 1.643894 & 0.030871 \\
    O & -2.663702 & -0.284774 & 0.377596 \\
    O & 1.789607 & -0.886123 & 1.055814 \\
    C & -0.503224 & 0.797470 & 0.176770 \\
    C & 0.848245 & 0.703034 & -0.506727 \\
    C & -1.400713 & -0.339391 & -0.281444 \\
    C & 1.457612 & -0.685407 & -0.301239 \\
    C & -0.685533 & -1.664957 & -0.047065 \\
    H & -0.353720 & 0.706587 & 1.262331 \\
    H & 0.712834 & 0.866781 & -1.582035 \\
    H & -1.621378 & -0.218434 & -1.344405 \\
    H & 2.341593 & -0.804418 & -0.933476 \\
    H & -0.546584 & -1.831553 & 1.028200 \\
    H & -1.262887 & -2.490207 & -0.460755 \\
    H & -1.952845 & 2.087387 & 0.174389 \\
    H & 1.328072 & 2.501166 & 0.042617 \\
    H & -2.531660 & -0.531975 & 1.300318 \\
    H & 2.170274 & -0.054861 & 1.368821 \\
    \bottomrule
    \end{tabular}
\end{center}
\end{table}

\begin{center}
   \textbf{$\alpha$-C5}
\end{center}

\begin{table}[H]
\caption{Cartesian coordinates in {\AA} of the $\alpha$-C5 structure optimized at revDSD-PBEP86-D3(BJ)/jun-cc-pVTZ level of theory. }
\begin{center}
    \begin{tabular}{cD{.}{.}{8} D{.}{.}{8} D{.}{.}{8}}
    \toprule
    \multicolumn{1}{c}{Atom} & \multicolumn{1}{c}{X} & \multicolumn{1}{c}{Y} & \multicolumn{1}{c}{Z} \\
    \midrule
    O & 0.712345 & -1.233724 & -0.640633 \\
    O & -1.233960 & 1.660130 & -1.058175 \\
    O & 0.708799 & 0.397392 & 1.746715 \\
    O & -1.606894 & -1.160150 & 1.185252 \\
    O & 2.576154 & -0.031793 & -0.168285 \\
    C & -0.982498 & 0.949946 & 0.152621 \\
    C & 0.521488 & 0.930045 & 0.443446 \\
    C & -1.519446 & -0.471555 & -0.047382 \\
    C & 1.244053 & 0.069676 & -0.590165 \\
    C & -0.655896 & -1.219764 & -1.055285 \\
    H & -1.505410 & 1.402395 & 1.000642 \\
    H & 0.921867 & 1.951347 & 0.384352 \\
    H & -2.534787 & -0.396980 & -0.443789 \\
    H & 1.168822 & 0.535273 & -1.583046 \\
    H & -0.966215 & -2.261786 & -1.108652 \\
    H & -0.745181 & -0.756665 & -2.045847 \\
    H & -1.084171 & 2.597204 & -0.900709 \\
    H & 1.626154 & 0.096169 & 1.789025 \\
    H & -0.764870 & -1.028804 & 1.642751 \\
    H & 2.986041 & -0.743066 & -0.673129 \\
    \bottomrule
    \end{tabular}
\end{center}
\end{table}

\begin{center}
   \textbf{$\alpha$-C6}
\end{center}

\begin{table}[H]
\caption{Cartesian coordinates in {\AA} of the $\alpha$-C6 structure optimized at revDSD-PBEP86-D3(BJ)/jun-cc-pVTZ level of theory. }
\begin{center}
    \begin{tabular}{cD{.}{.}{8} D{.}{.}{8} D{.}{.}{8}}
    \toprule
    \multicolumn{1}{c}{Atom} & \multicolumn{1}{c}{X} & \multicolumn{1}{c}{Y} & \multicolumn{1}{c}{Z} \\
    \midrule
    O & 0.555789 & -1.696079 & -0.707574 \\
    O & -1.177993 & 2.010336 & -0.130582 \\
    O & 1.672212 & 1.700713 & 0.084858 \\
    O & -2.626383 & -0.367894 & 0.452210 \\
    O & 1.811416 & -0.897187 & 1.042110 \\
    C & -0.514105 & 0.792617 & 0.186974 \\
    C & 0.835826 & 0.695506 & -0.501238 \\
    C & -1.406517 & -0.347037 & -0.266975 \\
    C & 1.447702 & -0.699880 & -0.306605 \\
    C & -0.693638 & -1.666283 & -0.014252 \\
    H & -0.371682 & 0.710270 & 1.271262 \\
    H & 0.703180 & 0.880029 & -1.571758 \\
    H & -1.593881 & -0.235554 & -1.343377 \\
    H & 2.321238 & -0.826088 & -0.956444 \\
    H & -0.544576 & -1.810475 & 1.058662 \\
    H & -1.276313 & -2.496619 & -0.408450 \\
    H & -0.628944 & 2.732934 & 0.191944 \\
    H & 2.374793 & 1.923159 & -0.533787 \\
    H & -2.999753 & 0.519211 & 0.399263 \\
    H & 2.120000 & -0.045509 & 1.377087 \\
    \bottomrule
    \end{tabular}
\end{center}
\end{table}

\begin{center}
   \textbf{$\alpha$-C7}
\end{center}

\begin{table}[H]
\caption{Cartesian coordinates in {\AA} of the $\alpha$-C7 structure optimized at revDSD-PBEP86-D3(BJ)/jun-cc-pVTZ level of theory. }
\begin{center}
    \begin{tabular}{cD{.}{.}{8} D{.}{.}{8} D{.}{.}{8}}
    \toprule
    \multicolumn{1}{c}{Atom} & \multicolumn{1}{c}{X} & \multicolumn{1}{c}{Y} & \multicolumn{1}{c}{Z} \\
    \midrule
    O & 0.664416 & -1.217190 & -0.734419 \\
    O & -1.213494 & 1.752644 & -0.936029 \\
    O & 0.850492 & 0.314607 & 1.753194 \\
    O & -1.505680 & -1.116795 & 1.176624 \\
    O & 2.608399 & -0.112956 & -0.301402 \\
    C & -0.943864 & 0.968925 & 0.222082 \\
    C & 0.565195 & 0.877080 & 0.481848 \\
    C & -1.526594 & -0.411843 & -0.068778 \\
    C & 1.273436 & 0.067072 & -0.612123 \\
    C & -0.691611 & -1.113503 & -1.130045 \\
    H & -1.434945 & 1.382323 & 1.109426 \\
    H & 0.992077 & 1.884859 & 0.486622 \\
    H & -2.554709 & -0.293250 & -0.427789 \\
    H & 1.219016 & 0.583270 & -1.576899 \\
    H & -1.047634 & -2.134899 & -1.287084 \\
    H & -0.787492 & -0.566131 & -2.076426 \\
    H & -0.958321 & 2.662199 & -0.754600 \\
    H & 0.250930 & -0.438857 & 1.861745 \\
    H & -1.608717 & -2.057112 & 0.997991 \\
    H & 2.637358 & -0.311265 & 0.645368 \\
    \bottomrule
    \end{tabular}
\end{center}
\end{table}

\begin{center}
   \textbf{$\alpha$-C8}
\end{center}

\begin{table}[H]
\caption{Cartesian coordinates in {\AA} of the $\alpha$-C8 structure optimized at revDSD-PBEP86-D3(BJ)/jun-cc-pVTZ level of theory. }
\begin{center}
    \begin{tabular}{cD{.}{.}{8} D{.}{.}{8} D{.}{.}{8}}
    \toprule
    \multicolumn{1}{c}{Atom} & \multicolumn{1}{c}{X} & \multicolumn{1}{c}{Y} & \multicolumn{1}{c}{Z} \\
    \midrule
    O & 0.653679 & -1.640440 & -0.745955 \\
    O & -1.113319 & 2.036544 & -0.080300 \\
    O & 1.681265 & 1.771943 & -0.032048 \\
    O & -2.582545 & -0.360218 & 0.478690 \\
    O & 1.780989 & -0.751610 & 1.100886 \\
    C & -0.525035 & 0.775816 & 0.219309 \\
    C & 0.822702 & 0.745703 & -0.481036 \\
    C & -1.372639 & -0.379581 & -0.277827 \\
    C & 1.495069 & -0.606776 & -0.264384 \\
    C & -0.608359 & -1.685679 & -0.080544 \\
    H & -0.376118 & 0.671108 & 1.300115 \\
    H & 0.645264 & 0.838977 & -1.560848 \\
    H & -1.582261 & -0.224703 & -1.342626 \\
    H & 2.402173 & -0.677805 & -0.868470 \\
    H & -0.480083 & -1.869162 & 0.990194 \\
    H & -1.147865 & -2.525384 & -0.520479 \\
    H & -1.972631 & 2.060069 & 0.354458 \\
    H & 1.164714 & 2.586038 & -0.042559 \\
    H & -3.242665 & -0.891107 & 0.023699 \\
    H & 2.358493 & -1.514687 & 1.203213 \\
    \bottomrule
    \end{tabular}
\end{center}
\end{table}

\begin{center}
   \textbf{$\alpha$-C9}
\end{center}

\begin{table}[H]
\caption{Cartesian coordinates in {\AA} of the $\alpha$-C9 structure optimized at revDSD-PBEP86-D3(BJ)/jun-cc-pVTZ level of theory. }
\begin{center}
    \begin{tabular}{cD{.}{.}{8} D{.}{.}{8} D{.}{.}{8}}
    \toprule
    \multicolumn{1}{c}{Atom} & \multicolumn{1}{c}{X} & \multicolumn{1}{c}{Y} & \multicolumn{1}{c}{Z} \\
    \midrule
    O & 0.610353 & -1.670055 & -0.719077 \\
    O & -1.169136 & 2.037086 & -0.016713 \\
    O & 1.714510 & 1.694025 & 0.024592 \\
    O & -2.608761 & -0.453717 & 0.465380 \\
    O & 1.831719 & -0.841028 & 1.037244 \\
    C & -0.529107 & 0.772434 & 0.189557 \\
    C & 0.822398 & 0.723153 & -0.504999 \\
    C & -1.394801 & -0.391081 & -0.260316 \\
    C & 1.475449 & -0.650908 & -0.313408 \\
    C & -0.632537 & -1.687152 & -0.017947 \\
    H & -0.384579 & 0.703914 & 1.269982 \\
    H & 0.683115 & 0.879501 & -1.584038 \\
    H & -1.594905 & -0.293322 & -1.338911 \\
    H & 2.353600 & -0.742025 & -0.959044 \\
    H & -0.473844 & -1.824685 & 1.054408 \\
    H & -1.188980 & -2.537319 & -0.408251 \\
    H & -1.332598 & 2.137726 & -0.962563 \\
    H & 1.210390 & 2.504982 & 0.160698 \\
    H & -2.951228 & 0.444784 & 0.534641 \\
    H & 2.201146 & -0.002723 & 1.344351 \\
    \bottomrule
    \end{tabular}
\end{center}
\end{table}

\begin{center}
   \textbf{$\alpha$-C10}
\end{center}

\begin{table}[H]
\caption{Cartesian coordinates in {\AA} of the $\alpha$-C10 structure optimized at revDSD-PBEP86-D3(BJ)/jun-cc-pVTZ level of theory. }
\begin{center}
    \begin{tabular}{cD{.}{.}{8} D{.}{.}{8} D{.}{.}{8}}
    \toprule
    \multicolumn{1}{c}{Atom} & \multicolumn{1}{c}{X} & \multicolumn{1}{c}{Y} & \multicolumn{1}{c}{Z} \\
    \midrule
    O & 0.675759 & -1.200616 & -0.749617 \\
    O & -1.111170 & 1.828778 & -0.878678 \\
    O & 0.850399 & 0.286777 & 1.755893 \\
    O & -1.512613 & -1.147446 & 1.155700 \\
    O & 2.612343 & -0.088238 & -0.304595 \\
    C & -0.942011 & 0.962694 & 0.239261 \\
    C & 0.564080 & 0.876011 & 0.496346 \\
    C & -1.520391 & -0.420485 & -0.075878 \\
    C & 1.276617 & 0.089177 & -0.610395 \\
    C & -0.678296 & -1.100890 & -1.145892 \\
    H & -1.425352 & 1.364458 & 1.135651 \\
    H & 0.972217 & 1.887900 & 0.518051 \\
    H & -2.549987 & -0.308026 & -0.442086 \\
    H & 1.215547 & 0.618576 & -1.567295 \\
    H & -1.028195 & -2.121719 & -1.320663 \\
    H & -0.775413 & -0.538746 & -2.083859 \\
    H & -2.037464 & 2.081365 & -0.935471 \\
    H & 0.267344 & -0.481286 & 1.843904 \\
    H & -1.601293 & -2.086021 & 0.960111 \\
    H & 2.644849 & -0.289584 & 0.641388 \\
    \bottomrule
    \end{tabular}
\end{center}
\end{table}

\begin{center}
   \textbf{$\alpha$-C11}
\end{center}

\begin{table}[H]
\caption{Cartesian coordinates in {\AA} of the $\alpha$-C11 structure optimized at revDSD-PBEP86-D3(BJ)/jun-cc-pVTZ level of theory. }
\begin{center}
    \begin{tabular}{cD{.}{.}{8} D{.}{.}{8} D{.}{.}{8}}
    \toprule
    \multicolumn{1}{c}{Atom} & \multicolumn{1}{c}{X} & \multicolumn{1}{c}{Y} & \multicolumn{1}{c}{Z} \\
    \midrule
    O & 0.594340 & -1.681807 & -0.694181 \\
    O & -1.212889 & 2.000170 & -0.074349 \\
    O & 1.661062 & 1.771926 & -0.074785 \\
    O & -2.628249 & -0.427735 & 0.446022 \\
    O & 1.972637 & -0.782915 & 1.001316 \\
    C & -0.535419 & 0.784356 & 0.197012 \\
    C & 0.810397 & 0.722707 & -0.511366 \\
    C & -1.409644 & -0.361022 & -0.272980 \\
    C & 1.465578 & -0.648923 & -0.316151 \\
    C & -0.670969 & -1.670064 & -0.037039 \\
    H & -0.388220 & 0.682431 & 1.284100 \\
    H & 0.645749 & 0.876958 & -1.580505 \\
    H & -1.596708 & -0.238466 & -1.347495 \\
    H & 2.338739 & -0.742984 & -0.959579 \\
    H & -0.564674 & -1.840321 & 1.042810 \\
    H & -1.232611 & -2.507319 & -0.446735 \\
    H & -0.587605 & 2.716863 & 0.087094 \\
    H & 2.124486 & 1.443714 & 0.706946 \\
    H & -3.032071 & 0.446519 & 0.401927 \\
    H & 1.238046 & -0.836827 & 1.622399 \\
    \bottomrule
    \end{tabular}
\end{center}
\end{table}

\begin{center}
   \textbf{$\alpha$-C12}
\end{center}

\begin{table}[H]
\caption{Cartesian coordinates in {\AA} of the $\alpha$-C12 structure optimized at revDSD-PBEP86-D3(BJ)/jun-cc-pVTZ level of theory. }
\begin{center}
    \begin{tabular}{cD{.}{.}{8} D{.}{.}{8} D{.}{.}{8}}
    \toprule
    \multicolumn{1}{c}{Atom} & \multicolumn{1}{c}{X} & \multicolumn{1}{c}{Y} & \multicolumn{1}{c}{Z} \\
    \midrule
    O & 0.631004 & -1.654310 & -0.724049 \\
    O & -1.187713 & 2.023683 & -0.122173 \\
    O & 1.681010 & 1.795940 & -0.055273 \\
    O & -2.588840 & -0.352809 & 0.496843 \\
    O & 1.845386 & -0.697724 & 1.034064 \\
    C & -0.547592 & 0.803712 & 0.201625 \\
    C & 0.809830 & 0.775126 & -0.493090 \\
    C & -1.376959 & -0.385776 & -0.257125 \\
    C & 1.470323 & -0.597084 & -0.326859 \\
    C & -0.610756 & -1.679236 & -0.019991 \\
    H & -0.401475 & 0.720293 & 1.287698 \\
    H & 0.649285 & 0.948555 & -1.559599 \\
    H & -1.586050 & -0.267744 & -1.326895 \\
    H & 2.347062 & -0.668933 & -0.973562 \\
    H & -0.449907 & -1.818628 & 1.053165 \\
    H & -1.157266 & -2.538966 & -0.409108 \\
    H & -2.070759 & 1.979600 & 0.261870 \\
    H & 1.969716 & 1.555537 & 0.832990 \\
    H & -3.253365 & -0.877589 & 0.041347 \\
    H & 2.436911 & -1.450829 & 1.129436 \\
    \bottomrule
    \end{tabular}
\end{center}
\end{table}

\begin{center}
   \textbf{$\alpha$-C13}
\end{center}

\begin{table}[H]
\caption{Cartesian coordinates in {\AA} of the $\alpha$-C13 structure optimized at revDSD-PBEP86-D3(BJ)/jun-cc-pVTZ level of theory. }
\begin{center}
    \begin{tabular}{cD{.}{.}{8} D{.}{.}{8} D{.}{.}{8}}
    \toprule
    \multicolumn{1}{c}{Atom} & \multicolumn{1}{c}{X} & \multicolumn{1}{c}{Y} & \multicolumn{1}{c}{Z} \\
    \midrule
    O & 0.646509 & -1.634318 & -0.764699 \\
    O & -1.095449 & 2.051001 & -0.080559 \\
    O & 1.703342 & 1.758579 & -0.010975 \\
    O & -2.642659 & -0.386954 & 0.366324 \\
    O & 1.737081 & -0.777624 & 1.118674 \\
    C & -0.518550 & 0.785058 & 0.213774 \\
    C & 0.835339 & 0.745544 & -0.473209 \\
    C & -1.378252 & -0.362357 & -0.292474 \\
    C & 1.487270 & -0.615589 & -0.253024 \\
    C & -0.630124 & -1.681663 & -0.131222 \\
    H & -0.368680 & 0.678303 & 1.296930 \\
    H & 0.669296 & 0.844108 & -1.554067 \\
    H & -1.605551 & -0.191936 & -1.347238 \\
    H & 2.408509 & -0.691288 & -0.834273 \\
    H & -0.517183 & -1.912284 & 0.935656 \\
    H & -1.178491 & -2.492181 & -0.608787 \\
    H & -2.007040 & 2.032082 & 0.231920 \\
    H & 1.208638 & 2.585227 & -0.047877 \\
    H & -2.494523 & -0.627920 & 1.288231 \\
    H & 2.320330 & -1.535550 & 1.226303 \\
    \bottomrule
    \end{tabular}
\end{center}
\end{table}

\begin{center}
   \textbf{$\alpha$-C14}
\end{center}

\begin{table}[H]
\caption{Cartesian coordinates in {\AA} of the $\alpha$-C14 structure optimized at revDSD-PBEP86-D3(BJ)/jun-cc-pVTZ level of theory. }
\begin{center}
    \begin{tabular}{cD{.}{.}{8} D{.}{.}{8} D{.}{.}{8}}
    \toprule
    \multicolumn{1}{c}{Atom} & \multicolumn{1}{c}{X} & \multicolumn{1}{c}{Y} & \multicolumn{1}{c}{Z} \\
    \midrule
    O & 0.692478 & -1.173410 & -0.767809 \\
    O & -1.151464 & 1.827659 & -0.858421 \\
    O & 0.787224 & 0.146274 & 1.786272 \\
    O & -1.569732 & -1.268726 & 0.979764 \\
    O & 2.629338 & -0.087467 & -0.258113 \\
    C & -0.932924 & 0.962549 & 0.250922 \\
    C & 0.562889 & 0.828579 & 0.561846 \\
    C & -1.525052 & -0.391578 & -0.150160 \\
    C & 1.294726 & 0.100593 & -0.571362 \\
    C & -0.653527 & -1.056042 & -1.202490 \\
    H & -1.447773 & 1.333711 & 1.146739 \\
    H & 1.002699 & 1.823659 & 0.679305 \\
    H & -2.527114 & -0.233239 & -0.562818 \\
    H & 1.249064 & 0.675352 & -1.502857 \\
    H & -1.008059 & -2.068981 & -1.389781 \\
    H & -0.711256 & -0.473914 & -2.130394 \\
    H & -0.832976 & 2.706715 & -0.630606 \\
    H & 0.271858 & -0.673339 & 1.741689 \\
    H & -2.228604 & -0.934974 & 1.597384 \\
    H & 2.652742 & -0.374244 & 0.665261 \\
    \bottomrule
    \end{tabular}
\end{center}
\end{table}

\begin{center}
   \textbf{$\alpha$-C15}
\end{center}

\begin{table}[H]
\caption{Cartesian coordinates in {\AA} of the $\alpha$-C15 structure optimized at revDSD-PBEP86-D3(BJ)/jun-cc-pVTZ level of theory. }
\begin{center}
    \begin{tabular}{cD{.}{.}{8} D{.}{.}{8} D{.}{.}{8}}
    \toprule
    \multicolumn{1}{c}{Atom} & \multicolumn{1}{c}{X} & \multicolumn{1}{c}{Y} & \multicolumn{1}{c}{Z} \\
    \midrule
    O & 0.631696 & -1.647864 & -0.739087 \\
    O & -1.179736 & 2.032149 & -0.118384 \\
    O & 1.689964 & 1.794078 & -0.043330 \\
    O & -2.649248 & -0.393306 & 0.380371 \\
    O & 1.812816 & -0.710737 & 1.052379 \\
    C & -0.544783 & 0.807672 & 0.195248 \\
    C & 0.817969 & 0.776961 & -0.488762 \\
    C & -1.381159 & -0.377583 & -0.272173 \\
    C & 1.467964 & -0.598596 & -0.316650 \\
    C & -0.624391 & -1.680731 & -0.064148 \\
    H & -0.394705 & 0.720618 & 1.283606 \\
    H & 0.665673 & 0.951015 & -1.556196 \\
    H & -1.605348 & -0.249027 & -1.333765 \\
    H & 2.358183 & -0.671341 & -0.944264 \\
    H & -0.476409 & -1.861072 & 1.008152 \\
    H & -1.176861 & -2.517813 & -0.487799 \\
    H & -2.105568 & 1.937211 & 0.135053 \\
    H & 1.969263 & 1.554082 & 0.848010 \\
    H & -2.501855 & -0.565891 & 1.317662 \\
    H & 2.410102 & -1.458684 & 1.152852 \\
    \bottomrule
    \end{tabular}
\end{center}
\end{table}

\begin{center}
   \textbf{$\alpha$-C16}
\end{center}

\begin{table}[H]
\caption{Cartesian coordinates in {\AA} of the $\alpha$-C16 structure optimized at revDSD-PBEP86-D3(BJ)/jun-cc-pVTZ level of theory. }
\begin{center}
    \begin{tabular}{cD{.}{.}{8} D{.}{.}{8} D{.}{.}{8}}
    \toprule
    \multicolumn{1}{c}{Atom} & \multicolumn{1}{c}{X} & \multicolumn{1}{c}{Y} & \multicolumn{1}{c}{Z} \\
    \midrule
    O & 0.700355 & -1.144968 & -0.797011 \\
    O & -1.033646 & 1.913637 & -0.763651 \\
    O & 0.786937 & 0.084521 & 1.786903 \\
    O & -1.583526 & -1.307189 & 0.935992 \\
    O & 2.632297 & -0.059854 & -0.266290 \\
    C & -0.929797 & 0.953424 & 0.282483 \\
    C & 0.562802 & 0.815658 & 0.590416 \\
    C & -1.521933 & -0.392587 & -0.160779 \\
    C & 1.295848 & 0.130636 & -0.567033 \\
    C & -0.644357 & -1.025906 & -1.228654 \\
    H & -1.439041 & 1.303396 & 1.190083 \\
    H & 0.985432 & 1.809850 & 0.742263 \\
    H & -2.524787 & -0.229701 & -0.579548 \\
    H & 1.239969 & 0.731974 & -1.480284 \\
    H & -0.993675 & -2.035615 & -1.442460 \\
    H & -0.704522 & -0.421332 & -2.142350 \\
    H & -1.962097 & 2.124817 & -0.901292 \\
    H & 0.314563 & -0.755559 & 1.690460 \\
    H & -2.174787 & -0.944894 & 1.604386 \\
    H & 2.664225 & -0.359458 & 0.652598 \\
    \bottomrule
    \end{tabular}
\end{center}
\end{table}

\begin{center}
   \textbf{$\alpha$-C17}
\end{center}

\begin{table}[H]
\caption{Cartesian coordinates in {\AA} of the $\alpha$-C17 structure optimized at revDSD-PBEP86-D3(BJ)/jun-cc-pVTZ level of theory. }
\begin{center}
    \begin{tabular}{cD{.}{.}{8} D{.}{.}{8} D{.}{.}{8}}
    \toprule
    \multicolumn{1}{c}{Atom} & \multicolumn{1}{c}{X} & \multicolumn{1}{c}{Y} & \multicolumn{1}{c}{Z} \\
    \midrule
    O & 0.671650 & -1.273376 & -0.608408 \\
    O & -1.107387 & 1.739396 & -1.047223 \\
    O & 0.844352 & 0.356046 & 1.680688 \\
    O & -1.620013 & -1.108458 & 1.221391 \\
    O & 2.610852 & -0.110267 & -0.334026 \\
    C & -0.952118 & 0.968042 & 0.140740 \\
    C & 0.556896 & 0.919244 & 0.397090 \\
    C & -1.528847 & -0.445741 & -0.026956 \\
    C & 1.261670 & 0.017698 & -0.619473 \\
    C & -0.688029 & -1.229786 & -1.025778 \\
    H & -1.446627 & 1.451201 & 0.992690 \\
    H & 0.970597 & 1.928338 & 0.311651 \\
    H & -2.553860 & -0.366803 & -0.404522 \\
    H & 1.177906 & 0.455430 & -1.619489 \\
    H & -1.030887 & -2.262737 & -1.069781 \\
    H & -0.765770 & -0.777005 & -2.022666 \\
    H & -2.047447 & 1.858570 & -1.213975 \\
    H & 0.650464 & 1.012641 & 2.355950 \\
    H & -0.722614 & -1.158861 & 1.576272 \\
    H & 2.675185 & -0.344256 & 0.600760 \\
    \bottomrule
    \end{tabular}
\end{center}
\end{table}

\begin{center}
   \textbf{$\alpha$-C18}
\end{center}

\begin{table}[H]
\caption{Cartesian coordinates in {\AA} of the $\alpha$-C18 structure optimized at revDSD-PBEP86-D3(BJ)/jun-cc-pVTZ level of theory. }
\begin{center}
    \begin{tabular}{cD{.}{.}{8} D{.}{.}{8} D{.}{.}{8}}
    \toprule
    \multicolumn{1}{c}{Atom} & \multicolumn{1}{c}{X} & \multicolumn{1}{c}{Y} & \multicolumn{1}{c}{Z} \\
    \midrule
    O & 0.727543 & -1.234101 & -0.627242 \\
    O & -1.277330 & 1.776951 & -0.987162 \\
    O & 0.693725 & 0.417849 & 1.740875 \\
    O & -1.593504 & -1.157026 & 1.196954 \\
    O & 2.580175 & -0.012142 & -0.163103 \\
    C & -0.989196 & 0.953982 & 0.137997 \\
    C & 0.515140 & 0.945576 & 0.432338 \\
    C & -1.514404 & -0.478425 & -0.041149 \\
    C & 1.250093 & 0.075116 & -0.582983 \\
    C & -0.639470 & -1.241613 & -1.031828 \\
    H & -1.503606 & 1.410115 & 0.984303 \\
    H & 0.899910 & 1.968883 & 0.365448 \\
    H & -2.530934 & -0.422545 & -0.439712 \\
    H & 1.182539 & 0.525435 & -1.587882 \\
    H & -0.942849 & -2.286442 & -1.065294 \\
    H & -0.731997 & -0.821307 & -2.046481 \\
    H & -1.002716 & 1.327223 & -1.791770 \\
    H & 1.618325 & 0.145931 & 1.806809 \\
    H & -0.765195 & -0.983813 & 1.666673 \\
    H & 2.998658 & -0.723552 & -0.660921 \\
    \bottomrule
    \end{tabular}
\end{center}
\end{table}

\begin{center}
   \textbf{$\alpha$-C19}
\end{center}

\begin{table}[H]
\caption{Cartesian coordinates in {\AA} of the $\alpha$-C19 structure optimized at revDSD-PBEP86-D3(BJ)/jun-cc-pVTZ level of theory. }
\begin{center}
    \begin{tabular}{cD{.}{.}{8} D{.}{.}{8} D{.}{.}{8}}
    \toprule
    \multicolumn{1}{c}{Atom} & \multicolumn{1}{c}{X} & \multicolumn{1}{c}{Y} & \multicolumn{1}{c}{Z} \\
    \midrule
    O & 0.657653 & -1.278471 & -0.605168 \\
    O & -1.216459 & 1.659350 & -1.080877 \\
    O & 0.860911 & 0.371456 & 1.680199 \\
    O & -1.606542 & -1.102944 & 1.227133 \\
    O & 2.606356 & -0.128041 & -0.334150 \\
    C & -0.954846 & 0.969580 & 0.138538 \\
    C & 0.558374 & 0.917788 & 0.395357 \\
    C & -1.535842 & -0.439120 & -0.022361 \\
    C & 1.256864 & 0.006038 & -0.618085 \\
    C & -0.703406 & -1.223159 & -1.026515 \\
    H & -1.457540 & 1.456600 & 0.983558 \\
    H & 0.985429 & 1.924518 & 0.299072 \\
    H & -2.560379 & -0.346213 & -0.388214 \\
    H & 1.175931 & 0.440980 & -1.619776 \\
    H & -1.051995 & -2.253855 & -1.075282 \\
    H & -0.777318 & -0.764309 & -2.020054 \\
    H & -0.977108 & 2.585008 & -0.973228 \\
    H & 0.545376 & 0.975593 & 2.358971 \\
    H & -0.698096 & -1.215921 & 1.535376 \\
    H & 2.673491 & -0.359971 & 0.600881 \\
    \bottomrule
    \end{tabular}
\end{center}
\end{table}

\begin{center}
   \textbf{$\alpha$-C20}
\end{center}

\begin{table}[H]
\caption{Cartesian coordinates in {\AA} of the $\alpha$-C20 structure optimized at revDSD-PBEP86-D3(BJ)/jun-cc-pVTZ level of theory. }
\begin{center}
    \begin{tabular}{cD{.}{.}{8} D{.}{.}{8} D{.}{.}{8}}
    \toprule
    \multicolumn{1}{c}{Atom} & \multicolumn{1}{c}{X} & \multicolumn{1}{c}{Y} & \multicolumn{1}{c}{Z} \\
    \midrule
    O & 0.673605 & -1.210460 & -0.740942 \\
    O & -1.244528 & 1.879805 & -0.830556 \\
    O & 0.844178 & 0.300413 & 1.753508 \\
    O & -1.508484 & -1.126728 & 1.162943 \\
    O & 2.609525 & -0.103687 & -0.285129 \\
    C & -0.948393 & 0.971591 & 0.222720 \\
    C & 0.559176 & 0.881320 & 0.490011 \\
    C & -1.526396 & -0.414663 & -0.078314 \\
    C & 1.278077 & 0.074103 & -0.600699 \\
    C & -0.681814 & -1.115615 & -1.133738 \\
    H & -1.437706 & 1.373232 & 1.111043 \\
    H & 0.974292 & 1.891285 & 0.504559 \\
    H & -2.552781 & -0.297104 & -0.442831 \\
    H & 1.241448 & 0.588619 & -1.572586 \\
    H & -1.030934 & -2.138335 & -1.294902 \\
    H & -0.778439 & -0.583090 & -2.092822 \\
    H & -0.710621 & 1.664469 & -1.601272 \\
    H & 0.246915 & -0.455843 & 1.853946 \\
    H & -1.664758 & -2.060656 & 0.989278 \\
    H & 2.634318 & -0.277746 & 0.667116 \\
    \bottomrule
    \end{tabular}
\end{center}
\end{table}

\begin{center}
   \textbf{$\alpha$-C21}
\end{center}

\begin{table}[H]
\caption{Cartesian coordinates in {\AA} of the $\alpha$-C21 structure optimized at revDSD-PBEP86-D3(BJ)/jun-cc-pVTZ level of theory. }
\begin{center}
    \begin{tabular}{cD{.}{.}{8} D{.}{.}{8} D{.}{.}{8}}
    \toprule
    \multicolumn{1}{c}{Atom} & \multicolumn{1}{c}{X} & \multicolumn{1}{c}{Y} & \multicolumn{1}{c}{Z} \\
    \midrule
    O & 0.701749 & -1.163660 & -0.777028 \\
    O & -1.191833 & 1.943417 & -0.743905 \\
    O & 0.788590 & 0.140622 & 1.782995 \\
    O & -1.559437 & -1.287937 & 0.970542 \\
    O & 2.629035 & -0.072016 & -0.248803 \\
    C & -0.940499 & 0.961835 & 0.254741 \\
    C & 0.555884 & 0.836347 & 0.566627 \\
    C & -1.523402 & -0.399524 & -0.151710 \\
    C & 1.297044 & 0.111783 & -0.563754 \\
    C & -0.645419 & -1.057969 & -1.203514 \\
    H & -1.450672 & 1.323867 & 1.150819 \\
    H & 0.978721 & 1.835284 & 0.687702 \\
    H & -2.527306 & -0.248294 & -0.563091 \\
    H & 1.265620 & 0.684894 & -1.502209 \\
    H & -0.991941 & -2.072654 & -1.395551 \\
    H & -0.709325 & -0.488435 & -2.143381 \\
    H & -0.688031 & 1.731412 & -1.535715 \\
    H & 0.276533 & -0.680683 & 1.735044 \\
    H & -2.211341 & -0.956722 & 1.597165 \\
    H & 2.651262 & -0.326913 & 0.684479 \\
    \bottomrule
    \end{tabular}
\end{center}
\end{table}

\begin{center}
   \textbf{$\alpha$-C22}
\end{center}

\begin{table}[H]
\caption{Cartesian coordinates in {\AA} of the $\alpha$-C22 structure optimized at revDSD-PBEP86-D3(BJ)/jun-cc-pVTZ level of theory. }
\begin{center}
    \begin{tabular}{cD{.}{.}{8} D{.}{.}{8} D{.}{.}{8}}
    \toprule
    \multicolumn{1}{c}{Atom} & \multicolumn{1}{c}{X} & \multicolumn{1}{c}{Y} & \multicolumn{1}{c}{Z} \\
    \midrule
    O & -0.753184 & 0.627476 & 1.292478 \\
    O & 0.943918 & 2.025116 & -0.558302 \\
    O & -0.362353 & -1.280275 & -1.303876 \\
    O & 2.076892 & -1.358692 & 0.202497 \\
    O & -2.398671 & -0.527988 & 0.098208 \\
    C & 0.943304 & 0.614061 & -0.733194 \\
    C & -0.456718 & 0.125135 & -1.097403 \\
    C & 1.441773 & -0.132080 & 0.517985 \\
    C & -1.415834 & 0.483061 & 0.056837 \\
    C & 0.269526 & -0.361228 & 1.491140 \\
    H & 1.630639 & 0.401030 & -1.552352 \\
    H & -0.808371 & 0.620008 & -2.007660 \\
    H & 2.207748 & 0.492115 & 0.981479 \\
    H & -1.876716 & 1.455929 & -0.125622 \\
    H & -0.143214 & -1.360869 & 1.336553 \\
    H & 0.580249 & -0.269010 & 2.530063 \\
    H & 0.496578 & 2.209746 & 0.275911 \\
    H & -1.240304 & -1.650651 & -1.148498 \\
    H & 1.444786 & -1.857107 & -0.332298 \\
    H & -3.036515 & -0.299988 & 0.782204 \\
    \bottomrule
    \end{tabular}
\end{center}
\end{table}

\begin{center}
   \textbf{$\alpha$-C23}
\end{center}

\begin{table}[H]
\caption{Cartesian coordinates in {\AA} of the $\alpha$-C23 structure optimized at revDSD-PBEP86-D3(BJ)/jun-cc-pVTZ level of theory. }
\begin{center}
    \begin{tabular}{cD{.}{.}{8} D{.}{.}{8} D{.}{.}{8}}
    \toprule
    \multicolumn{1}{c}{Atom} & \multicolumn{1}{c}{X} & \multicolumn{1}{c}{Y} & \multicolumn{1}{c}{Z} \\
    \midrule
    O & 0.670206 & -1.290957 & -0.573630 \\
    O & -1.248779 & 1.765073 & -1.038622 \\
    O & 0.853388 & 0.416446 & 1.665792 \\
    O & -1.600964 & -1.077337 & 1.258954 \\
    O & 2.608905 & -0.121126 & -0.328676 \\
    C & -0.961203 & 0.973985 & 0.109784 \\
    C & 0.553350 & 0.936157 & 0.365411 \\
    C & -1.532812 & -0.445948 & -0.007067 \\
    C & 1.261368 & -0.001281 & -0.615839 \\
    C & -0.691317 & -1.262955 & -0.980761 \\
    H & -1.454391 & 1.478364 & 0.943738 \\
    H & 0.964339 & 1.942933 & 0.244178 \\
    H & -2.558368 & -0.375514 & -0.376973 \\
    H & 1.189249 & 0.399678 & -1.635901 \\
    H & -1.033918 & -2.296633 & -0.991696 \\
    H & -0.772278 & -0.863960 & -2.005157 \\
    H & -0.982531 & 1.287339 & -1.830192 \\
    H & 0.635440 & 1.085200 & 2.321892 \\
    H & -0.701298 & -1.102859 & 1.610366 \\
    H & 2.675390 & -0.291098 & 0.620040 \\
    \bottomrule
    \end{tabular}
\end{center}
\end{table}

\begin{center}
   \textbf{$\alpha$-C24}
\end{center}

\begin{table}[H]
\caption{Cartesian coordinates in {\AA} of the $\alpha$-C24 structure optimized at revDSD-PBEP86-D3(BJ)/jun-cc-pVTZ level of theory. }
\begin{center}
    \begin{tabular}{cD{.}{.}{8} D{.}{.}{8} D{.}{.}{8}}
    \toprule
    \multicolumn{1}{c}{Atom} & \multicolumn{1}{c}{X} & \multicolumn{1}{c}{Y} & \multicolumn{1}{c}{Z} \\
    \midrule
    O & -1.088057 & -1.114682 & -1.006980 \\
    O & 1.069811 & 2.192695 & -0.186584 \\
    O & 2.371431 & -0.568911 & 0.078284 \\
    O & -1.727418 & 0.651243 & 1.290020 \\
    O & -0.058084 & -1.716292 & 1.005202 \\
    C & 0.499951 & 0.990378 & 0.315841 \\
    C & 1.088613 & -0.220773 & -0.414796 \\
    C & -1.027135 & 1.069159 & 0.131069 \\
    C & 0.128109 & -1.398947 & -0.377633 \\
    C & -1.448462 & 0.276413 & -1.094850 \\
    H & 0.698670 & 0.893025 & 1.388823 \\
    H & 1.235034 & 0.053396 & -1.462156 \\
    H & -1.286214 & 2.117719 & -0.025971 \\
    H & 0.555003 & -2.265717 & -0.886934 \\
    H & -2.531785 & 0.290976 & -1.201717 \\
    H & -0.994938 & 0.714822 & -1.991283 \\
    H & 2.027841 & 2.110565 & -0.116079 \\
    H & 2.242634 & -0.943270 & 0.958180 \\
    H & -1.436239 & -0.249655 & 1.484111 \\
    H & -0.497924 & -2.571662 & 1.055706 \\
    \bottomrule
    \end{tabular}
\end{center}
\end{table}

\begin{center}
   \textbf{$\alpha$-C25}
\end{center}

\begin{table}[H]
\caption{Cartesian coordinates in {\AA} of the $\alpha$-C25 structure optimized at revDSD-PBEP86-D3(BJ)/jun-cc-pVTZ level of theory. }
\begin{center}
    \begin{tabular}{cD{.}{.}{8} D{.}{.}{8} D{.}{.}{8}}
    \toprule
    \multicolumn{1}{c}{Atom} & \multicolumn{1}{c}{X} & \multicolumn{1}{c}{Y} & \multicolumn{1}{c}{Z} \\
    \midrule
    O & -0.790993 & 0.024305 & 1.447163 \\
    O & 0.895588 & 2.088336 & 0.117790 \\
    O & -0.402033 & -0.745253 & -1.634717 \\
    O & 2.140669 & -1.317813 & -0.229978 \\
    O & -2.473095 & -0.406979 & -0.109825 \\
    C & 0.944880 & 0.804252 & -0.489386 \\
    C & -0.469001 & 0.502705 & -0.953943 \\
    C & 1.426739 & -0.307543 & 0.465449 \\
    C & -1.408382 & 0.452279 & 0.264768 \\
    C & 0.251673 & -0.925850 & 1.241855 \\
    H & 1.612660 & 0.802687 & -1.357682 \\
    H & -0.824786 & 1.286854 & -1.629096 \\
    H & 2.142752 & 0.135081 & 1.162170 \\
    H & -1.784019 & 1.452526 & 0.485219 \\
    H & -0.121730 & -1.794780 & 0.692804 \\
    H & 0.563411 & -1.252032 & 2.232630 \\
    H & 1.788269 & 2.337679 & 0.374016 \\
    H & -1.286539 & -1.130973 & -1.589790 \\
    H & 1.540384 & -1.650141 & -0.910203 \\
    H & -3.066949 & -0.482725 & 0.644015 \\
    \bottomrule
    \end{tabular}
\end{center}
\end{table}

\begin{center}
   \textbf{$\alpha$-C26}
\end{center}

\begin{table}[H]
\caption{Cartesian coordinates in {\AA} of the $\alpha$-C26 structure optimized at revDSD-PBEP86-D3(BJ)/jun-cc-pVTZ level of theory. }
\begin{center}
    \begin{tabular}{cD{.}{.}{8} D{.}{.}{8} D{.}{.}{8}}
    \toprule
    \multicolumn{1}{c}{Atom} & \multicolumn{1}{c}{X} & \multicolumn{1}{c}{Y} & \multicolumn{1}{c}{Z} \\
    \midrule
    O & -0.725601 & 0.897405 & 1.163568 \\
    O & 0.960730 & 1.918497 & -0.819225 \\
    O & -0.502816 & -1.442105 & -1.194406 \\
    O & 1.819136 & -1.436052 & 0.399461 \\
    O & -2.397446 & -0.490824 & 0.309225 \\
    C & 0.923732 & 0.502161 & -0.815357 \\
    C & -0.484394 & -0.017485 & -1.106615 \\
    C & 1.414401 & -0.063261 & 0.528992 \\
    C & -1.453529 & 0.479676 & 0.006297 \\
    C & 0.264416 & -0.047369 & 1.549162 \\
    H & 1.601375 & 0.175385 & -1.608244 \\
    H & -0.831394 & 0.364750 & -2.067153 \\
    H & 2.254819 & 0.548138 & 0.868033 \\
    H & -1.970493 & 1.386750 & -0.306656 \\
    H & -0.168224 & -1.050426 & 1.618928 \\
    H & 0.607417 & 0.244146 & 2.539974 \\
    H & 0.427970 & 2.209263 & -0.067183 \\
    H & 0.253877 & -1.791444 & -0.698854 \\
    H & 2.669146 & -1.465004 & -0.052396 \\
    H & -2.064271 & -1.319253 & -0.070316 \\
    \bottomrule
    \end{tabular}
\end{center}
\end{table}

\begin{center}
   \textbf{$\alpha$-C27}
\end{center}

\begin{table}[H]
\caption{Cartesian coordinates in {\AA} of the $\alpha$-C27 structure optimized at revDSD-PBEP86-D3(BJ)/jun-cc-pVTZ level of theory. }
\begin{center}
    \begin{tabular}{cD{.}{.}{8} D{.}{.}{8} D{.}{.}{8}}
    \toprule
    \multicolumn{1}{c}{Atom} & \multicolumn{1}{c}{X} & \multicolumn{1}{c}{Y} & \multicolumn{1}{c}{Z} \\
    \midrule
    O & -1.103581 & -1.099843 & -0.999306 \\
    O & 1.079001 & 2.194725 & -0.184261 \\
    O & 2.357262 & -0.586032 & 0.097855 \\
    O & -1.590847 & 0.539532 & 1.319210 \\
    O & -0.059592 & -1.848119 & 0.959417 \\
    C & 0.512877 & 0.990603 & 0.313495 \\
    C & 1.087157 & -0.227383 & -0.417428 \\
    C & -1.004631 & 1.075380 & 0.125505 \\
    C & 0.122891 & -1.410341 & -0.368868 \\
    C & -1.429003 & 0.284936 & -1.100029 \\
    H & 0.711723 & 0.890125 & 1.385990 \\
    H & 1.248393 & 0.049144 & -1.463080 \\
    H & -1.279369 & 2.127131 & 0.000330 \\
    H & 0.552002 & -2.260458 & -0.899754 \\
    H & -2.512104 & 0.324330 & -1.237699 \\
    H & -0.962447 & 0.734312 & -1.985417 \\
    H & 2.037368 & 2.100720 & -0.136298 \\
    H & 2.190918 & -1.080794 & 0.911063 \\
    H & -2.537831 & 0.435920 & 1.175979 \\
    H & -0.642348 & -1.201697 & 1.389501 \\
    \bottomrule
    \end{tabular}
\end{center}
\end{table}

\begin{center}
   \textbf{$\alpha$-C28}
\end{center}

\begin{table}[H]
\caption{Cartesian coordinates in {\AA} of the $\alpha$-C28 structure optimized at revDSD-PBEP86-D3(BJ)/jun-cc-pVTZ level of theory. }
\begin{center}
    \begin{tabular}{cD{.}{.}{8} D{.}{.}{8} D{.}{.}{8}}
    \toprule
    \multicolumn{1}{c}{Atom} & \multicolumn{1}{c}{X} & \multicolumn{1}{c}{Y} & \multicolumn{1}{c}{Z} \\
    \midrule
    O & -0.792343 & -0.067142 & 1.445870 \\
    O & 1.001105 & 2.030512 & 0.332547 \\
    O & -0.408780 & -0.630367 & -1.688652 \\
    O & 2.140265 & -1.307232 & -0.303187 \\
    O & -2.473078 & -0.400725 & -0.136731 \\
    C & 0.951121 & 0.827682 & -0.429764 \\
    C & -0.467549 & 0.563430 & -0.917671 \\
    C & 1.426438 & -0.339626 & 0.451547 \\
    C & -1.408788 & 0.433198 & 0.294402 \\
    C & 0.248215 & -1.006272 & 1.180244 \\
    H & 1.627206 & 0.877663 & -1.288582 \\
    H & -0.827073 & 1.393118 & -1.538674 \\
    H & 2.135738 & 0.070266 & 1.170644 \\
    H & -1.788766 & 1.415760 & 0.579193 \\
    H & -0.129153 & -1.835988 & 0.575676 \\
    H & 0.557470 & -1.396827 & 2.147966 \\
    H & 0.997970 & 2.775155 & -0.276208 \\
    H & -1.291808 & -1.021711 & -1.659702 \\
    H & 1.539626 & -1.605096 & -0.998431 \\
    H & -3.055186 & -0.543173 & 0.616788 \\
    \bottomrule
    \end{tabular}
\end{center}
\end{table}

\begin{center}
   \textbf{$\alpha$-C29}
\end{center}

\begin{table}[H]
\caption{Cartesian coordinates in {\AA} of the $\alpha$-C29 structure optimized at revDSD-PBEP86-D3(BJ)/jun-cc-pVTZ level of theory. }
\begin{center}
    \begin{tabular}{cD{.}{.}{8} D{.}{.}{8} D{.}{.}{8}}
    \toprule
    \multicolumn{1}{c}{Atom} & \multicolumn{1}{c}{X} & \multicolumn{1}{c}{Y} & \multicolumn{1}{c}{Z} \\
    \midrule
    O & -1.093325 & -1.133332 & -0.989039 \\
    O & 1.182316 & 2.105700 & -0.265334 \\
    O & 2.297673 & -0.677050 & 0.158540 \\
    O & -1.715142 & 0.645990 & 1.295326 \\
    O & -0.165004 & -1.738187 & 1.002189 \\
    C & 0.500718 & 0.981458 & 0.294033 \\
    C & 1.079311 & -0.233148 & -0.423942 \\
    C & -1.034517 & 1.054604 & 0.124669 \\
    C & 0.116529 & -1.409246 & -0.360367 \\
    C & -1.460113 & 0.250001 & -1.091784 \\
    H & 0.712617 & 0.898456 & 1.364016 \\
    H & 1.230781 & 0.030194 & -1.477595 \\
    H & -1.310981 & 2.099132 & -0.043114 \\
    H & 0.549744 & -2.282263 & -0.855239 \\
    H & -2.544376 & 0.258435 & -1.191542 \\
    H & -1.015339 & 0.685039 & -1.995028 \\
    H & 0.980023 & 2.880110 & 0.269426 \\
    H & 2.911242 & 0.065464 & 0.153264 \\
    H & -1.463399 & -0.275392 & 1.459601 \\
    H & 0.685981 & -1.846141 & 1.447102 \\
    \bottomrule
    \end{tabular}
\end{center}
\end{table}

\begin{center}
   \textbf{$\alpha$-C30}
\end{center}

\begin{table}[H]
\caption{Cartesian coordinates in {\AA} of the $\alpha$-C30 structure optimized at revDSD-PBEP86-D3(BJ)/jun-cc-pVTZ level of theory. }
\begin{center}
    \begin{tabular}{cD{.}{.}{8} D{.}{.}{8} D{.}{.}{8}}
    \toprule
    \multicolumn{1}{c}{Atom} & \multicolumn{1}{c}{X} & \multicolumn{1}{c}{Y} & \multicolumn{1}{c}{Z} \\
    \midrule
    O & -0.716315 & 0.882139 & 1.186186 \\
    O & 0.960837 & 1.928242 & -0.796426 \\
    O & -0.521923 & -1.416455 & -1.219673 \\
    O & 1.866790 & -1.409476 & 0.228238 \\
    O & -2.387750 & -0.499623 & 0.324561 \\
    C & 0.921295 & 0.512400 & -0.814508 \\
    C & -0.492020 & 0.005891 & -1.103548 \\
    C & 1.421893 & -0.073184 & 0.513120 \\
    C & -1.452784 & 0.480964 & 0.026787 \\
    C & 0.276731 & -0.068152 & 1.547119 \\
    H & 1.596570 & 0.190100 & -1.608456 \\
    H & -0.842542 & 0.413250 & -2.052465 \\
    H & 2.265748 & 0.535357 & 0.848645 \\
    H & -1.976314 & 1.391464 & -0.264138 \\
    H & -0.164274 & -1.069646 & 1.607914 \\
    H & 0.626881 & 0.214224 & 2.539749 \\
    H & 0.430699 & 2.210785 & -0.039897 \\
    H & 0.265498 & -1.779322 & -0.786337 \\
    H & 2.190124 & -1.815472 & 1.038897 \\
    H & -2.056179 & -1.316872 & -0.080805 \\
    \bottomrule
    \end{tabular}
\end{center}
\end{table}

\begin{center}
   \textbf{$\alpha$-C31}
\end{center}

\begin{table}[H]
\caption{Cartesian coordinates in {\AA} of the $\alpha$-C31 structure optimized at revDSD-PBEP86-D3(BJ)/jun-cc-pVTZ level of theory. }
\begin{center}
    \begin{tabular}{cD{.}{.}{8} D{.}{.}{8} D{.}{.}{8}}
    \toprule
    \multicolumn{1}{c}{Atom} & \multicolumn{1}{c}{X} & \multicolumn{1}{c}{Y} & \multicolumn{1}{c}{Z} \\
    \midrule
    O & -0.297846 & -1.714178 & -0.104410 \\
    O & 0.972715 & 2.119832 & -0.087491 \\
    O & -1.751787 & 1.594591 & 0.286453 \\
    O & 2.587474 & -0.377542 & -0.643176 \\
    O & -2.185061 & -0.699944 & -0.855365 \\
    C & 0.486977 & 0.794455 & -0.223234 \\
    C & -0.798593 & 0.585175 & 0.567073 \\
    C & 1.550202 & -0.158569 & 0.303291 \\
    C & -1.336990 & -0.826006 & 0.254460 \\
    C & 0.902436 & -1.504327 & 0.636651 \\
    H & 0.299590 & 0.557804 & -1.277680 \\
    H & -0.575386 & 0.667350 & 1.634707 \\
    H & 1.966710 & 0.278986 & 1.219630 \\
    H & -1.888017 & -1.240339 & 1.110127 \\
    H & 1.578414 & -2.311188 & 0.360279 \\
    H & 0.692978 & -1.574047 & 1.711060 \\
    H & 0.217021 & 2.705399 & -0.220910 \\
    H & -2.264848 & 1.288888 & -0.473107 \\
    H & 2.938914 & 0.485514 & -0.889188 \\
    H & -2.393531 & -1.584815 & -1.172455 \\
    \bottomrule
    \end{tabular}
\end{center}
\end{table}

\begin{center}
   \textbf{$\alpha$-C32}
\end{center}

\begin{table}[H]
\caption{Cartesian coordinates in {\AA} of the $\alpha$-C32 structure optimized at revDSD-PBEP86-D3(BJ)/jun-cc-pVTZ level of theory. }
\begin{center}
    \begin{tabular}{cD{.}{.}{8} D{.}{.}{8} D{.}{.}{8}}
    \toprule
    \multicolumn{1}{c}{Atom} & \multicolumn{1}{c}{X} & \multicolumn{1}{c}{Y} & \multicolumn{1}{c}{Z} \\
    \midrule
    O & -1.023485 & -1.189862 & -0.993109 \\
    O & 0.956070 & 2.244818 & -0.220068 \\
    O & 2.388418 & -0.457148 & 0.124267 \\
    O & -1.734986 & 0.444398 & 1.227729 \\
    O & 0.030912 & -1.822184 & 0.995379 \\
    C & 0.456860 & 1.015952 & 0.289713 \\
    C & 1.105832 & -0.180389 & -0.411781 \\
    C & -1.067981 & 1.011097 & 0.092757 \\
    C & 0.206654 & -1.412696 & -0.346557 \\
    C & -1.446340 & 0.169556 & -1.108861 \\
    H & 0.666481 & 0.940099 & 1.363997 \\
    H & 1.261657 & 0.083830 & -1.461465 \\
    H & -1.398264 & 2.040525 & -0.075598 \\
    H & 0.692518 & -2.250159 & -0.848680 \\
    H & -2.529337 & 0.134904 & -1.221080 \\
    H & -1.013637 & 0.629850 & -2.005000 \\
    H & 1.918864 & 2.196941 & -0.194457 \\
    H & 2.239455 & -0.948428 & 0.942984 \\
    H & -1.635302 & 1.051098 & 1.968597 \\
    H & -0.668013 & -1.259953 & 1.365497 \\
    \bottomrule
    \end{tabular}
\end{center}
\end{table}

\begin{center}
   \textbf{$\alpha$-C33}
\end{center}

\begin{table}[H]
\caption{Cartesian coordinates in {\AA} of the $\alpha$-C33 structure optimized at revDSD-PBEP86-D3(BJ)/jun-cc-pVTZ level of theory. }
\begin{center}
    \begin{tabular}{cD{.}{.}{8} D{.}{.}{8} D{.}{.}{8}}
    \toprule
    \multicolumn{1}{c}{Atom} & \multicolumn{1}{c}{X} & \multicolumn{1}{c}{Y} & \multicolumn{1}{c}{Z} \\
    \midrule
    O & -0.213261 & -1.159374 & -1.215035 \\
    O & -2.008327 & 1.039934 & -0.676414 \\
    O & 1.396630 & 1.693665 & 0.303328 \\
    O & 0.131694 & -0.473867 & 1.907662 \\
    O & 1.890375 & -0.896907 & -0.183431 \\
    C & -0.900856 & 0.949823 & 0.205460 \\
    C & 0.441536 & 1.060688 & -0.536286 \\
    C & -0.969777 & -0.360829 & 0.989512 \\
    C & 0.923498 & -0.333980 & -1.024225 \\
    C & -0.873877 & -1.525997 & 0.003922 \\
    H & -0.978806 & 1.803850 & 0.882516 \\
    H & 0.262490 & 1.686857 & -1.416582 \\
    H & -1.911601 & -0.419628 & 1.541346 \\
    H & 1.386171 & -0.263106 & -2.006782 \\
    H & -0.357927 & -2.365039 & 0.478724 \\
    H & -1.867448 & -1.853577 & -0.297593 \\
    H & -1.858474 & 0.401579 & -1.384554 \\
    H & 2.193290 & 1.148256 & 0.284441 \\
    H & 0.310099 & 0.397823 & 2.280317 \\
    H & 1.522173 & -0.902853 & 0.718987 \\
    \bottomrule
    \end{tabular}
\end{center}
\end{table}

\begin{center}
   \textbf{$\alpha$-C34}
\end{center}

\begin{table}[H]
\caption{Cartesian coordinates in {\AA} of the $\alpha$-C34 structure optimized at revDSD-PBEP86-D3(BJ)/jun-cc-pVTZ level of theory. }
\begin{center}
    \begin{tabular}{cD{.}{.}{8} D{.}{.}{8} D{.}{.}{8}}
    \toprule
    \multicolumn{1}{c}{Atom} & \multicolumn{1}{c}{X} & \multicolumn{1}{c}{Y} & \multicolumn{1}{c}{Z} \\
    \midrule
    O & -0.907866 & -1.197205 & -1.073168 \\
    O & 0.796097 & 2.287734 & -0.287302 \\
    O & 2.418808 & -0.272402 & 0.145270 \\
    O & -1.729746 & 0.298731 & 1.359408 \\
    O & 0.179105 & -1.757329 & 0.921131 \\
    C & 0.367511 & 1.075488 & 0.325381 \\
    C & 1.133318 & -0.040641 & -0.394612 \\
    C & -1.162622 & 0.890367 & 0.204641 \\
    C & 0.330227 & -1.333596 & -0.440400 \\
    C & -1.489663 & 0.117766 & -1.061480 \\
    H & 0.626457 & 1.062604 & 1.389243 \\
    H & 1.290860 & 0.288090 & -1.424409 \\
    H & -1.614835 & 1.881825 & 0.119318 \\
    H & 0.880901 & -2.101410 & -0.987178 \\
    H & -2.564227 & -0.034937 & -1.150312 \\
    H & -1.133532 & 0.680969 & -1.931793 \\
    H & 0.518619 & 3.022720 & 0.268321 \\
    H & 2.292414 & -0.711870 & 0.994308 \\
    H & -1.286815 & -0.552300 & 1.484783 \\
    H & -0.133660 & -2.668236 & 0.913834 \\
    \bottomrule
    \end{tabular}
\end{center}
\end{table}

\begin{center}
   \textbf{$\alpha$-C35}
\end{center}

\begin{table}[H]
\caption{Cartesian coordinates in {\AA} of the $\alpha$-C35 structure optimized at revDSD-PBEP86-D3(BJ)/jun-cc-pVTZ level of theory. }
\begin{center}
    \begin{tabular}{cD{.}{.}{8} D{.}{.}{8} D{.}{.}{8}}
    \toprule
    \multicolumn{1}{c}{Atom} & \multicolumn{1}{c}{X} & \multicolumn{1}{c}{Y} & \multicolumn{1}{c}{Z} \\
    \midrule
    O & -1.219955 & -0.974755 & -1.018148 \\
    O & 1.546620 & 1.853390 & -0.226290 \\
    O & 2.203025 & -1.016228 & 0.160752 \\
    O & -1.624242 & 0.990374 & 1.220641 \\
    O & -0.433265 & -1.599079 & 1.086108 \\
    C & 0.654358 & 0.875025 & 0.311492 \\
    C & 1.035714 & -0.428855 & -0.376753 \\
    C & -0.835385 & 1.212962 & 0.065540 \\
    C & -0.107167 & -1.423880 & -0.286749 \\
    C & -1.360144 & 0.449966 & -1.142971 \\
    H & 0.806266 & 0.766300 & 1.388793 \\
    H & 1.175871 & -0.210177 & -1.443360 \\
    H & -0.912896 & 2.280655 & -0.156034 \\
    H & 0.183666 & -2.383446 & -0.717940 \\
    H & -2.426349 & 0.634854 & -1.267028 \\
    H & -0.834121 & 0.789479 & -2.043196 \\
    H & 1.498028 & 2.642622 & 0.322177 \\
    H & 2.906219 & -0.361955 & 0.085538 \\
    H & -1.485301 & 0.065823 & 1.471877 \\
    H & -1.013100 & -2.365082 & 1.151316 \\
    \bottomrule
    \end{tabular}
\end{center}
\end{table}

\begin{center}
   \textbf{$\alpha$-C36}
\end{center}

\begin{table}[H]
\caption{Cartesian coordinates in {\AA} of the $\alpha$-C36 structure optimized at revDSD-PBEP86-D3(BJ)/jun-cc-pVTZ level of theory. }
\begin{center}
    \begin{tabular}{cD{.}{.}{8} D{.}{.}{8} D{.}{.}{8}}
    \toprule
    \multicolumn{1}{c}{Atom} & \multicolumn{1}{c}{X} & \multicolumn{1}{c}{Y} & \multicolumn{1}{c}{Z} \\
    \midrule
    O & -0.696531 & 0.820720 & 1.223497 \\
    O & 0.920000 & 1.934354 & -0.823509 \\
    O & -0.485862 & -1.442183 & -1.067063 \\
    O & 2.086970 & -1.324824 & 0.332491 \\
    O & -2.458393 & -0.422080 & 0.312262 \\
    C & 0.913956 & 0.514036 & -0.820690 \\
    C & -0.502882 & 0.000387 & -1.072758 \\
    C & 1.451671 & -0.067660 & 0.500574 \\
    C & -1.441857 & 0.494180 & 0.058966 \\
    C & 0.303705 & -0.153855 & 1.523126 \\
    H & 1.576470 & 0.198540 & -1.628346 \\
    H & -0.883425 & 0.365619 & -2.027891 \\
    H & 2.229012 & 0.611782 & 0.853474 \\
    H & -1.911879 & 1.434089 & -0.227370 \\
    H & -0.128313 & -1.160050 & 1.509678 \\
    H & 0.646608 & 0.050487 & 2.535637 \\
    H & 0.474242 & 2.216466 & -0.014743 \\
    H & -0.581619 & -1.755764 & -1.969987 \\
    H & 1.426954 & -1.918537 & -0.044905 \\
    H & -2.125084 & -1.293054 & 0.057732 \\
    \bottomrule
    \end{tabular}
\end{center}
\end{table}

\begin{center}
   \textbf{$\alpha$-C37}
\end{center}

\begin{table}[H]
\caption{Cartesian coordinates in {\AA} of the $\alpha$-C37 structure optimized at revDSD-PBEP86-D3(BJ)/jun-cc-pVTZ level of theory. }
\begin{center}
    \begin{tabular}{cD{.}{.}{8} D{.}{.}{8} D{.}{.}{8}}
    \toprule
    \multicolumn{1}{c}{Atom} & \multicolumn{1}{c}{X} & \multicolumn{1}{c}{Y} & \multicolumn{1}{c}{Z} \\
    \midrule
    O & -1.154795 & -1.040821 & -1.026609 \\
    O & 1.315208 & 2.043062 & -0.189578 \\
    O & 2.289568 & -0.739444 & 0.223010 \\
    O & -1.685036 & 0.885179 & 1.224065 \\
    O & -0.333683 & -1.641819 & 1.074747 \\
    C & 0.596772 & 0.923381 & 0.316059 \\
    C & 1.059918 & -0.355376 & -0.372554 \\
    C & -0.905801 & 1.163226 & 0.074273 \\
    C & -0.021546 & -1.426446 & -0.293694 \\
    C & -1.383381 & 0.377570 & -1.138180 \\
    H & 0.757330 & 0.817507 & 1.392638 \\
    H & 1.202586 & -0.129824 & -1.436219 \\
    H & -1.036090 & 2.226219 & -0.137021 \\
    H & 0.319911 & -2.366637 & -0.738371 \\
    H & -2.459260 & 0.494602 & -1.258869 \\
    H & -0.881377 & 0.753583 & -2.037155 \\
    H & 2.249863 & 1.882975 & -0.017349 \\
    H & 2.718224 & -1.396071 & -0.334384 \\
    H & -1.475849 & -0.021750 & 1.486199 \\
    H & -0.921202 & -2.403985 & 1.120026 \\
    \bottomrule
    \end{tabular}
\end{center}
\end{table}

\begin{center}
   \textbf{$\alpha$-C38}
\end{center}

\begin{table}[H]
\caption{Cartesian coordinates in {\AA} of the $\alpha$-C38 structure optimized at revDSD-PBEP86-D3(BJ)/jun-cc-pVTZ level of theory. }
\begin{center}
    \begin{tabular}{cD{.}{.}{8} D{.}{.}{8} D{.}{.}{8}}
    \toprule
    \multicolumn{1}{c}{Atom} & \multicolumn{1}{c}{X} & \multicolumn{1}{c}{Y} & \multicolumn{1}{c}{Z} \\
    \midrule
    O & -0.225120 & -1.303217 & -1.058483 \\
    O & -2.025680 & 0.901439 & -0.796546 \\
    O & 1.426504 & 1.667549 & 0.124308 \\
    O & 0.135628 & -0.103995 & 1.997487 \\
    O & 1.849561 & -0.923504 & -0.150802 \\
    C & -0.929063 & 0.956510 & 0.102473 \\
    C & 0.416466 & 1.012405 & -0.636906 \\
    C & -0.935924 & -0.241116 & 1.055572 \\
    C & 0.868008 & -0.425997 & -1.041401 \\
    C & -0.830367 & -1.524408 & 0.226314 \\
    H & -1.046165 & 1.880367 & 0.671682 \\
    H & 0.274131 & 1.600941 & -1.544574 \\
    H & -1.863086 & -0.251957 & 1.630389 \\
    H & 1.254234 & -0.443253 & -2.062715 \\
    H & -0.271739 & -2.285664 & 0.777105 \\
    H & -1.820727 & -1.917759 & 0.001472 \\
    H & -1.872207 & 0.155794 & -1.389354 \\
    H & 1.227883 & 1.490206 & 1.054218 \\
    H & 0.871658 & -0.632295 & 1.658157 \\
    H & 2.424150 & -0.166911 & 0.039588 \\
    \bottomrule
    \end{tabular}%
\end{center}
\end{table}

\newpage
\subsection{$\beta$ Anomers}

\begin{center}
   \textbf{$\beta$-C1}
\end{center}

\begin{table}[H]
\caption{Cartesian coordinates in {\AA} of the $\beta$-C1 structure optimized at revDSD-PBEP86-D3(BJ)/jun-cc-pVTZ level of theory. }
\begin{center}
    %
\end{center}
\end{table}

\newpage

\begin{center}
   \textbf{$\beta$-C2}
\end{center}

\begin{table}[H]
\caption{Cartesian coordinates in {\AA} of the $\beta$-C2 structure optimized at revDSD-PBEP86-D3(BJ)/jun-cc-pVTZ level of theory. }
\begin{center}
    %
\end{center}
\end{table}

\begin{center}
   \textbf{$\beta$-C3}
\end{center}

\begin{table}[H]
\caption{Cartesian coordinates in {\AA} of the $\beta$-C3 structure optimized at revDSD-PBEP86-D3(BJ)/jun-cc-pVTZ level of theory. }
\begin{center}
    %
\end{center}
\end{table}

\begin{center}
   \textbf{$\beta$-C4}
\end{center}

\begin{table}[H]
\caption{Cartesian coordinates in {\AA} of the $\beta$-C4 structure optimized at revDSD-PBEP86-D3(BJ)/jun-cc-pVTZ level of theory. }
\begin{center}
    %
\end{center}
\end{table}

\begin{center}
   \textbf{$\beta$-C5}
\end{center}

\begin{table}[H]
\caption{Cartesian coordinates in {\AA} of the $\beta$-C5 structure optimized at revDSD-PBEP86-D3(BJ)/jun-cc-pVTZ level of theory. }
\begin{center}
    %
\end{center}
\end{table}

\section{Energy Table $\alpha$ anomer}

\begin{table*}[ht!]
\centering
\footnotesize
\caption{Zero-point corrected energies (in SI kcal mol$^{-1}$) relative to the most thermodynamically stable $\alpha$ conformer computed using the two different levels of theory discussed in Section 2.}
\label{tab:alfa_energies_SI}
%
\end{table*}

\newpage

\section{Energy Table $\beta$ anomer}

\begin{table}[ht!]
\centering
\footnotesize
\caption{Zero-point corrected energies (in SI kcal mol$^{-1}$) relative to the most thermodynamically stable $\beta$ conformer computed using the two different levels of theory discussed in Section 2.}
\label{tab:beta_energies_SI}
%%
\end{table}

\newpage

\section{Enthalpies and Free Energies for $\alpha$-anomer }

\begin{table}[h]
\small
\centering
%
\caption{Relative enthalpies and Free energies $\alpha$-type conformers expressed in kcal mol$^{-1}$.}
\label{tab:alpha_HG_SI}
\end{table}

\newpage

\section{Enthalpies and Entropies for $\beta$-anomer}

\begin{table}[h]
\scriptsize
\centering
%
\caption{Relative enthalpies and Free energies $\beta$-type conformers expressed in kcal mol$^{-1}$.}
\label{tab:beta_HG_SI}
\end{table}